\documentclass[11pt]{article}
\title{The two D6 R4 type invariants and their higher order generalisation}

\global\arraycolsep=1pt
\usepackage[colorlinks=true,backref=true,linkcolor=black,anchorcolor=black,citecolor=black,filecolor=black,menucolor=black,pagecolor=black,urlcolor=black]{hyperref}

\usepackage{xcolor}
\usepackage{tikz}
\usetikzlibrary{arrows}
\usepackage{pgfplots}
\usepackage{amsmath}
\usepackage{amsthm}
\usepackage{amssymb}
\usepackage{mathrsfs}
\usepackage[Symbol]{upgreek}
\usepackage{dsfont}
\usepackage[vcentermath]{youngtab}
\usepackage{textcomp}
\usepackage{bbold}

\usepackage{latexsym}
\usepackage{graphicx}

\newcommand{\eprint}[1]{{\href{http://arxiv.org/abs/#1}{\texttt{[#1}]}}}
\newcommand{\eprintN}[1]{{\href{http://arxiv.org/abs/#1}{\texttt{#1 [hep-th]}}}}

\def\DJo{$\;$\kern-.4em \hbox{D\kern-.8em\raise.15ex\hbox{--}\kern.35em okovi\'c}}

\def\DEVII#1#2#3#4#5#6#7{{\tiny $ { \left[ \begin{array}{ccccccc}  & & \mathfrak{#2} \hspace{-0.7mm}&&&& \vspace{ -1.5mm} \\ \mathfrak{#1}\hspace{-0.7mm} &  \mathfrak{#3} \hspace{-0.7mm}& \mathfrak{#4} \hspace{-0.7mm} & \mathfrak{#5}\hspace{-0.7mm}&\mathfrak{#6}\hspace{-0.7mm}& #7 \hspace{-0.8mm} \end{array}\right] }$}}
\def\DEVIII#1#2#3#4#5#6#7#8{{\tiny $ { \left[ \begin{array}{ccccccc}  & & \mathfrak{#2} \hspace{-0.7mm}&&&& \vspace{ -1.5mm} \\ \mathfrak{#1}\hspace{-0.7mm} &  \mathfrak{#3} \hspace{-0.7mm}& \mathfrak{#4} \hspace{-0.7mm} & \mathfrak{#5}\hspace{-0.7mm}&\mathfrak{#6}\hspace{-0.7mm}&\mathfrak{#7}\hspace{-0.7mm}&\mathfrak{#8} \end{array}\right] }$}}
\def\DSOXVI#1#2#3#4#5#6#7#8{{\tiny $ {  \vspace{-2mm} \left[ \begin{array}{ccccccccc}  && \mathfrak{#8} \hspace{-1.2mm}&&&&&& \vspace{ -1.3mm} \\ \cdot \hspace{-1.0mm}& \mathfrak{#7}\hspace{-1.2mm} &\mathfrak{#6}\hspace{-1.2mm} &  \mathfrak{#5} \hspace{-1.2mm}& \mathfrak{#4} \hspace{-1.2mm} & \mathfrak{#3}\hspace{-1.2mm}&\mathfrak{#2}\hspace{-1.2mm}&\mathfrak{#1}  \hspace{-1.2mm}\end{array}\right] }$}}

\def\DSOXII#1#2#3#4#5#6#7{{\tiny $ {   \left[ \begin{array}{ccccccc}  && & \mathfrak{#6} \hspace{-1.2mm}&&& \vspace{ -1.5mm} \\  \mathfrak{#1}\hspace{-1.2mm} &\mathfrak{#2}\hspace{-1.2mm} &  \mathfrak{#3} \hspace{-1.2mm}& \mathfrak{#4} \hspace{-1.2mm} & \mathfrak{#5}\hspace{-1.2mm}&\cdot \hspace{-0.5mm}& \mathfrak{#7} \end{array}\right] }$}}

\def\DSON#1#2#3#4#5#6#7#8{{\tiny $ {   \left[ \begin{array}{cccccccc}  &&&&& & \mathfrak{#7} \hspace{-1.2mm}& \vspace{ -1.5mm} \\  \mathfrak{#1}\hspace{-1.2mm} &\mathfrak{#2}\hspace{-1.2mm} &  \mathfrak{#3} \hspace{-1.2mm}& \mathfrak{#4} \hspace{-1.2mm} & \cdot \hspace{-0.8mm}\cdot\hspace{-0.8mm}\cdot \hspace{-1.2mm} & \mathfrak{#5}\hspace{-1.2mm}& \mathfrak{#6}\hspace{-1.2mm} & \mathfrak{#8} \end{array}\right] }$}}

\def\WSOXVI#1#2#3#4#5#6#7#8{{\tiny $ {   \biggl[ \begin{array}{ccc}  &&\mathfrak{#7}  \vspace{ -1.5mm} \\  \mathfrak{#1}\hspace{0.2mm}  \mathfrak{#2}\hspace{0.2mm}  \mathfrak{#3}\hspace{0.2mm}  \mathfrak{#4}\hspace{0.2mm}\mathfrak{#5}\hspace{-0.6mm} &\mathfrak{#6} \hspace{-0.9mm}&\vspace{-1.5mm}\\ && \mathfrak{#8}  \end{array}\biggr] }$}}
\def\WSOXII#1#2#3#4#5#6{{\tiny $ {   \biggl[ \begin{array}{ccc}  &&\mathfrak{#5}  \vspace{ -1.5mm} \\  \mathfrak{#1}\hspace{0.2mm}  \mathfrak{#2}\hspace{0.2mm}  \mathfrak{#3}\hspace{-0.6mm} &\mathfrak{#4} \hspace{-0.9mm}&\vspace{-1.5mm}\\ && \mathfrak{#6}  \end{array}\biggr] }$}}
\def\WSOXIIT#1#2#3#4#5#6{{\tiny $ {   \biggl[ \begin{array}{ccc}  &&{#5}  \vspace{ -1.5mm} \\  {#1}\hspace{0.2mm}  \mathfrak{#2}\hspace{0.2mm}  \mathfrak{#3}\hspace{-0.6mm} &\mathfrak{#4} \hspace{-0.9mm}&\vspace{-1.5mm}\\ && \mathfrak{#6}  \end{array}\biggr] }$}}

\def\DSON#1#2#3#4#5#6#7#8{{\tiny $ {   \left[ \begin{array}{cccccccc}  &&&&& & \mathfrak{#7} \hspace{-1.2mm}& \vspace{ -1.5mm} \\  \mathfrak{#1}\hspace{-1.2mm} &\mathfrak{#2}\hspace{-1.2mm} &  \mathfrak{#3} \hspace{-1.2mm}& \mathfrak{#4} \hspace{-1.2mm} & \cdot \hspace{-0.8mm}\cdot\hspace{-0.8mm}\cdot \hspace{-1.2mm} & \mathfrak{#5}\hspace{-1.2mm}& \mathfrak{#6}\hspace{-1.2mm} & \mathfrak{#8} \end{array}\right] }$}}

\def\SU{SU_{\scriptscriptstyle \rm c}(8)}
\def\Sp{Sp_{\scriptscriptstyle \rm c}(4)}

\newfont{\bbbold}{msbm10 scaled \magstep1}

\def\cD{{\cal D}}
\def\cE{{\cal E}}
\def\cF{{\cal F}}

\def\cL{{\cal L}}
\def\cM{{\cal M}}
\def\cN{{\cal N}}
\def\cO{{\cal O}}

\def\cV{{\cal V}}

\newfont{\goth}{eufm10 scaled \magstep1}

\def\gl{\mbox{\goth l}}

\def\adt{{\dot \alpha}}
\def\bdt{{\dot \beta}}
\def\cdt{\dot\gamma}

\def\be{\begin{equation}}\def\ee{\end{equation}}
\def\bea{\begin{eqnarray}}\def\eea{\end{eqnarray}}
\def\barr{\begin{array}}\def\earr{\end{array}}

\def\O{\Omega}

\def\hi{\hat{\imath}}\def\hj{\hat{\jmath}}
\def\hp{\hat{p}}\def\hq{\hat{q}}
\def\hk{\hat{k}}\def\hl{\hat{l}}

\def\nn{\nonumber}
\def\bd{\begin{document}}
\def\ed{\end{document}}
\def\ba{\begin{array}}
\def\ea{\end{array}}
\def\bea{\begin{eqnarray}}
\def\eea{\end{eqnarray}}
\def\ft#1#2{{\frac{\scriptstyle #1}{\scriptstyle #2}}}
\def\fft#1#2{\frac{#1}{#2}}
\def\sst#1{{\scriptscriptstyle #1}}
\def\oneone{\rlap 1\mkern4mu{\rm l}}
\def\DSOXVI#1#2#3#4#5#6#7#8{{\tiny $ {  \vspace{-2mm} \left[ \begin{array}{ccccccccc}  && \mathfrak{#8} \hspace{-0.7mm}&&&&&& \vspace{ -1.5mm} \\ \cdot \hspace{-0.5mm}& \mathfrak{#7}\hspace{-0.7mm} &\mathfrak{#6}\hspace{-0.7mm} &  \mathfrak{#5} \hspace{-0.7mm}& \mathfrak{#4} \hspace{-0.7mm} & \mathfrak{#3}\hspace{-0.7mm}&\mathfrak{#2}\hspace{-0.7mm}&\mathfrak{#1} \end{array}\right] }$}}

\def\DEVI#1#2#3#4#5#6{{\tiny $ { \left[ \begin{array}{cccccc}  & & \mathfrak{#2} \hspace{-0.7mm}&&& \vspace{ -1.5mm} \\ \mathfrak{#1}\hspace{-0.7mm} &  \mathfrak{#3} \hspace{-0.7mm}& \mathfrak{#4} \hspace{-0.7mm} & \mathfrak{#5}\hspace{-0.7mm}& #6 \end{array}\right] }$}}
\def\DSOXII#1#2#3#4#5#6#7{{\tiny $ {   \left[ \begin{array}{ccccccc}  && & \mathfrak{#6} \hspace{-0.7mm}&&& \vspace{ -1.5mm} \\  \mathfrak{#1}\hspace{-0.7mm} &\mathfrak{#2}\hspace{-0.7mm} &  \mathfrak{#3} \hspace{-0.7mm}& \mathfrak{#4} \hspace{-0.7mm} & \mathfrak{#5}\hspace{-0.7mm}&\cdot \hspace{-0.5mm}& \mathfrak{#7} \end{array}\right] }$}}

\def\DSON#1#2#3#4#5#6#7#8{{\tiny $ {   \left[ \begin{array}{cccccccc}  &&&&& & \mathfrak{#7} \hspace{-0.7mm}& \vspace{ -1.5mm} \\  \mathfrak{#1}\hspace{-0.7mm} &\mathfrak{#2}\hspace{-0.7mm} &  \mathfrak{#3} \hspace{-0.7mm}& \mathfrak{#4} \hspace{-0.7mm} & \cdot \hspace{-0.8mm}\cdot\hspace{-0.8mm}\cdot \hspace{-0.7mm} & \mathfrak{#5}\hspace{-0.7mm}& \mathfrak{#6}\hspace{-0.7mm} & \mathfrak{#8} \end{array}\right] }$}}

\def\DlacedLeft{{\fontsize{0.004pt}{0.0005pt}\selectfont  \scriptscriptstyle \mbox{$= \hspace{-2.2mm}  \langle$} \fontsize{12pt}{14.5pt}\selectfont }}

\def\DSpIV#1#2#3#4{{\tiny $ { \left[   \mathfrak{#1}\,  \mbox{-} \mathfrak{#2} \, \mbox{-} \mathfrak{#3} \hspace{-0.2mm}\DlacedLeft \hspace{0.2mm} \mathfrak{#4} \right] }$}}

\def\DSOX#1#2#3#4#5{{\tiny $ {   \biggl[ \begin{array}{ccc}  &&\mathfrak{#3}  \vspace{ -1.5mm} \\  \mathfrak{#1}\hspace{0.2mm}\mathfrak{#2}\hspace{-0.6mm} &\mathfrak{#4} \hspace{-0.9mm}&\vspace{-1.5mm}\\ && \mathfrak{#5}  \end{array}\biggr] }$}}

\def\DSOVIII#1#2#3#4{{\tiny $ {   \biggl[ \begin{array}{ccc}  &&\mathfrak{#2}  \vspace{ -1.5mm} \\  \mathfrak{#1}\hspace{-0.6mm} &\mathfrak{#4} \hspace{-0.9mm}&\vspace{-1.5mm}\\ && \mathfrak{#3}  \end{array}\biggr] }$}}

\def\Evector#1{E_{\scriptscriptstyle [#1\mathfrak{0}\mathfrak{0}\mathfrak{0}]}}
\def\Etensor#1{E_{\scriptscriptstyle [\mathfrak{0}\mathfrak{0}#1\mathfrak{0}]}}
\def\EA#1{E_{\scriptscriptstyle [#1\mathfrak{0}]}}

\def\EiEVII#1{{E_{\fontsize{6.35pt}{6pt}\selectfont   \left[ \begin{array}{cccccc}  & & \mathfrak{0} \hspace{-0.6mm}&&& \vspace{ -1.0mm} \\ #1 \hspace{-0.6mm} &  \mathfrak{0} \hspace{-0.6mm}& \mathfrak{0} \hspace{-0.6mm} & \mathfrak{0}\hspace{-0.6mm}&\mathfrak{0}\hspace{-0.6mm}&\mathfrak{0} \end{array}\right] \fontsize{12.35pt}{12pt}\selectfont }}}

\def\EiEVI#1{{E_{\fontsize{6.35pt}{6pt}\selectfont   \left[ \begin{array}{cccccc}  & & \mathfrak{0} \hspace{-0.6mm}&& \vspace{ -1.0mm} \\ #1 \hspace{-0.6mm} &  \mathfrak{0} \hspace{-0.6mm}& \mathfrak{0} \hspace{-0.6mm} & \mathfrak{0}\hspace{-0.6mm}&\mathfrak{0} \end{array}\right] \fontsize{12.35pt}{12pt}\selectfont }}}

\def\sp{{\mathfrak{sp}}}
\def\gl{{\mathfrak{gl}}}
\def\sl{{\mathfrak{sl}}}

\newcommand{\eq}[1]{(\ref{#1})}
\newcommand{\w}[1]{\\[0.#1cm]}
\def\eqs#1#2{(\ref{#1}-\ref{#2})}
\def\det{{\rm det\,}}
\def\tr{{\rm tr}}

\newcommand{\hoch}[1]{$\, ^{#1}$}
\newcommand{\imperial}{\it\small Theoretical Physics Group, Imperial College London\\ Prince Consort Road, London SW7 2AZ, UK}
\newcommand{\kings}
{\it\small Department of Mathematics, King's College, University of London\\ Strand, London WC2R 2LS, UK}
\newcommand{\uu}
{\it\small Department of Theoretical Physics, Uppsala, Sweden}
\newcommand{\hip}
{\it\small HIP-Helsinki Institute of Physics, P.O. Box 64 FIN-00014
University of Helsinki, Suomi-Finland}
\newcommand{\stock}
{\it\small Department of Theoretical Physics, Stockholm, Sweden}
\newcommand{\cpht}
{\it\small Centre de Physique Th{\'e}orique, Ecole Polytechnique, CNRS\\ 91128 Palaiseau Cedex, France}
\makeatletter
\renewcommand\theequation{\thesection.\arabic{equation}}
\@addtoreset{equation}{section} \makeatother

\newcommand{\sa}{/ \hspace{-1.2ex}}
\newcommand{\saa}{/ \hspace{-1.4ex}}
\newcommand{\saaa}{\, / \hspace{-1.6ex}}
\newcommand{\Scal}[1]{\Bigl ({#1} \Bigr )}
\newcommand{\scal}[1]{\bigl ({#1} \bigr )}

\newcommand{\CR}{\nonumber \\*}

\newcommand{\trace}{\hbox {tr}~}
\newcommand{\traceS}{\hbox {tr}_{\scriptscriptstyle \mathfrak{S}}~}

\DeclareMathAlphabet{\mathpzc}{OT1}{pzc}{m}{it}
\def\BRST{\,\mathpzc{s}\,}
\def\aBRST{{\scriptstyle (\mathpzc{s})}}
\def\q{{{\scriptscriptstyle (Q)}}}
\def\qs{{\scriptscriptstyle (Q\mathpzc{s})}}
\def\Qsla{{\mathcal{S}_{\q}}}
\def\Slav{{\mathcal{S}_\aBRST}}
\def\epsilonb{{\overline{\epsilon}}}
\def\bulletup{{\scriptstyle \bullet}}

\newcommand{\grad}[3]{{\scriptscriptstyle (#1 , #2, #3 )}}
\newcommand{\gra}[2]{{\scriptscriptstyle (#1 , #2 )}}
\newcommand{\ord}[1]{{\scriptscriptstyle (#1)}}

\def\cL{{\cal L}}
\def\cN{\mathcal{N}}
\def\cO{\mathcal{O}}

\def\ie{{\it i.e.}\ }
\def\eg{{\it e.g.}\ }

\newcommand{\sfrac}[2]{{\scriptstyle \frac{#1}{#2}}}
\newcommand{\stfrac}[2]{{\scriptscriptstyle \frac{#1}{#2}}}

 \def\balpha{{\overline{\alpha}}}
 \def\bbeta{{\overline{\beta}}}
 \def\bgamma{{\overline{\gamma}}}
 \def\bdelta{{\overline{\delta}}}
 \def\bepsilon{{\overline{\epsilon}}}
 \def\bvarepsilon{{\overline{\varepsilon}}}
 \def\bzeta{{\overline{\zeta}}}
 \def\bareta{{\overline{\eta}}}
 \def\btheta{{\overline{\theta}}}
 \def\bvartheta{{\overline{\vartheta}}}
 \def\biota{{\overline{\iota}}}
 \def\bkappa{{\overline{\kappa}}}
 \def\blambda{{\overline{\lambda}}}
 \def\bmu{{\overline{\mu}}}
 \def\bnu{{\overline{\nu}}}
 \def\bxi{{\overline{\xi}}}
 \def\bpi{{\overline{\pi}}}
 \def\brho{{\overline{\rho}}}
 \def\bvarrho{{\overline{\varrho}}}
 \def\bsigma{{\overline{\sigma}}}
 \def\bvarsigma{{\overline{\varsigma}}}
 \def\btau{{\overline{\tau}}}
 \def\bphi{{\overline{\phi}}}
 \def\bvarphi{{\overline{\varphi}}}
 \def\bchi{{\overline{\chi}}}
 \def\bpsi{{\overline{\psi}}}
 \def\bomega{{\overline{\omega}}}

\def\thalf{{\textrm{\tiny\textonehalf}}}
\def\tquarter{{\textrm{\tiny\textonequarter}}}
\def\Ko{{\scriptscriptstyle K}}
\def\tKo{\scriptscriptstyle k }
\def\N{{\mathcal{N}}}
\def\csN{{\fontsize{9.35pt}{9pt}\selectfont \mbox{$\cN$} \fontsize{12.35pt}{12pt}\selectfont }}
\def\cssN{{\fontsize{6.35pt}{6pt}\selectfont \mbox{$\cN$} \fontsize{12.35pt}{12pt}\selectfont }}
\def\csssN{{\fontsize{4.35pt}{4pt}\selectfont \mbox{$\cN$} \fontsize{12.35pt}{12pt}\selectfont }}

\def\ai{{\hat{\imath}}}
\def\aj{{\hat{\jmath}}}
\def\ak{{\hat{k}}}

\def\inv{{ \scriptscriptstyle \rm{\mbox{\tiny-1}}}}
\def\un{{\mathpzc{1}}}
\def\deux{{\mathpzc{2}}}
\def\trois{{\mathpzc{3}}}
\def\quatre{{\mathpzc{4}}}
\def\cinq{{\mathpzc{5}}}

\colorlet{rouge}{red!70!black}
\newcommand{\red}[1]{ {\color{red} #1 }} 
\newcommand{\blue}[1]{{\color{blue} #1 }}
\newcommand{\green}[1]{{\color{green} #1 }}
\newcommand{\bleu}[1]{ {\color{cyan} #1 }} 

 \def\xshift{- 1}
  \def\xmin{1}
 \def\ymin{0}

\renewcommand{\thefootnote}{\arabic{footnote}}

\begin{document}

\renewcommand{\thefootnote}{\arabic{footnote}}
\setcounter{footnote}{0}

\allowdisplaybreaks[1]
\renewcommand{\thefootnote}{\fnsymbol{footnote}}
\def\corr{$\spadesuit $}
\def\trefle{ $\clubsuit$}
\begin{titlepage}
\begin{flushright}
CPHT-RR007.0315\\
\end{flushright}

\bigskip
\bigskip
\centerline{\Large The two $\nabla^6 R^4$ type invariants and their higher order generalisation}
\centerline{\Large \bf }
\bigskip
\bigskip
\centerline{{\bf Guillaume Bossard and Valentin Verschinin}}
\bigskip
\centerline{Centre de Physique Th\'eorique, Ecole Polytechnique, CNRS}
\centerline{91128 Palaiseau cedex, France \footnote{email: bossard@cpht.polytechnique.fr,  valentin.verschinin@cpht.polytechnique.fr}}
\bigskip
\bigskip

\begin{abstract}
We show that there are two distinct classes of $\nabla^6 R^4$ type supersymmetry invariants in maximal supergravity. The second class includes a coupling in $F^{2} \nabla^4 R^4$ that generalises to 1/8 BPS protected  $F^{2k} \nabla^4 R^4$ couplings. We work out the supersymmetry constraints on the corresponding threshold functions, and argue that the functions in the second class satisfy to homogeneous differential equations for arbitrary $k\ge1$, such that the corresponding exact threshold functions in type II string theory should be proportional to Eisenstein series, which we identify. This analysis explains in particular that the exact $\nabla^6 R^4$ threshold function is the sum of an Eisenstein function and a solution to an inhomogeneous Poisson equation in string theory. 
\end{abstract}

\end{titlepage}
\renewcommand{\thefootnote}{\arabic{footnote}}
\setcounter{footnote}{0}

\tableofcontents

\section{Introduction}
The determination of the exact string theory low energy effective action is a very difficult problem in general. In the case of type II string theory on $\mathds{R}^{1,10-d} \times T^{d-1}$, the lowest order non-perturbative corrections could nonetheless have been computed  \cite{Green:1997tv,Green:1997as,Kiritsis:1997em}. Although there is no non-perturbative formulation of the theory,  the constraints following from supersymmetry and $U$-duality have permitted to determine the non-perturbative low energy effective action from perturbative computations in string theory \cite{D'Hoker:2005jc,Green:2008uj,Gomez:2013sla,Green:2014yxa,D'Hoker:2014gfa} and in eleven-dimensional supergravity \cite{Green:1997as,Green:1999pu,Green:2005ba,Green:2008bf,Basu:2014uba}. The four-graviton amplitude allows in particular to determine the $\nabla^{2k}R^4$ type correction in the effective action,
\be \cL\sim  \frac{1}{\kappa^2}  R +\sum_{p,q}  \kappa^{2 \frac{d-3+4p+6q}{9-d}}  E_{(p,q)} \nabla^{4p+6q} R^4 +\dots    \ee 
where the dots stand for other terms including the supersymmetric completion, $(p,q)$ labels the different invariant combinations of derivatives compatible with supersymmetry according to the notations used in \cite{Green:2010wi}, and $E_{(p,q)}$ are automorphic functions of the scalar fields defined on $E_{d(d)}(\mathds{Z}) \backslash E_{d(d)}/ K_d$. For $(p,q) = (0,0)$, $(1,0)$ and $(0,1)$, the complete effective action at this order is determined by these functions $\cE_\gra{p}{q}$, which have been extensively studied  \cite{Review,Obers:1999um,Basu:2007ck,Green:2010kv,Green:2011vz,Fleig:2013psa,Minimal,D4R4,Gustafsson:2014iva,Pioline:2015yea}. 

$E_{(0,0)}$ is an Eisenstein series associated to the minimal unitary representation \cite{Obers:1999um,Green:2010kv}, $E_{(1,0)}$ is an (or a sum of two) Eisenstein series associated to the next to minimal unitary representation(s) \cite{Green:2010kv}, and both are therefore relatively well understood.  They are nonetheless very complicated functions, and the explicit expansion of $E_{(1,0)}$ in Fourier modes is not yet determined \cite{Green:2011vz,Fleig:2013psa,Gustafsson:2014iva}. $E_{(0,1)}$ is not even an Eisenstein series, and was shown in \cite{Green:2005ba} to satisfy to an inhomogeneous Poisson equation in type IIB. A proposal for this function in eight dimensions \cite{Basu:2007ck}, suggested a split of the function into the sum of an Eisenstein series and an inhomogeneous solution, which was subsequently generalised in seven and six dimensions \cite{Green:2010wi,Green:2010kv}, and  recently clarified in \cite{Pioline:2015yea}. 

In this paper we extend the analysis carried out in \cite{Minimal,D4R4} to the study of $E_{(0,1)}$. We show that this function indeed splits into the sum of two functions that are associated to two distinct supersymmetry invariants, and therefore satisfy to inequivalent tensorial differential equations. In particular, the second satisfies to a homogeneous equation, which is solved by the Eisenstein function appearing in \cite{Green:2010wi,Basu:2007ck,Pioline:2015yea}. One can distinguish the two functions by looking at specific higher point couplings that we identify. The new class of invariants generalises to an infinite class admitting a coupling in $F^{2k} \nabla^4 R^4$, and we identify a unique Eisenstein function solving the corresponding tensorial differential equations in all dimensions greater than four. This function turns out to be compatible with perturbative string theory, and only admits three perturbative contributions in four dimensions, at 1-loop, $(k+2)$-loop, and $2k$-loop.  However, the only amplitude that seems to unambiguously distinguish it from others is the $(k+2)$-loop four-graviton amplitude in a non-trivial Ramond--Ramond background, which makes an explicit check extremely challenging.

We start with the analysis of the supersymmetry invariants in four dimensions. The two $\nabla^6 R^4$ type invariants in the linear approximation are associated to two distinct classes of chiral primary operators of $SU(2,2|8)$ discussed in \cite{Drummond:2003ex}. We identify the corresponding representations of $E_{7(7)}$ associated to nilpotent coadjoint orbits \cite{E7Djo} that are summarised in figure \ref{ClosureDiag}. 
\begin{figure}[htbp]
\begin{center}
 \begin{tikzpicture}
 \draw (\xmin,\ymin) node{\textbullet};
 \draw (\xmin,\ymin - 1) node{\textbullet};
  \draw (\xmin,\ymin - 2)  node{\textbullet};
  \draw (\xmin,\ymin + 2)  node{\textbullet};
  \draw (\xmin,\ymin + 4)  node{\textbullet};
  \draw (\xmin + 1,\ymin + 2.5)  node{\textbullet};
   \draw (\xmin - 1,\ymin + 3.5)  node{\textbullet};
    \draw (\xmin - 1 ,\ymin + 1) node{\textbullet};
   
  \draw (\xmin + 0.3,\ymin - 2) node{$R$};
  \draw (\xmin + 0.4,\ymin - 1) node{$R^4$};
  \draw (\xmin + 0.7,\ymin) node{$ \nabla^4 R^4$};
  \draw (\xmin - 1.9,\ymin + 1) node{$ {F}^{2k} \nabla^4 R^4$};
  \draw (\xmin + 0.7 + 1,\ymin + 2.5) node{$ \nabla^6 R^4$};
  \draw[-,draw=black,very thick](\xmin,\ymin) -- (\xmin,\ymin + 2 );
   \draw[-,draw=black,very thick](\xmin - 1,\ymin + 1) -- (\xmin - 1,\ymin + 3.5);
    \draw[-,draw=black,very thick](\xmin - 1,\ymin + 3.5) -- (\xmin,\ymin + 2);
    \draw[-,draw=black,very thick](\xmin,\ymin + 2) -- (\xmin + 1,\ymin + 2.5);
    \draw[-,draw=black,very thick](\xmin + 1,\ymin + 2.5) -- (\xmin,\ymin + 4);
    \draw[-,draw=black,very thick](\xmin - 1,\ymin + 3.5) -- (\xmin,\ymin + 4);
     \draw[dashed,draw=black,very thick](\xmin,\ymin + 4) -- (\xmin,\ymin + 4.5);
  \draw[-,draw=black,very thick](\xmin,\ymin) -- (\xmin - 1,\ymin + 1);
\draw[-,draw=black,very thick] (\xmin,\ymin - 1) -- (\xmin,\ymin);
\draw[-,draw=black,very thick] (\xmin,\ymin - 2) -- (\xmin,\ymin - 1);
\draw[<-,draw=black,thick] (\xmin - 3-1,\ymin + 5) -- (\xmin - 3 - 1,\ymin - 2);
\draw (\xmin - 3 + 0.2 - 1,\ymin - 2) node{$0$};
\draw (\xmin - 3- 1,\ymin - 2) node{-};
\draw (\xmin - 3- 1,\ymin - 1) node{-};
\draw (\xmin - 3- 1,\ymin) node{-};
\draw (\xmin - 3- 1,\ymin + 2) node{-};
\draw (\xmin - 3- 1,\ymin + 1) node{-};
\draw (\xmin - 3- 1,\ymin + 2.5) node{-};
\draw (\xmin - 3- 1 ,\ymin + 3.5) node{-};
\draw (\xmin - 3 - 1,\ymin + 4) node{-};
\draw (\xmin - 3 + 0.3 - 1,\ymin - 1) node{$34$};
\draw (\xmin - 3 + 0.3 - 1,\ymin) node{$52$};
\draw (\xmin - 3 + 0.3 - 1,\ymin + 1) node{$54$};
\draw (\xmin - 3 + 0.3 - 1,\ymin + 2) node{$64$};
\draw (\xmin - 3 + 0.3 - 1,\ymin + 2.5) node{$66$};
\draw (\xmin - 3 + 0.3 - 1,\ymin + 3.5) node{$70$};
\draw (\xmin - 3 + 0.3 - 1,\ymin + 4) node{$76$};
\draw (\xmin - 3 + 0.5 - 1,\ymin + 5) node{dim};
\end{tikzpicture}
\end{center}
\caption{\small Closure diagram of nilpotent orbits  of $E_{7(7)}$ of  dimension smaller than 76.}
\label{ClosureDiag}
\end{figure}
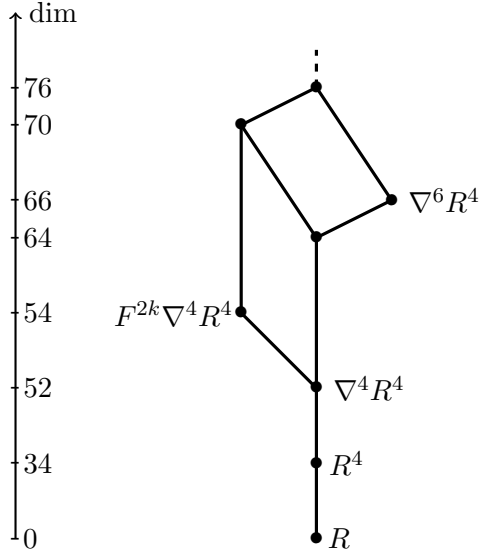
In the linearised approximation, the $F^2 \nabla^4 R^4$ type invariant does not carry a $\nabla^6 R^4$ coupling, but we explain that the structure of the linearised invariant allows for this mixing at the non-linear level, and that the latter must occur because the two classes of invariants merge in one single $E_{8(8)}$ representation in three dimensions. We conclude that the exact threshold function in four dimensions takes the form
\be E_{\gra{0}{1}} = \hat{\cE}_{\grad{8}{1}{1}}  + \frac{32}{189\pi}\hat{{E}}_{\mbox{\DEVII000000{5}}}\ ,  \ee
where $\hat{\cE}_{\grad{8}{1}{1}}$ is the solution to the inhomogeneous differential equation \eqref{E811Equation} that is consistent with perturbative string theory. The explicit relation between the tensorial differential equations and the associated nilpotent orbits permits us to determine the wavefront set of the associated functions, extending the results of \cite{Green:2011vz,Fleig:2013psa} to the $\nabla^6 R^4$ threshold function. It appears, as can be seen in figure \ref{ClosureDiag}, that the two functions admit distinct wavefront sets. In particular we show that although $\hat{\cE}_{\grad{8}{1}{1}}$ is not an Eisenstein series, it admits the same wavefront set as $\hat{{E}}{\mbox{\DEVII{\mathnormal{6}}00000{\mathfrak{0}}}}$.

\begin{figure}[htbp]
\center
 \begin{tikzpicture}
\draw (\xmin - 1,\ymin + 5) node{$IIA$}; \draw (\xmin,\ymin + 5) node{$IIB$};   \draw (\xmin + 2,\ymin + 5) node{$IIA$}; \draw (\xmin + 3,\ymin + 5) node{$IIB$};
 \draw (\xmin + 5,\ymin + 5) node{$IIA$}; \draw (\xmin + 6,\ymin + 5) node{$IIB$};
 
\draw[-,draw=black, thick](\xmin - 1,\ymin + 4) -- (\xmin - 1,\ymin + 4.7); \draw[-,draw=black, thick](\xmin,\ymin + 4) -- (\xmin,\ymin + 4.7);
\draw[-,draw=black, thick](\xmin + 2,\ymin + 4) -- (\xmin + 2,\ymin + 4.7); \draw[-,draw=black, thick](\xmin + 3,\ymin + 4) -- (\xmin + 3,\ymin + 4.7);
\draw[-,draw=black, thick](\xmin + 5,\ymin + 4) -- (\xmin  + 5,\ymin + 4.7); \draw[-,draw=black, thick](\xmin + 6,\ymin + 4) -- (\xmin + 6,\ymin + 4.7);
\draw[dashed,draw=black, thick](\xmin + 7,\ymin + 4) -- (\xmin + 7,\ymin + 4.5);
\draw[-,draw=black, thick](\xmin,\ymin + 4) -- (\xmin - 0.9,\ymin + 4.7);
\draw[-,draw=black, thick](\xmin + 3,\ymin + 4) -- (\xmin + 2.1,\ymin + 4.7);

\draw[dashed,draw=black, thick](\xmin + 2.5,\ymin - 1) -- (\xmin + 2.5,\ymin - 1.3);
\draw[dashed,draw=black, thick](\xmin - 0.5,\ymin - 1) -- (\xmin - 0.5,\ymin - 1.3);
\draw[dashed,draw=black, thick](\xmin + 3 + \xshift + 7.5/2,\ymin - 1) -- (\xmin + \xshift  + 3 + 7.5/2 ,\ymin - 1.3);

\draw[<-,draw=black,thick] (\xmin - 3,\ymin + 6) -- (\xmin - 3,\ymin - 2);               
\draw (\xmin - 3,\ymin + 5) node{-};
\draw (\xmin - 3,\ymin + 4) node{-};   \draw (\xmin + \xshift,\ymin + 4) node{\color{rouge}  \textbullet};\draw (\xmin + 1 + \xshift,\ymin + 4) node{ \textbullet};  \draw (\xmin + \xshift + 3,\ymin + 4) node{\textbullet};\draw (\xmin + \xshift + 4,\ymin + 4) node{$\circ$};
\draw (\xmin - 3,\ymin + 3) node{-};                \draw (\xmin + \xshift + 0.5,\ymin + 3) node{\textbullet};   \draw (\xmin + \xshift + 3,\ymin + 3) node{\textbullet};\draw (\xmin + 4 + \xshift ,\ymin + 3) node{$\circ$};
\draw (\xmin - 3,\ymin + 2) node{-};                \draw (\xmin + \xshift + 0.5,\ymin + 2) node{\textbullet};   \draw (\xmin + \xshift + 3,\ymin + 2) node{\textbullet};\draw (\xmin + 4 + \xshift,\ymin + 2) node{\color{rouge} \textbullet};
\draw (\xmin - 3,\ymin + 1) node{-};                \draw (\xmin + \xshift + 0.5,\ymin + 1) node{\textbullet};   \draw (\xmin + \xshift + 3.5,\ymin + 1) node{\textbullet};
\draw (\xmin - 3,\ymin) node{-};			    \draw (\xmin + \xshift + 0.5,\ymin) node{\textbullet};         \draw (\xmin + \xshift + 3.5,\ymin) node{\textbullet};
\draw (\xmin - 3,\ymin - 1) node{-};                 \draw (\xmin + \xshift + 0.5 ,\ymin - 1) node{\textbullet};   \draw (\xmin + \xshift + 3.5,\ymin - 1) node{\textbullet};
\draw (\xmin - 3 + 0.3,\ymin + 5) node{$10$};
\draw (\xmin - 3 + 0.3,\ymin + 4) node{$8$};
\draw (\xmin - 3 + 0.3,\ymin + 3) node{$7$};
\draw (\xmin - 3 + 0.3,\ymin + 2) node{$6$};
\draw (\xmin - 3 + 0.3,\ymin + 1) node{$5$};
\draw (\xmin - 3 + 0.3,\ymin) node{$4$};
\draw (\xmin - 3 + 0.3,\ymin - 1) node{$3$};
\draw (\xmin - 3 + 0.5,\ymin + 6) node{dim};

\draw (\xmin + \xshift + 6/2 + 7.5/2 ,\ymin - 1) node{\textbullet}; 
\draw (\xmin + \xshift + 6 ,\ymin ) node{\color{rouge}  \textbullet}; 
\draw (\xmin + \xshift + 7.5 ,\ymin ) node{ \textbullet}; 
\draw (\xmin + \xshift + 6 ,\ymin +1) node{$\circ$}; 
\draw (\xmin + \xshift + 7.5 ,\ymin +1) node{\textbullet}; 
\draw (\xmin + \xshift + 6 ,\ymin +2) node{$\circ$}; 
\draw (\xmin + \xshift + 7.5 ,\ymin +2) node{\color{rouge} \textbullet}; 
\draw (\xmin + \xshift + 6 ,\ymin +3) node{$\circ$}; 
\draw (\xmin + \xshift + 7.5 ,\ymin +3) node{$\circ$}; 
\draw (\xmin + \xshift + 6 ,\ymin +4) node{$\circ$}; 
\draw (\xmin + \xshift + 7 ,\ymin +4) node{$\circ$};
\draw (\xmin + \xshift + 8 ,\ymin +4) node{$\circ$}; 

\draw (\xmin + 0.5 + \xshift,\ymin - 1.6) node{$R^4$};
\draw (\xmin + 3.5 + \xshift,\ymin - 1.6) node{$\nabla^4 R^4$};
\draw (\xmin + 6/2 + 7.5/2 + \xshift,\ymin - 1.6) node{$\nabla^6 R^4$};

\draw[-,draw=black, thick](\xmin + 8 + \xshift,\ymin  + 4) -- (\xmin + 0.5 + 7 + \xshift,\ymin + 3);
\draw[-,draw=black, thick](\xmin + 0.5 + 6.5 + \xshift,\ymin  + 4) -- (\xmin + 0.5 + 7 \xshift,\ymin + 3);
\draw[-,draw=black, thick](\xmin + 0.5 + \xshift,\ymin -1) -- (\xmin + 0.5 + \xshift,\ymin + 3);
\draw[-,draw=black, thick](\xmin + 3.5 + \xshift ,\ymin -1) -- (\xmin + 3.5 + \xshift,\ymin + 1);
\draw[-,draw=black, thick](\xmin + 3 + 7.5/2 + \xshift ,\ymin -1) -- (\xmin + 6 + \xshift,\ymin);
\draw[-,draw=black, thick](\xmin + 3 + 7.5/2 + \xshift ,\ymin -1) -- (\xmin + 7.5 + \xshift,\ymin);
\draw[-,draw=black, thick](\xmin + 6 + \xshift,\ymin) -- (\xmin + 6 + \xshift,\ymin + 4);
\draw[-,draw=black, thick](\xmin + 7.5 + \xshift,\ymin) -- (\xmin + 7.5 + \xshift,\ymin + 3);

\draw[-,draw=black, thick](\xmin + 3.5 + \xshift,\ymin + 1) -- (\xmin + \xshift + 4,\ymin + 2);  \draw[-,draw=black, thick](\xmin + \xshift + 3.5,\ymin + 1) -- (\xmin + \xshift + 3,\ymin + 2);
\draw[-,draw=black, thick](\xmin + \xshift + 0.5,\ymin + 3) -- (\xmin + \xshift ,\ymin + 4); \draw[-,draw=black, thick](\xmin + \xshift + 0.5,\ymin + 3) -- (\xmin + \xshift + 1,\ymin + 4);

\draw[-,draw=black, thick](\xmin + \xshift + 4,\ymin + 4) -- (\xmin + \xshift + 4,\ymin + 2);  \draw[-,draw=black, thick](\xmin + \xshift + 3,\ymin + 4) -- (\xmin + \xshift + 3,\ymin + 2);

\end{tikzpicture}

\caption{\label{DimensionMultiplets}\small Each node corresponds to an inequivalent supersymmetry invariant, white if it cannot be written in harmonic superspace in the linearised approximation, and red if the corresponding harmonic superspace is chiral. For $\nabla^6 R^4$, the links to 10 dimensions are valid for the homogeneous solution, while all the eight-dimensional invariants uplift to type IIA for the inhomogeneous solution.}
\end{figure}
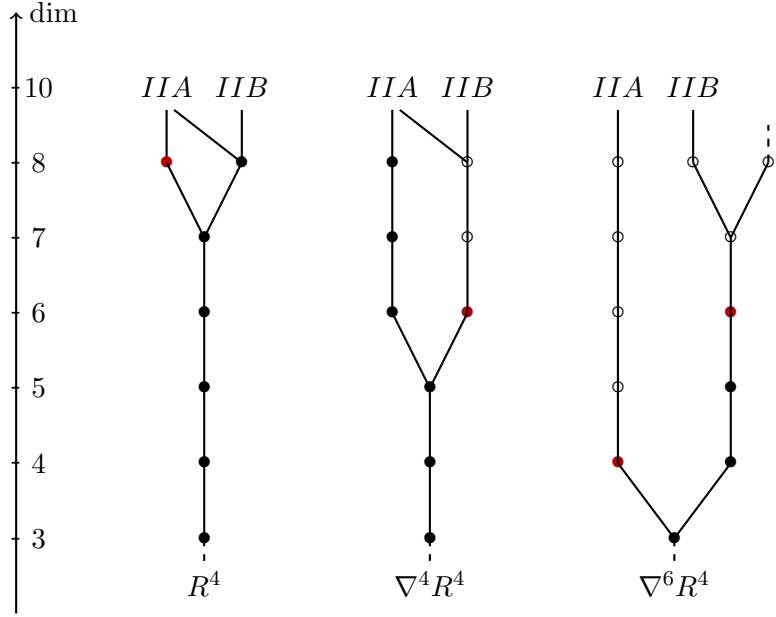

We then consider the uplift of our results in higher dimensions, and exhibit that this general structure extends to all dimensions lower than eight, and is in perfect agreement with the exact threshold functions proposed in \cite{Green:2010wi,Basu:2007ck,Pioline:2015yea}. In each dimension, the supersymmetry invariants transform in irreducible representations of $E_{d(d)}$, defined by the representation of $E_{d(d)}$ on the associated function on $E_{d(d)}/ K_d$ satisfying to the relevant differential equations implied by supersymmetry. The inequivalent invariants are summarised in figure \ref{DimensionMultiplets}.
The tensorial differential equations satisfied by Eisenstein functions relevant to our analysis are reviewed in the appendices. 

%
\section{$\cN=8$ supergravity in four dimensions}
Maximal supergravity includes 70 scalar fields parametrising the symmetric space $E_{7(7)} / SU_{\scriptscriptstyle \rm c}(8)$ \cite{Cremmer:1979up}, and can be defined in superspace by promoting these fields to superfields $\phi^\upmu$ \cite{Brink:1979nt,Howe:1981gz}. One defines the Maurer--Cartan form 
\be d \cV \,  \cV^{-1} = \left( \begin{array}{cc}\  2 \delta_{[i}^{[k} \omega^{l]}{}_{j]} \ &\  P_{ijkl} \ \\ \ P^{ijkl}\ & -2 \delta^{[i}_{[k} \omega^{j]}{}_{l]} \ \end{array}\right) \ , \ee
with  
\be P^{ijkl} = \frac{1}{24} \varepsilon^{ijklpqrs} P_{pqrs} \ . \label{ComplexSelfual}\ee
The metric on $E_{7(7)} / SU_{\scriptscriptstyle \rm c}(8)$ is defined as
\be G_{\upmu\upnu}(\phi) d\phi^\upmu d\phi^\upnu = \frac{1}{3} P_{ijkl} P^{ijkl} \ , \ee
and the derivative in tangent frame is defined such that for any function 
\be d \cE = 3 P^{ijkl} \cD_{ijkl} \cE \ . \ee
The superfields satisfy to
\be D_\alpha^i \cE = \frac{1}{4} \varepsilon^{ijklpqrs} \chi_{\alpha jkl} \,  \cD_{pqrs} \cE \ , \qquad \bar D_{\adt i} \cE = 6 \bar \chi_{\adt}^{jkl} \,  \cD_{ijkl} \cE \ , \ee
where $\chi_{\alpha ijk}$ is the Dirac superfield in Weyl components, and $\bar \chi_\adt^{ijk}$ its complex conjugate. The expansion of the scalar fields include the 28 Maxwell field strengths $F_{\alpha\beta ij}$, the 8 Rarita--Schwinger field strengths $\rho_{\alpha\beta\gamma i}$ and the Weyl tensor $C_{\alpha\beta\gamma\delta}$, satisfying to $\cN=8$ supergravity classical (two derivatives) field equations. The supervielbeins are the solutions to the Bianchi identities defined such that the Riemann tensor is valued in $\mathfrak{sl}(2,\mathds{C}) \oplus \mathfrak{su}(8)$ and the $\mathfrak{su}(8)$ component is identified with the scalar field curvature \cite{Brink:1979nt,Howe:1981gz},
\be R^i{}_j = \frac{1}{3} P_{jklp} \wedge P^{iklp} \ . \ee
The covariant derivative on $E_{7(7)} / SU_{\scriptscriptstyle \rm c}(8)$ in tangent frame satisfies to 
\be [ \cD^{ijkl} , \cD_{pqrs} ]  \cD_{tuvw}= - 24 \delta^{ijkl}_{qrs][t} \cD_{uvw][p} + 3 \delta^{ijkl}_{pqrs} \cD_{tuvw} \ ,  \label{Comut} \ee
and the Laplace operator is
\be \Delta = \frac{1}{3} \cD^{ijkl} \cD_{ijkl} \ . \ee
In the linearised approximation, the scalar superfield $W_{ijkl}$ satisfies to the reality constraint \eqref{ComplexSelfual} and to
\be D_\alpha^p W_{ijkl} = 2 \delta^p_{[i} \chi_{\alpha jkl]} \ , \qquad \bar D_{\adt p } W_{ijkl} = \frac{1}{12} \varepsilon_{ijklpqrs} \bar \chi_\adt^{qrs} \ . \ee 
In this approximation the superfield $W^{ijkl}$ transforms in the minimal unitary representation of the superconformal group $SU(2,2|8)$ \cite{Gunaydin:1984vz}. This property permits a complete classification of supersymmetry invariants in the linearised approximation in terms of irreducible representations of $SU(2,2|8)$ of Lorentz invariant top component \cite{Drummond:2003ex,Drummond:2010fp}. In our analysis, we rely on the assumption of absence of supersymmetry anomaly, such that there is no algebraic obstruction to the extension of a linearised invariant to a full non-linear invariant. This implies a bijective correspondence between the set of linearised invariants and the non-linear invariants, such that one can deduce the explicit gradient expansion of the functions (or tensor functions) of the scalar fields on $E_{7(7)} / SU_{\scriptscriptstyle \rm c}(8)$ that determine the invariants.
\subsection{The standard $\nabla^6 R^4$ type invariant}\label{811D6R4}
One can define a $\nabla^6 R^4$ type invariant in harmonic superspace, using the harmonic variables $u^1{}_i,\, u^r{}_i,\, u^8{}_i$ parametrising $SU(8)/ S(U(1) \times U(6)\times U(1))$, such that $r=2$ to $7$ of $SU(6)$ \cite{Drummond:2003ex,Drummond:2010fp,Hartwell:1994rp}. In this case the harmonic superspace integral can be defined at the non-linear level \cite{Bossard:2011tq}, but we will only consider its linearised approximation. The superfield  in the ${\bf 20}$ of $SU(6)$
\be W_{rst} = u^i{}_8 u^j{}_r u^k{}_s u^l{}_t W_{ijkl} \ , \ee
satisfies to the G-analyticity constraints 
\be u^1{}_i D_\alpha^i W_{rst} = 0   \ , \qquad u^i{}_8 \bar D_{\adt i} W_{rst} = 0 \  . \ee
One can therefore integrate any function of $W_{rst}$ on the associated analytic superspace. To understand the most general integrand, we must decompose monomials of $W_{rst}$ in irreducible representations of $SU(6)$. At quadratic order we have the representation $[0,0,2,0,0]$ and the combination 
\be W^{rtu} W_{stu} = \frac{1}{6} \varepsilon^{rtuvwx} W_{stu} W_{vwx} \ee
in the $[1,0,0,0,1]$. Because one obtains the $[0,0,2,0,0]$ by simply adding the Dynkin labels of $W_{rst}$, we will say that this representation is freely generated, whereas we shall consider the $[1,0,0,0,1]$ as a new generator at order two. At cubic order, we have the two elements freely generated by the ones already discussed, \ie $[0,0,3,0,0]$ and $[1,0,1,0,1]$, and the additional combination 
\be W_{u[rs} W_{t]vw} W^{uvw} \ , \ee
in the $[0,0,1,0,0]$. At quartic order we have the four elements freely generated by the ones already discussed, and the two additional elements 
\be W_{vw[r} W^{vw[t} W_{s]xy} W^{u]xy}  \ , \quad  W_{urs} W_{tvw} W^{uvw} W^{rst} \ , \ee
that decompose into the $[0,1,0,1,0]$ and the singlet representation. One checks that these elements freely generate the general polynomials in $W_{rst}$, such that the latter are labeled by five integers. 

To integrate such a function in analytic superspace, one needs to consider these generating monomials with additional harmonic variables in order to compensate for the $S(U(1)\times U(6)\times U(1))$ representation, \ie 
\bea \label{Uintegral811}  \int du\,  u^8{}_i u^r{}_j u^s{}_k u^t{}_l W_{rst} &=& W_{ijkl} \, , \\
\int du\,  u^8{}_i u^s{}_j u_1{}^k u_r{}^l  W^{rtu} W_{stu}  &=& W_{ijpq} W^{klpq} - \frac{1}{28} \delta_{ij}^{kl} W_{pqrs} W^{pqrs}  \, , \CR
\int du\,  u_1{}^q u^8{}_p u^8{}_i u^r{}_j u^s{}_k u^t{}_l W_{u[rs} W_{t]vw} W^{uvw}   &=& W_{po[ij} W_{kl]mn} W^{qomn} - \frac{|W|^2\hspace{-1.7mm}}{108} \scal{ \delta_p^q W_{ijkl} - \delta^p_{[i} W_{jkl]p}}    \, , \CR
\int du u_1{}^k u_1{}^l u^8{}_i u^8{}_j  W_{urs} W_{tvw} W^{uvw} W^{rst} &=& W_{npq(i} W_{j)mp^\prime q^\prime} W^{np^\prime q^\prime(k} W^{l)pqm}  - \delta_{(i}^{(k} \delta_{j)}^{l)} ( \dots ) \ ,\nn 
\eea
which are respectively in the $[0,0,0,1,0,0,0]$, the $[0,1,0,0,0,1,0]$, the $[1,0,0,1,0,0,1]$ and the $[2,0,0,0,0,0,2]$ irreducible representations of $SU(8)$, whereas  
\be \int du u_1{}^m u_1{}^n u^8{}_k u^8{}_l u^r{}_i u^s{}_j  u_t{}^p u_u{}^q  W_{vwr} W^{vwt} W_{sxy} W^{uxy}  = W_{i^\prime j^\prime [ij} W_{k]lk^\prime l^\prime} W^{i^\prime j^\prime [pq} W^{m]nk^\prime l^\prime} + \dots \label{Uintegral811B}
\ee
gives rise to the fourth order monomial in the $[1,0,1,0,1,0,1]$ irreducible representation. 

One obtains in this way that the harmonic superspace integral of a general monomial of order $n_1 + 2n_2 +3n_3+ 4n_4+4n_4^\prime + 4$ in the $[n_2,n_4,n_1+n_3,n_4,n_2]$ of $SU(6)$ gives rise to a term in $\nabla^6 R^4$ with a monomial of order $n_1 + 2n_2 +3n_3+ 4n_4+4n_4^\prime $ in the $[n_3+n_4 +2n_4^\prime,n_2,n_4,n_1+n_3,n_4,n_2,n_3+n_4 +2n_4^\prime]$ of $SU(8)$, \ie 
\bea &&\int du  D^{14} \bar D^{14} \, F(u)_{\scriptscriptstyle [n_3+n_4 +2n_4^\prime,n_2,n_4,n_1+n_3,n_4,n_2,n_3+n_4 +2n_4^\prime]}^{[ n_2,n_4,n_1+n_3,n_4,n_2]}  W^{n_1 + 2n_2 +3n_3+ 4n_4+4n_4^\prime + 4}|_{[n_2,n_4,n_1+n_3,n_4,n_2]} \CR
&&\sim \nabla^6 R^4  \, W^{n_1 + 2n_2 +3n_3+ 4n_4+4n_4^\prime }|_{[n_3+n_4 +2n_4^\prime,n_2,n_4,n_1+n_3,n_4,n_2,n_3+n_4 +2n_4^\prime]} +\dots\label{811Linear} \eea
where the function $F(u)$ is the function of the harmonic variable defined as a product of the generating functions defined in (\ref{Uintegral811},\ref{Uintegral811B}). One needs at least one quartic singlet in the G-analytic superfield to get a non-vanishing integral \cite{Drummond:2003ex}. 

Referring to the one to one correspondence between linearised and non-linear invariants \cite{Drummond:2003ex}, one deduces that the non-linear invariant must admit the same gradient expansion, \ie 
\bea&&  \cL_\grad811[\cE_\grad811] \CR
&=& \hspace{-5mm}\sum_{n_1,n_2,n_3,n_4,n_4^\prime} \hspace{-5mm} \cD^{n_1 + 2 n_2 + 3 n_3 + 4 n_4 + 4 n^\prime_4 }_{\scriptscriptstyle [n_3+n_4 +2n_4^\prime,n_2,n_4,n_1+n_3,n_4,n_2,n_3+n_4 +2n_4^\prime]} \hspace{2mm} 
\cE_{\grad{8}{1}{1}} \, \cL^{\scriptscriptstyle [n_3+n_4 +2n_4^\prime,n_2,n_4,n_1+n_3,n_4,n_2,n_3+n_4 +2n_4^\prime]}_\grad811 \hspace{5mm} \  \label{811GradExpand} \eea
where each $ \cL^{\scriptscriptstyle [n_3+n_4 +2n_4^\prime,n_2,n_4,n_1+n_3,n_4,n_2,n_3+n_4 +2n_4^\prime]}_\grad811$ is an $E_{7(7)}$ invariant superform in the corresponding representation of $SU(8)$. Note that although the irreducible representation remains unchanged under the substitution 
\be (n_1,n_3,n_4^\prime) \rightarrow(n_1+2, n_3-2,n_4^\prime +1)\ee
the corresponding superforms and the tensor structure of the derivative are different, and are really labelled by the five integers $n_1,n_2,n_3,n_4,n_4^\prime$ without any further identification. Of course the mass dimension implies that these integers are bounded from above, and the maximal weight terms in $\chi^{14} \bar \chi^{14}$ can only be in representations like $[2,6,0,8,0,6,2]$, $[2,6,1,6,1,6,2]$, \dots $[2,10,0,0,0,10,2]$, \dots $[11,1,0,0,0,1,11]$.

This gradient expansion implies in particular that the third order derivative of $\cE_\grad811$ in the $[0,2,0,0,0,0,0]$ and its complex conjugate must vanish, \ie 
\bea  \Scal{ 4 \cD_{ijpq} \cD^{pqmn} \cD_{mnkl} - \cD_{ijkl} \scal{ \Delta + 24} } \cE_\grad811 &=& 0 \ , \label{CubicC}\CR
  \Scal{ 4 \cD^{ijpq} \cD_{pqmn} \cD^{mnkl} - \cD^{ijkl} \scal{ \Delta + 24} } \cE_\grad811 &=& 0 \ .  \eea
These equations imply all the higher order constraints on the function such that  its gradient expansion is in agreement with \eqref{811GradExpand}. 
Defining the covariant derivative in tangent frame as a Lie algebra generator in the fundamental representation of $E_{7(7)}$, this equation reads equivalently 
\be {\bf D}_{56}^{\; 3} \cE_\grad811  = {\bf D}_{56} \Scal{ 6 + \tfrac{1}{4} \Delta } \cE_\grad811 \ . \label{Cubic56} \ee
This implies in particular that all the Casimir operators are determined by the quadratic one such that 
\be \tr \scal{ {\bf D}_{56}^{\; 2+2n}} \, \cE_\grad811= 6 \Delta \scal{ 6 + \tfrac{1}{4} \Delta}^{n} \cE_\grad811 \ , \ee
but the quadratic Casimir is not a priori determined by equation \eqref{CubicC} alone. We will need to consider the other invariants to finally conclude that supersymmetry moreover implies \cite{Green:2010kv}
\be  \Delta  \cE_\grad811 = - 60  \cE_\grad811 - (\cE_\grad844)^2  \ .\label{Laplace811} \ee
Equation \eqref{Cubic56} defines a qantization of the algebraic condition ${\bf Q}_{56}^{\; 3}=0$ associated to the complex nilpotent orbit of $E_{7}$ of Dynkin label \DEVII{2}00000{\mathfrak{0}}, while the condition that the fourth order derivative does not vanish generically in the ${\scriptstyle [2,0,0,0,0,0,2]}$ distinguishes its real form of  $SU(8)$ Dynkin label [$\scriptstyle \mathfrak{2}\mathfrak{0}\mathfrak{0}\mathfrak{0}\mathfrak{0}\mathfrak{0}\mathfrak{2}$] \cite{E7Djo}, which defines the graded decomposition of $SU(8)$ associated to the $(8,1,1)$ harmonic superspace we consider in this section. The property that the linearised structure does not permit to determine the eigenvalue of the Laplace operator in this case, implies that the quantization of the associated nilpotent orbit is not unique, and depends on one free parameter. This property follows from the fact that a nilpotent element of this kind can be obtained as the appropriate limit of a semi-simple element satisfying to the characteristic equation ${\bf Q}_{56}^{\; 3}=\frac{1}{24} \tr({\bf Q}_{56}^{\; 2}) {\bf Q}_{56} $.

\subsection{$F^2 \nabla^4 R^4$ type invariant and its relation to $\nabla^6 R^4$}
\label{F2D4R4} 
Although the $\nabla^6 R^4$ type invariant provides the unique supesymmetric invariant preserving $SU(8)$ one can write at this order, there is another class of invariants that can be defined form the chiral harmonic superspace defined in terms of the harmonic variables $u^{\hat{r}}{}_i,\, u^r{}_i$ parametrising $SU(8)/S(U(2)\times U(6))$ \cite{Drummond:2003ex,Hartwell:1994rp},  with $\hat{r},\hat{s}$ equal to  $1, 2$ of $SU(2)$, and $r ,s$ running from $3$ to $8$ of $SU(6)$. One defines the superfield 
\be
W^{rs} = u^1{}_i u^2{}_j u^r{}_k u^s{}_l W^{i j k l}
\ee
that satisfies to the G-analiticity constraint 
\be 
u^{\hat{r}}{}_{i} \bar D^i_\alpha W^{rs} = 0 \ . 
\ee
Similarly as in the preceding section, the most general function of $W^{rs}$ is freely generated by the three monomials 
\be W^{rs} \ , \qquad \frac{1}{2} \varepsilon_{rstuvw} W^{tu} W^{vw} \ , \qquad \frac{1}{2} \varepsilon_{rstuvw} W^{rs} W^{tu} W^{vw} \ . \ee
One must supplement them with harmonic variables to preserve $S(U(2)\times U(6))$ invariance, using 
\bea \int du\, u^i{}_1 u^j{}_2 u^k{}_r u^l{}_s \, W^{rs} &=& W^{ijkl} \ , \\
 \int du\, u^i{}_1 u^j{}_2 u^r{}_k u^s{}_l \,  \frac{1}{2} \varepsilon_{rstuvw} W^{tu} W^{vw}  &=& W^{ijpq} W_{klpq} - \frac{1}{28} \delta^{ij}_{kl}W^{pqrs} W_{pqrs} \ , \CR
  \int du\, u^i{}_1 u^j{}_2 u^k{}_1 u^l{}_2 \,  \frac{1}{2} \varepsilon_{rstuvw} W^{rs} W^{tu} W^{vw}  &=& W^{ijpq} W_{pqrs} W^{rskl}  - \frac{1}{12} W^{ijkl} W_{pqrs} W^{pqrs}  \ . \nn
\eea
One only gets a non-trivial integral if the cubic $SU(6)$ singlet in $W^{rs}$ appears at least quadratically, which can be understood from the property that the associated chiral primary operator of $SU(2,2|8)$ is otherwise in a short representation \cite{Drummond:2003ex}. Because the $U(1)$ weight of the measure is compensated by a single factor of this cubic $SU(6)$ singlet, it appears that there is no $SU(8)$ invariant that exists in this class. 

 For a general monomial, one gets an invariant of the form 
\bea&&  \int du \bar D^{16}  D^{12}  \, F(u)^{[0,n_1,0,n_2,0]}_{[0,n_2+2n_3+2,0,n_1,0,n_2,0]} \, W^{n_1 + 2n_2 + 3n_3 + 6} |_{[0,n_1,0,n_2,0]}  \\
&\sim&W^{n_1 + 2n_2 + 3n_3}_{\scriptscriptstyle [0,n_2+2n_3,0,n_1,0,n_2,0]} \bar F^{2}_{\scriptscriptstyle[0,2,0,0,0,0,0]} \nabla^4 R^4+\dots + W^{n_1+2n_2+n_3-22}_{\scriptscriptstyle [0,n_2+2n_3-8,0,n_1-8,0,n_2-4,0]}\bar \chi^{16}_{\scriptscriptstyle[0,8,0,4,0,0,0]}  \chi^{12}_{\scriptscriptstyle[0,2,0,4,0,4,0]} \ , \nn\eea
where all terms are projected to the $\scriptstyle [0,n_2+2n_3+2,0,n_1,0,n_2,0]$ irreducible representation, and the term in $\bar F^2$ is
\be \bar F_{\adt\bdt}^{ij} \bar F^{\adt\bdt  kl}  -   \bar F_{\adt\bdt}^{[ij} \bar F^{kl]\adt\bdt} \ . \ee
For a generic function $\cF[W]$ of $W_{rs}$, one obtains
\be D^{16} \bar D^{12}  \cF[W] = \sum_{n_1,n_2,n_3} \frac{\partial^{n_1+2n_2+3n_3+6} \cF[W]}{\partial W^{n_1+2n_2+3n_3+6}}{}\Big|_{[0,n_2,0,n_1,0]} \cL_{\grad820\, {\rm \scriptscriptstyle lin}}^{[0,n_1,0,n_2,0]\ord{n_1+2n_2+3n_3+3}} \ , \ee  
where the densities $ \cL_{\grad820\, {\rm \scriptscriptstyle lin}}^{[0,n_1,0,n_2,0]\ord{n_1+2n_2+3n_3+3}}$ are of order $n_1+2n_2+3n_3+6$ in the fields and only depend on the scalar fields through their space-time derivative. The number $n_1+2n_2+3n_3+3$ is the $U(1)$ weight of the density. These densities determine by construction covariant superforms in the linearised approximation \cite{Voronov,Gates:1997kr,Gates:1997ag}, such that 
\bea d^\ord{0}  \cL_{\grad820\, {\rm \scriptscriptstyle lin}}^{ij,kl} &=& 0 \ ,  \CR
\Scal{ d^\ord{0}  \cL_{\grad820\, {\rm \scriptscriptstyle lin}}^{ij,kl,pqrs} + 3 P^{pqrs} \wedge   \cL_{\grad820\, {\rm \scriptscriptstyle lin}}^{ij,kl} }{}_{\scriptscriptstyle [0,2,0,1,0,0,0]} &=& 0 \ ,  \CR
\Scal{ d^\ord{0}  \cL_{\grad820\, {\rm \scriptscriptstyle lin}}^{ij,kl,pqrs,mntu} + 3 P^{pqrs} \wedge   \cL_{\grad820\, {\rm \scriptscriptstyle lin}}^{ij,kl,mntu} }{}_{\scriptscriptstyle [0,2,0,2,0,0,0]} &=& 0 \  , \CR
\Scal{ d^\ord{0}  \cL_{\grad820\, {\rm \scriptscriptstyle lin}}^{ij,kl,pq}{}_{rs}+18 P_{rsmn} \wedge   \cL_{\grad820\, {\rm \scriptscriptstyle lin}}^{ij,kl,pqmn} }{}_{\scriptscriptstyle [0,3,0,0,0,1,0]} &=& 0 \  ,
\eea
where $d^\ord{0}$ is the superspace exterior derivative in the linear approximation. At the next order, because 
\be d = \sum_{n=0}^{\infty} d^\ord{n} \ee
satisfies to $d^2=0$, one has 
\be \{ d^\ord{0},d^\ord{1}\} = 0\ , \ee
and therefore 
\be d^\ord{0} \Scal{ d^\ord{1}  \cL_{\grad820\, {\rm \scriptscriptstyle lin}}^{ij,kl} } = 0 \ . \ee
We assume in this paper that the structure of superconformal multiplets implies the absence of supersymmetry anomaly, or equivalently that the fifth cohomology of $d^\ord{0}$ is empty. Nevertheless, even if $d^\ord{1}  \cL_{\grad820\, {\rm \scriptscriptstyle lin}}^{ij,kl} $ only depends on the covariant superfields, nothing prevents its $d^\ord{0}$ antecedent to depend explicitly on the scalar fields.  This implies in this case that
\be d^\ord{1}  \cL_{\grad820\, {\rm \scriptscriptstyle lin}}^{ij,kl}  =-d^\ord{0}  \cL_{\grad820\, \ord{1}}^{ij,kl} +  P_{pqrs} \wedge  \cM^{ij,kl,pqrs} + P^{pqij} \wedge \cM_{pq}{}^{kl} + P^{pqkl} \wedge \cM_{pq}{}^{ij} -2 P^{i]pq[k} \wedge \cM_{pq}{}^{l][j} \ , \label{d1F2R4} \ee
where $ \cL_{\grad820\, \ord{1}}^{ij,kl}$ is the covariant correction to the superform, whereas $\cM^{ij,kl,pqrs}$ and $\cM^{ij}{}_{kl}$ are superforms of order six in the fields in the $[0,2,0,1,0,0,0]$ and the $[0,1,0,0,0,1,0]$, respectively, that must satisfy to
\bea
d^\ord{0} \cM^{ij,kl,pqrs} &=& \scal{ P^{pqrs} \wedge \cN^{ij,kl} }{}_{\scriptscriptstyle [0,2,0,1,0,0,0]}\ ,  \CR
d^\ord{0} \cM^{ij}{}_{kl} &=& P^{ijpq} \wedge \cN_{klpq} - \frac{1}{28} \delta^{ij}_{kl} P^{pqrs} \wedge \cN_{pqrs} \ . \eea
In order to have such corrections that could not be reabsorbed in a covariant correction as $ \cL_{\grad820\, \ord{1}}^{ij,kl}$, one must have a corresponding short multiplet associated to a linearised invariant of the same dimension. The only candidate for a superform $ \cM^{ij,kl,pqrs}$ is $  \cL_{\grad820\, {\rm \scriptscriptstyle lin}}^{ij,kl,pqrs} $, but it is of order seven in the fields, and therefore  $ \cM^{ij,kl,pqrs}=0$ at this order. However, there is a candidate for $ \cM^{ij}{}_{kl} $ which is $  \cL_{\grad811\, {\rm \scriptscriptstyle lin}}^{\hspace{10mm}ij}{}_{kl} $, the superform that appears in the $\nabla^6 R^4$ type invariant discussed in the last section. Following \eqref{811Linear}, we have 
\bea d^\ord{0}  \cL_{\grad811\, {\rm \scriptscriptstyle lin}}   &=&  0 \ , \CR
 d^\ord{0}  \cL_{\grad811\, {\rm \scriptscriptstyle lin}}^{ijkl}   &=&  - 3 P^{ijkl} \wedge   \cL_{\grad811\, {\rm \scriptscriptstyle lin}}  \ , \CR
 d^\ord{0}  \cL_{\grad811\, {\rm \scriptscriptstyle lin}}^{ijkl,pqrs}     &=&  - 3 \scal{ P^{ijkl} \wedge \cL_{\grad811\, {\rm \scriptscriptstyle lin}}^{pqrs}  }{}_{\scriptscriptstyle [0,0,0,1,0,0,0]}  \ , \CR
   d^\ord{0}   \cL_{\grad811\, {\rm \scriptscriptstyle lin}}^{\hspace{10mm}ij}{}_{kl}  &=&  - 18 \scal{ P_{klpq} \wedge \cL_{\grad811\, {\rm \scriptscriptstyle lin}}^{ijpq}  }{}_{\scriptscriptstyle [0,1,0,0,0,1,0]}   \ ,\label{Linear811}
\eea
and therefore 
\bea && d^\ord{0} \Bigl(\Scal{ W^{ijpq} W_{pqrs} W^{rskl}  - \tfrac{1}{12} W^{ijkl} W_{pqrs} W^{pqrs} }   \cL_{\grad811\, {\rm \scriptscriptstyle lin}}  \Bigr . \CR
&&  \qquad + W^{ijpq} W_{pqrs}   \cL_{\grad811\, {\rm \scriptscriptstyle lin}}^{rskl} + W^{ijpq} W^{klrs}  \cL_{\grad811\, {\rm \scriptscriptstyle lin}}{}_{pqrs}+  W^{klpq} W_{pqrs}   \cL_{\grad811\, {\rm \scriptscriptstyle lin}}^{rsij}     \CR
&& \hspace{10mm} \Bigl . + 6 W^{pqij} \cL_{\grad811\, {\rm \scriptscriptstyle lin}}^{\hspace{10mm}kl}{}_{pq} + 6 W^{pqkl} \cL_{\grad811\, {\rm \scriptscriptstyle lin}}^{\hspace{10mm}ij}{}_{pq} -12 W^{i]pq[k} \cL_{\grad811\, {\rm \scriptscriptstyle lin}}^{\hspace{10mm}l][j}{}_{pq} \Bigr)\CR
&=& 18 \Scal{ P^{pqij}\wedge  \cL_{\grad811\, {\rm \scriptscriptstyle lin}}^{\hspace{10mm}kl}{}_{pq} + P^{pqkl} \wedge \cL_{\grad811\, {\rm \scriptscriptstyle lin}}^{\hspace{10mm}ij}{}_{pq} -2 P^{i]pq[k}\wedge  \cL_{\grad811\, {\rm \scriptscriptstyle lin}}^{\hspace{10mm}l][j}{}_{pq}} \ , \label{d0Cohom}
 \eea
such that $  \cL_{\grad811\, {\rm \scriptscriptstyle lin}}^{\hspace{10mm}ij}{}_{kl} $ is indeed a consistent candidate. Moreover, the structure of the linearised $(8,1,1)$ invariant does not permit to have the tensor function $W^{ijpq} W_{pqrs} W^{rskl}$, such that \eqref{d0Cohom} is not the exterior derivative of a superform that does not depend on the naked scalar fields (uncovered by a space-time derivative). It follows that such a correction, if it appeared in \eqref{d1F2R4}, could not be reabsorbed in a redefinition of $ \cL_{\grad820\, \ord{1}}^{ij,kl}$.

If this mixing between the $(8,2,0)$ and the $(8,1,1)$ superforms was not appearing at the non-linear level, then the action of the exterior derivative in the function of the scalar fields should not introduce lower derivative terms such that it should satisfy then to 
\be \cD^{ijpq} \Scal{  4 \cD_{pqrs} \cD^{rsmn} \cD_{mnkl} - \cD_{pqkl} \scal{ \Delta + 24} } \cE_{\grad{8}{2}{0}} = 0 \ . \label{D4Consistency} \ee
If the mixing did appear, then the unicity of the linearised invariants \eqref{Linear811} would imply that the corresponding non-linear superform should be the same as in \eqref{811GradExpand}, such that once again the exterior derivative acting on $D^3_{\scriptscriptstyle [0,2,0,0,0,0,0]} \cE_\grad820$ should not generate lower derivative terms and one would conclude again that \eqref{D4Consistency} must be satisfied. Therefore this equation must be satisfied in either cases.

Using moreover the property that the gradient expansion of the linearised invariant is inconsistent with the presence of the third order derivative in the $[1,0,0,1,0,0,1]$ of $SU(8)$, one requires
\be \Scal{ 36 \cD_{jr[kl} \cD^{irmn} \cD_{pq]mn} - \delta^i_j \cD_{klpq} ( \Delta + 42)   + \delta^i_{[k} \cD_{lpq]j} ( \Delta-120)} \cE_{\grad{8}{2}{0}} = 0 \ . \label{CubicR} 
\ee
Using this equation one computes independently of \eqref{D4Consistency} that 
\be \cD^{ijpq} \Scal{  4 \cD_{pqrs} \cD^{rsmn} \cD_{mnkl} - \cD_{pqkl} \scal{ \Delta + 24} } \cE_{\grad{8}{2}{0}}  = \frac{1}{12} \Scal{ 28 \cD^{ijpq} \cD_{klpq} - 3 \delta^{ij}_{kl} \Delta} \scal{ \Delta + 60 } \cE_\grad820 \ee
and we conclude that \eqref{D4Consistency}  and \eqref{CubicR} imply together
\be  \Delta \cE_\grad820 = -60  \cE_\grad820 \ .\label{Laplace820}  \ee
This eigenvalue is such that the structure of the invariant is consistent with the mixing between the $(8,2,0)$ and the $(8,1,1)$ superforms. Only in this case can they reduce to the same invariant for a function $\cE_{\grad{8}{2}{2}}$ satisfying to both \eqref{CubicC} and \eqref{CubicR}, as for the $\nabla^4 R^4$ type invariant. 

We are going to argue now that this chiral invariant must indeed include a $\nabla^6 R^4$ coupling, because the two classes of invariants reduce to one single class in three dimensions. But before to do this, let us mention that \eqref{CubicR} can be rewritten as
\be {\bf D}_{133}^{\; 3} \cE_\grad820 =\frac{1}{3}   {\bf D}_{133} \Delta \cE_\grad820 \ , \ee
which defines a qantization of the algebraic eqation ${\bf Q}_{133}^{\; 3}=0$ associated to the complex nilpotent orbit of $E_{7}$ of Dynkin label \DEVII{0}00000{\mathfrak{2}} with the real form defined with the $SU(8)$ Dynkin label [$\scriptstyle \mathfrak{0}\mathfrak{2}\mathfrak{0}\mathfrak{0}\mathfrak{0}\mathfrak{0}\mathfrak{0}$] \cite{E7Djo}, which defines the graded decomposition of $SU(8)$ associated to the $(8,2,0)$ harmonic superspace we consider in this section. In this case the choice of real form moreover implies that the complex charge in the ${\bf 70}$ defining the nilpotent orbit through the Kostant--Sekiguchi correspondence satisfies to 
\be Q^{ijpq} Q_{pqmn} Q^{mnkl} = 0 \ , \ee
such that it admits a unique quantization, with the eigenvalue of the Laplace operator $-60$. However, we will see in the following that the constraint \eqref{D4Consistency} can be relaxed while keeping the property that the associated representation of $E_{7(7)}$ is a highest weight representation.

\subsection{Dimensional reduction to three dimensions}
In  three dimensions, the duality group is $E_{8(8)}$, of maximal compact subgroup $Spin(16)/\mathds{Z}_2$. We denote $i, j$ the $SO(16)$ vector indices and $A, B$ the positive chirality Weyl spinor indices. The covariant derivative in tangent frame is a chiral Weyl spinor, \ie in the  \WSOXVI00000001 representation. In the linearised approximation, the covariant fields all descend from the Weyl spinor scalar field, satisfying to \cite{Greitz:2011vh}
\be
D_{\alpha}^{i} W^{A} = \Gamma^{i A \dot{A}} \chi_{\alpha \dot A} \ . 
\ee
Both four-dimensional $(8,1,1)$ and $(8,2,0)$ harmonic superspaces descend to the same $(16,2)$ harmonic superspace in three dimensions, defined through the introduction of harmonic variables parametrising $SO(16)/(U(2) \times SO(12))$ \cite{Howe:1994ms}. The Weyl spinor representation decomposes with respect to $U(2) \times Spin(12)$ as
\be {\bf 128} \cong {\bf 32}_+^{\ord{-1}} \oplus \scal{ {\bf 2}\otimes {\bf 32}_-}^\ord{0} \oplus {\bf 32}_+^{\ord{1}}\ , \label{128inD6} \ee
such that the grad $1$ Weyl spinor $W$ of $Spin(12)$ satisfies to a G-analyticity constraint with respect to the positive grad covariant derivative in the ${\bf 2}$ of $U(2)$. The general polynomial in the $Spin(12)$ Weyl spinor is parametrised by four integers, just as for the rank three antisymmetric tensor of $SU(8)$ in section \eqref{811D6R4}.\footnote{This property follows from the fact that the classification of duality orbits of the black hole charges are the same in the $\cN=2$ supergravity theories of duality group $SO^*(12)$ and $SU(3,3)$ \cite{Ferrara:1997uz}.} One computes in a similar way the general integral 
\bea && \int du F(u)_{{\mbox{\WSOXVI0{n_3\hspace{-0.5mm}\mbox{+}n_4\hspace{-0.5mm}\mbox{+}2n_4^\prime}0{n_2}0{n_4}0{n_1\hspace{-0.5mm}\mbox{+}n_3}} } }^{{\mbox{\WSOXII0{n_2}0{n_4}0{n_1\hspace{-0.5mm}\mbox{+}n_3}} }} D^{28}  W^{n_1 + 2n_2 +3n_3+ 4n_4+4n_4^\prime + 4}|_{{\mbox{\WSOXII0{n_2}0{n_4}0{n_1\hspace{-0.5mm}\mbox{+}n_3}} }}  \CR
&\sim &\nabla^{10} P^4  \, W^{n_1 + 2n_2 +3n_3+ 4n_4+4n_4^\prime }|_{{\mbox{\WSOXVI0{n_3\hspace{-0.5mm}\mbox{+}n_4\hspace{-0.5mm}\mbox{+}2n_4^\prime}0{n_2}0{n_4}0{n_1\hspace{-0.5mm}\mbox{+}n_3}} } } +\dots\eea
where $\nabla^{10} P^4 $ is a $Spin(16)$ invariant quartic term in the scalar field momentum, that replaces the $\nabla^6 R^4$ type term that vanishes modulo the equations of motion in three dimensions. In three dimensions it is not established if there is a one to one correspondence between non-linear and linear invariants defined as harmonic superspace integrals. Nevertheless, the class of invariants we discuss descends from four dimensions, and we can therefore assume they admit the same structure, \ie 
\be \cL_\gra{16}{2}[\cE_{\gra{16}{2}}] = \sum_{n_1,n_2,n_3,n_4,n_4^\prime} \cD_{{\mbox{\WSOXVI0{n_3\hspace{-0.5mm}\mbox{+}n_4\hspace{-0.5mm}\mbox{+}2n_4^\prime}0{n_2}0{n_4}0{n_1\hspace{-0.5mm}\mbox{+}n_3}} }} \cE_{\gra{16}{2}}\, \cL^{{\mbox{\WSOXVI0{n_3\hspace{-0.5mm}\mbox{+}n_4\hspace{-0.5mm}\mbox{+}2n_4^\prime}0{n_2}0{n_4}0{n_1\hspace{-0.5mm}\mbox{+}n_3}} }}\ .  \ee
This expansion implies that the fourth order derivative of the function $\cE_{\gra{16}{2}}$ restricted  to the \WSOXVI10001000 must vanish, \ie \be \scal{ \cD \Gamma_{i[jk}{}^r \cD} \scal{ \cD \Gamma_{lpq]r} \cD }\cE_{\gra{16}{2}} = - \delta_{i[j} \scal{ \cD \Gamma_{klpq]} \cD} ( \Delta+48  ) \cE_{\gra{16}{2}} \ ,\label{quarticE8}  \ee
where the Laplace operator $\Delta$ is defined as
\be
\Delta = \cD_{A} \cD^{A}\ . 
\ee
By dimensional reduction of the four-dimensional equation  \eqref{Laplace820}, one computes that   \be
\Delta \cE_{\gra{16}{2}} = - 198  \cE_{\gra{16}{2}} \ . 
\ee
One can understand that the two kinds of 1/8 BPS invariants discussed in the preceding section dimensionally reduce to this single class. If one consider the decomposition of \eqref{128inD6} with respect to $U(6)\subset Spin(12)$, one obtains for one embedding 
 \be {\bf 32}_+\cong {\bf  6}^\ord{-2} \oplus {\bf 20}^\ord{0} \oplus  \overline{\bf 6}^\ord{2}\ , \label{32inA5} \ee
such that the G-analytic superfield in the ${\bf 32}_+$ includes the four-dimensional $(8,1,1)$ G-analytic scalar $W^{rst}$ as well as some components of the vector fields. A generic spinor of non-zero quartic invariant can be represented by $W^{rst}$. For the other embedding  $U(6)\subset Spin(12)$, one gets
 \be {\bf 32}_+\cong  \overline{\bf  1}^\ord{-3} \oplus {\bf 15}^\ord{-1}  \oplus \overline{\bf 15}^\ord{1} \oplus {\bf 1}^\ord{3}\ , \label{32inA5p} \ee
such that  the G-analytic superfield in the ${\bf 32}_+$ includes the four-dimensional $(8,2,0)$ G-analytic scalar $W^{rs}$ as well as some components of the vector fields, and a Ehlers complex scalar parametrising the four-dimensional metric. The scalar field alone only parametrises a null spinor of $Spin(12)$ of vanishing quartic invariant, and only together with the Ehlers scalar field it can provide a representative of a generic spinor. One could have naively concluded that the function $\cE_\grad820$ should give rise to a  function on $E_{8(8)}/Spin_{\scriptscriptstyle \rm c}(16)$ satisfying moreover to 
\be 5\scal{ \cD \Gamma_{ijpq}  \cD} \scal{ \cD \Gamma^{klpq} \cD} \cE = - 20 \scal{\cD \Gamma_{ij}{}^{kl} \cD} \scal{ \Delta + 48 } \cE  + 28 \delta_{ij}^{kl} \Delta \scal{\Delta + 120} \cE \ , \ee
but this equation only admits solutions for functions satisfying to the Laplace equation
\be \Delta \cE = -210\,  \cE \ , \ee
excepted for the functions satisfying to the quadratic and cubic constraints that define the $R^4$ and $\nabla^4 R^4$ type invariants. We see therefore that this equation is incompatible with supersymmetry.

 It follows that both $(8,1,1)$ and $(8,2,0)$ type invariants dimensionally reduce to three-dimensional invariants depending of functions on $E_{8(8)}/Spin_{\scriptscriptstyle \rm c}(16)$ that belong to the same representation of $E_{8(8)}$. Being in the same representation, they both carry a quartic component in the linearised approximation and they must both include a $\nabla^6 R^4$ type term in their uplift to four dimensions. This proves that the mixing between the two different linearised structures must occur such that the non-linear $\bar F^2 \nabla^4 R^4$ type invariant cannot exist without including a $\nabla^6 R^4$ type term as well. 

Before to end this section on the three-dimensional theory, let us discuss the modification of the supersymmetry constraint due to the completion of the $R^4$ type invariant at the next order. As it is argued in \cite{Green:2005ba}, the appearance of a $R^4$ correction with threshold function $\cE_\gra{16}{8}$, will modify the Laplace equation with a non-zero right-hand-side, \ie 
 \be
\Delta \cE_{\gra{16}{2}} = - 198  \cE_{\gra{16}{2}} - \cE_{\gra{16}{8}}^{\; 2} \ . 
\ee
Because the function $ \cE_{\gra{16}{8}}$ satisfies to \cite{Minimal}
\be \scal{ \cD \Gamma_{ijkl} \cD}  \cE_{\gra{16}{8}} = 0 \ , \ee
the second derivative of its square must necessarily vanish in the  \WSOXVI10001000, and we get accordingly a modification of \eqref{quarticE8} to
\be \scal{ \cD \Gamma_{i[jk}{}^r \cD} \scal{ \cD \Gamma_{lpq]r} \cD } \cE_{\gra{16}{2}}= 150 \delta_{i[j} \scal{ \cD \Gamma_{klpq]} \cD}  \cE_{\gra{16}{2}} + \delta_{i[j} \scal{ \cD \Gamma_{klpq]} \cD} \cE_{\gra{16}{8}}^{\; 2}\ . \ee

\subsection{$E_{7(7)}$ Eisenstein series}
In this section we shall discuss some properties of Einstein series that solve the differential equations we have derived for the $\nabla^6 R^4$ type invariants.

\addtocontents{toc}{\protect\setcounter{tocdepth}{1}}
\subsubsection{Fundamental representation}
\addtocontents{toc}{\protect\setcounter{tocdepth}{2}}
As discussed in \cite{Obers:1999um,D4R4}, one can define the Eisenstein series 
\be E_{\mbox{\DEVII000000s}} = \sum_{\vspace{-2mm}\begin{array}{c}\scriptstyle \vspace{-4mm}  \Gamma\in \mathds{Z}^{56} \vspace{2mm}\\ \scriptscriptstyle I_4^{\prime\prime}(\Gamma)|_{\bf 133}=0\end{array}} |Z(\Gamma)_{ij} Z(\Gamma)^{ij}|^{-s} \   , \label{E56s} \ee
as a sum over the rank one integral charge vectors $\Gamma$ in the ${\bf 56}$ of $E_{7(7)}$ satisfying to the constraint that the quadratic tensor  $\Gamma\otimes \Gamma$ vanishes  in the adjoint representation. This formula is rather useful to identify the differential equations satisfied by the Eisenstein function, because one can simply consider the case of one charge $\Gamma$, with  $Z(\Gamma)_{ij} = \cV_{ij}{}^I\Gamma_I$, such that the quadratic constraint becomes 
\be Z_{[ij} Z_{kl]} = \frac{1}{24} \varepsilon_{ijklpqrs} Z^{pq} Z^{rs} \ , \qquad Z_{ik} Z^{jk} = \frac{1}{8} \delta_i^j Z_{kl} Z^{kl} \ ,\ee
and the differential operator acts on $Z_{ij} $ as an element of $\mathfrak{e}_{7(7)}$
\be \cD_{ijkl} Z^{pq} = 3 \delta^{pq}_{[ij} Z_{kl]} \ , \qquad \cD_{ijkl} Z_{pq} = \frac{1}{8} \varepsilon_{ijklpqrs} Z^{rs} \ . \ee
Using the definition $|Z|^2 = Z_{ij} Z^{ij}$, one computes that the function $|Z|^{-2s}$ satisfies to 
\bea \cD_{ijpq} \cD^{klpq} |Z|^{-2s} &=& 2s(s-2) Z_{ij} Z^{kl} |Z|^{-2s-2} + \frac{s(s-11)}{4} \delta_{ij}^{kl} |Z|^{-2s} \  , \CR
\cD_{ijpq} \cD^{pqrs} \cD_{rskl} |Z|^{-2s} &=& - 3 s(s-2)(s-4) Z_{ij} Z_{kl} |Z|^{-2s-2} + \frac{s^2-15s + 8}{4} \cD_{ijkl} |Z|^{-2s}  \ , \CR
 \cD_{jr[kl} \cD^{irmn} \cD_{pq]mn} |Z|^{-2s} &=&\frac{(s-2)(s-7)}{12}  \delta^i_j\cD_{klpq}  |Z|^{-2s} -\frac{ s^2-9s -40}{12} \delta^i_{[k} \cD_{pql]j} |Z|^{-2s}  \ , \hspace{10mm} \label{ConstraintsZs} 
\eea
and to the Laplace equation 
\be \Delta |Z|^{-2s} = 3s(s-9) |Z|^{-2s} \ . \ee
For $s\ne2,\, 4$, the function admits a generic gradient expansion in the irreducible representations $[0,n_2+2n_3,0,n_1,0,n_2,0]$ and their complex conjugate. To exhibit this property, it is convenient to consider a restricted set of indices as follows 
\bea &&  \scal{ \cD_{12ij} \cD^{ijkl} \cD_{kl12}}^{n_3} \scal{  \cD_{12pq} \cD^{78pq}}^{n_2} \scal{\cD_{1234}}^{n_1} |Z|^{-2s} \\ &=& \tfrac{(s+n_1+n_2+n_3-1)!(s+n_2+n_3-3)!(s+n_3-5)!}{(s-1)!(s-3)!(s-5)!} \scal{\mbox{-}3 Z_{12}^{\; 2}}^{n_3} \scal{2 Z_{12} Z^{78}}^{n_2} \scal{\mbox{-}6 Z_{[12} Z_{34]}}^{n_1} |Z|^{-2(s+n_1+n_2+n_3)} \, . \nn \label{GradExpandZs} \eea
One computes moreover that for $m\le n$ 
\bea &&  \scal{\cD^{78ij} \cD_{ijkl} \cD^{kl78}}^{m} \scal{\cD_{12pq} \cD^{pqrs} \cD_{rs12}}^n |Z|^{-2s} \\
&=& \tfrac{(s+n-1)!(s+n-3)!(s+n-5)!(s+n+m-1)!(s+n+m-3)!(s-n+m-5)!}{(s-1)!(s-3)!(s-5)!(s+n-1)!(s+n-3)!(s-n-5)!} \scal{\mbox{-}3 Z^{78\; 2}}^{m} \scal{\mbox{-}3 Z_{12}^{\; 2}}^{n}|Z|^{-2(s+n+m)} \CR
&=&\scal{\mbox{-}\tfrac{3}{2}}^{n+m} \tfrac{(s+n-5)!(s+n+m-1)!(s+n+m-3)!(s-n+m-5)!}{(s+n-m-5)! (s+2n-1)!(s+2n-3)!(s-n-5 )! } \scal{\cD_{12ij} \cD^{ijkl} \cD_{kl12}}^{n-m}\scal{\cD_{12pq} \cD^{78pq}}^{n+m} |Z|^{-2s} \nn  \eea
such that acting with a derivative operator in the conjugate representation $[0,0,0,0,0,2m,0]$ does not produce an independent tensor. One has in particular for $s$ an integer greater than $5$ 
\be \scal{\cD^{78ij} \cD_{ijkl} \cD^{kl78}} \scal{\cD_{12pq} \cD^{pqrs} \cD_{rs12}}^{s-4} |Z|^{-2s} = 0 \ . \ee
This equation is the equivalent on $E_{7(7)}/\SU$ of the equation on $SL(2)/SO(2)$ 
\be \bar \cD \cD^{s-1} E_{[s]} = 0 \ ,  \ee
for integral $s$, and we would like to see that the function $ E{\mbox{\DEVII000000s}}$ also decomposes somehow into a ``holomorphic'' part $ \cF_s$ and a ``anti-holomorphic'' part $\bar \cF_s$, satisfying respectively to 
\be \scal{\cD_{12pq} \cD^{pqrs} \cD_{rs12}}^{s-4} \bar \cF_s = 0 \ , \qquad \scal{\cD^{78ij} \cD_{ijkl} \cD^{kl78}}^{s-4} \cF_s = 0 \  , \ee
such that 
\be   \scal{\cD_{12pq} \cD^{pqrs} \cD_{rs12}}^{s-4} E_{\mbox{\DEVII000000s}} =  \scal{\cD_{12pq} \cD^{pqrs} \cD_{rs12}}^{s-4} \cF_s  \ , \label{D3sF}  \ee
and respectively for the complex conjugate. By consistency, this requires for instance that acting with further derivatives on this tensor does not permit to get back lower order tensors with $n_3<s-4$ in \eqref{GradExpandZs}.

Through representation theory, one obtains that 
\bea&&  \cD_{[0,0,0,1,0,0,0]} \cD^{n_1+2n_2+3n_3}_{[0,n_2+2n_3,0,n_1,0,n_2,0]} |Z|^{-2s} \\
&\sim& \Bigl( \cD^{n_1+1+2n_2+3n_3}_{[0,n_2+2n_3,0,n_1+1,0,n_2,0]} +\cD^{n_1-1+2(n_2+1)+3n_3}_{[0,n_2+2n_3+1,0,n_1-1,0,n_2+1,0]}  +\cD^{n_1+2(n_2-1)+3(n_3+1)}_{[0,n_2+2n_3+1,0,n_1,0,n_2-1,0]}  \Bigr . \CR
&& \Bigl . + \cD^{n_1-1+2n_2+3n_3}_{[0,n_2+2n_3,0,n_1-1,0,n_2,0]} +\cD^{n_1+1+2(n_2-1)+3n_3}_{[0,n_2+2n_3-1,0,n_1+1,0,n_2-1,0]}  +\cD^{n_1+2(n_2+1)+3(n_3-1)}_{[0,n_2+2n_3-1,0,n_1,0,n_2+1,0]} \Bigr) |Z|^{-2s} \nn
\eea
for some coefficients that are not specified. So the only way to reduce $n_3$, is to increase $n_2$ by $1$ unit. We will check this equation in the case $n_1=n_2=0$. The restriction of the derivative $\cD^{3n} |Z|^{-2s}$ to the $[0,2n,0,0,0,0,0]$ with two free indices reads 
\bea && \cD^{3n\, [0,2n,0,0,0,0,0]}_{ij1^{2n-1}2^{2n-1}} |Z|^{-2s} \CR
&=& \tfrac{(s+n-1)!(s+n-3)!(s+n-5)!}{(s-1)!(s-3)!(s-5)!} \frac{(-3)^n}{2n} \Scal{ Z_{ij} Z_{12}^{\; 2n-1} - (2n-1) Z_{1[i} Z_{j]2} Z_{12}^{\; 2n-2}} |Z|^{-2(s+n)} \ , \eea
and one computes that 
\bea && \cD^{78ij} \frac{1}{2n}\Scal{ Z_{ij} Z_{12}^{\; 2n-1} - (2n-1) Z_{1[i} Z_{j]2} Z_{12}^{\; 2n-2}} |Z|^{-2(s+n)} \CR
&=& \frac{(2n+5 ) ( n - s+4)}{8n} Z^{78} Z_{12}^{\; 2n-1} |Z|^{-2s} \ , \eea
such that 
\bea&&  \cD^{78ij} \cD^{3n\, [0,2n,0,0,0,0,0]}_{ij1^{2n-1}2^{2n-1}} |Z|^{-2s} \CR
& =& \frac{3(s+n-5)(2n+5)(s-n-4)}{16n}   \scal{ \cD_{12ij} \cD^{ijkl} \cD_{kl12}}^{n-1} \scal{  \cD_{12pq} \cD^{78pq}} |Z|^{-2s} \ . \eea
In particular we have that 
\be \cD^{78ij} \cD^{3(s-4)\, [0,2(s-4),0,0,0,0,0]}_{ij1^{2s-5}2^{2s-5}} |Z|^{-2s} = 0  \ , \label{ConsSF2k} \ee
consistently with the assumption that no lower order tensor is produced out of the tensor function \eqref{D3sF}. We conclude therefore that the tensor 
\be \cF \hspace{-0.4mm} E_{[0,2(s-4),0,0,0,0,0]} = \cD^{3(s-4)}_{[0,2(s-4),0,0,0,0,0]} E_{\mbox{\DEVII000000s}} \ , \label{FE} \ee
is an $E_{7(7)}(\mathds{Z})$ modular form that is in some sense holomorphic, such that its gradient expansion is restricted to derivative of this tensor in the symmetric representations $[0,n_2+2(n_3+s-4),0,n_1,0,n_2,0]$, \ie
\be  \cD^{n_1+2n_2+3n_3}_{[0,n_2+2n_3,0,n_1,0,n_2,0]}   \cF \hspace{-0.4mm} E_{[0,2(s-4),0,0,0,0,0]}  \in [0,n_2+2(n_3+s-4),0,n_1,0,n_2,0] \ . \ee

Using Langlands functional identity  \cite{Green:2010kv}, one computes that the only integer values of $s\ge5$ for which the function  diverges are  
\bea   E_{\mbox{\DEVII000000{5\mbox{+}\epsilon}}} &=& \frac{63}{16\pi \, \epsilon} E_{\mbox{\DEVII0000004}} + \hat{E}_{\mbox{\DEVII000000{5}}} + \mathcal{O}(\epsilon) \ , \label{DivergenceF2D4R4} \CR
E_{\mbox{\DEVII000000{7\mbox{+}\epsilon}}} &=&\frac{1\, 964\, 655 \zeta(5)}{2048 \pi^{5} \, \epsilon} E_{\mbox{\DEVII0000002}} + \hat{E}_{\mbox{\DEVII000000{7}}} + \mathcal{O}(\epsilon) \ , \CR
 E_{\mbox{\DEVII000000{9\mbox{+}\epsilon}}} &=& \frac{12\, 642\, 554\, 925 \zeta(5)\zeta(9)}{2\, 097\, 152 \pi^{9}\, \epsilon} +\hat{E}_{\mbox{\DEVII000000{9}}}+ \mathcal{O}(\epsilon) \ . \eea
However, according to \eqref{ConstraintsZs}, the function $E_{\mbox{\DEVII000000s}}$ satisfies to 
\be \cD^{3}_{[0,2,0,0,0,0,0]} E_{\mbox{\DEVII000000s}} = 0 \ ,\quad  \mbox{for}\, s=0,\, 2,\, 4 \ , \ee
and it follows that the tensor $ \cF \hspace{-0.4mm} E_{[0,2(s-4),0,0,0,0,0]}$ \eqref{FE} is finite for all $s$. However, we have argued in the preceding section that the $\bar F^2 \nabla^4 R^4$ type invariant must include a  $\nabla^6 R^4$ type term, which will be multiplied by the function itself. In this case the relevant Eisenstein function diverges, and one must regularise it such that the differential equation \eqref{Laplace820} will be modified to
\be \Delta \, \hat{E}_{\mbox{\DEVII000000{5}}}   = - 60  \, \hat{E}_{\mbox{\DEVII000000{5}}} + \frac{189}{16 \pi}  E_{\mbox{\DEVII0000004}} \ . \ee
Such a correction is reminiscent of a 1-loop logarithm divergence of the $\nabla^4 R^4$ type invariant form factor. 

\addtocontents{toc}{\protect\setcounter{tocdepth}{1}}
\subsubsection{Adjoint representation}
\addtocontents{toc}{\protect\setcounter{tocdepth}{2}}

One can also consider the Eisenstein series \footnote{We assume here that all the elements of $\mathds{Z}^{133}$ are in the $E_{7(7)}(\mathds{Z})$ orbit of a relative integer times a normalised representative of the continuous orbit. This property does not affect our conclusions in any case, which only requires the generating character to satisfy to the differential equations we discuss.}
\be E_{\mbox{\DEVII{\mathnormal{s}}00000{\mathfrak{0}}}} =  \sum_{{\bf Q}\in \mathds{Z}^{133} |{\bf Q}^2=0} \scal{  X(Q)_{ijkl} X(Q)^{ijkl} }^{-s} \ ,  \ee 
as the sum over integral charges ${\bf Q}\in \mathfrak{e}_{7(7)}$ satisfying to the constraint ${\bf Q}^2 = 0 $, and such that the adjoint action of the coset representative $\cV$ on ${\bf Q}$ decomposes into the anti-Hermitian traceless matrix $\Lambda^i{}_j$ and the complex-selfdual antisymmetric tensor $X_{ijkl}$ satisfying to the constraints 
\bea \Lambda^i{}_k \Lambda^k{}_j &=& - \frac{1}{48} \delta^i_j X^{klpq} X_{klpq} \ ,\CR
\Lambda^{[i}{}_{[k} \Lambda^{j]}{}_{l]} &=& - \frac{1}{2} X^{ijpq} X_{klpq}  + \frac{1}{48} \delta^{ij}_{kl} X^{pqrs} X_{pqrs} \ , \CR
\Lambda^{[i}{}_p X^{j]pkl} &=& \Lambda^{[k}{}_p X^{l]pij}  \ . \eea
The action of the derivative on these tensors is defined as the $\mathfrak{e}_{7(7)}$ action 
\be \cD_{ijkl} X^{pqrs}  = 12 \delta^{[pqr}_{[ijk} \Lambda^{s]}{}_{l]} \ , \qquad 
\cD_{ijkl} \Lambda^p{}_q = 2 \delta^p_{[i} X_{jkl]q} + \frac{1}{4} \delta^p_q X_{ijkl} \ . \ee
One computes for $|X|^2 = X_{ijkl} X^{ijkl} $ that
\bea \cD_{ijkl} |X|^2 &=& -24 X_{p[ijk} \Lambda^p{}_{l]} \ , \qquad \qquad\qquad  \cD_{ijpq} X^{klpq} = 10 \delta_{[i}^{[k} \Lambda^{j]}{}_{l]} \ , \\
\cD_{ijpq} \cD^{klpq} |X|^2 &=& 30 X_{ijpq} X^{klpq} + 3 \delta_{ij}^{kl} |X|^2 \ , \qquad \cD_{ijpq} |X|^2 \cD^{klpq} |X|^2 = 12 X_{ijpq} X^{klpq} |X|^2 \ , \nn \eea
which permits to derive that 
\be \cD_{ijpq} \cD^{klpq} |X|^{-2s} = 6s(2s-3) X_{ijpq} X^{klpq} |X|^{-2s-2} - 3 s\delta_{ij}^{kl} |X|^{-2s} \ . \ee
One gets therefore a solution to the equation 
\be \cD_{ijpq} \cD^{klpq} \cE_\grad844 = - \frac{9}{2} \delta_{ij}^{kl} \cE_\grad844 \ , \label{R4Quadra} \ee
for $s=\frac{3}{2}$. One computes in general that 
\be \cD_{ijpq} \cD^{pqrs}\cD_{rskl} |X|^{-2s} = \scal{ s^2 - \tfrac{17}{2} s+6} \cD_{ijkl} |X|^{-2s} \ , \ee
and the function satisfies to  \eqref{CubicC} and its complex conjugate for all $s$. The restriction of the third order derivative to the $[1,0,0,1,0,0,1]$ gives
\be \cD_{1k[12} \cD_{34]ij} \cD^{8ijk} |X|^{-2s} = - \frac{3}{4} s(2s-3)(2s-5) \Lambda^8{}_1 X_{1234} |X|^{-2s-2} \ , \ee
showing that the function solves the cubic equation \eqref{CubicR} for $s=\frac{5}{2}$. These functions satisfy the same equations as their analog Eisenstein functions defined in the fundamental representation, consistently with the property that \footnote{We are grateful to Axel Kleinschmidt who provided the explicit coefficients.}
\be  E_{\mbox{\DEVII{\frac{3}{2}}00000{\mathfrak{0}}}}  = \frac{2}{\pi}   E_{\mbox{\DEVII{0}000002}} \ , \qquad  E_{\mbox{\DEVII{\frac{5}{2}}00000{\mathfrak{0}}}} = \frac{8}{15\pi}   E_{\mbox{\DEVII{0}000004}} \ . \ee 
One can also consider the restriction of the fourth order derivative to the $[2,0,0,0,0,0,0,2]$ to vanish, which defines a further restriction on solutions to \eqref{CubicC}. In this case one obtains 
\be \cD^{8kij} \cD_{1lij} \cD_{1kpq} \cD^{8lpq}   |X|^{-2s} = - \frac{9}{2} s(2s-3)(2s-5)(s-4) \Lambda^8{}_1 \Lambda^8{}_1 |X|^{-2s-2} \ , \ee
and this further restriction distinguishes the value $s=4$. 
We have in general 
\be \Delta  |X|^{-2s} = 2s(2s-17) |X|^{-2s}  \ , \ee
and the function $E{\mbox{\DEVII{4}00000{\mathfrak{0}}}}$ does not solve the 1/8 BPS equation. 
Using the same normalisation as   \cite{Green:2010kv} with a factor of $2\zeta(2s)$ in the definition of the Eisenstein function, one computes using Langlands formula 
\be  E_{\mbox{\DEVII{\mathnormal{s}}00000{\mathfrak{0}}}}   = \pi^{\frac{33}{2}} \tfrac{\Gamma(s-8) \Gamma(s-\stfrac{13}{2}) \Gamma(s-\stfrac{11}{2}) \Gamma(2s-\stfrac{17}{2}) \zeta(2s-16) \zeta(2s-13)\zeta(2s-11) \zeta(4s-17) }{ \Gamma(s-\stfrac{5}{2}) \Gamma(s-\stfrac{3}{2}) \Gamma(s) \Gamma(2s-8)\zeta(17-2s)\zeta(2s-5)\zeta(2s-3)\zeta(4s-16)} E_{\mbox{{{\tiny $ { \left[ \begin{array}{cccccc}  & & \mathfrak{0} \hspace{-0.7mm}&&& \vspace{ -1.5mm} \\ \stfrac{17}{2}\mbox{-}s\hspace{-0.7mm} &  \mathfrak{0} \hspace{-0.7mm}& \mathfrak{0} \hspace{-0.7mm} & \mathfrak{0}\hspace{-0.7mm}&\mathfrak{0}\hspace{-0.7mm}& \mathfrak{0}  \end{array}\right] }$}}
}} \ . \ee
 The function is singular for various values of $s$, \ie $\frac{9}{2}$, $\frac{11}{2}$, $6$, $\frac{13}{2}$, $7$, and $\frac{17}{2}$, and in particular for $s=6$, which is the relevant value to solve equation \eqref{CubicC} with \eqref{Laplace811}. One should therefore consider the regularised series
\be E_{\mbox{{{\tiny $ { \left[ \begin{array}{cccccc}  & & \mathfrak{0} \hspace{-0.7mm}&&& \vspace{ -1.5mm} \\ 6\mbox{+}\epsilon\hspace{-0.7mm} &  \mathfrak{0} \hspace{-0.7mm}& \mathfrak{0} \hspace{-0.7mm} & \mathfrak{0}\hspace{-0.7mm}&\mathfrak{0}\hspace{-0.7mm}& \mathfrak{0}  \end{array}\right] }$}}}}= \frac{\pi^5}{8 \zeta(9)\, \epsilon}  E_{\mbox{{{\tiny $ { \left[ \begin{array}{cccccc}  & & \mathfrak{0} \hspace{-0.7mm}&&& \vspace{ -1.5mm} \\ \frac{5}{2}\hspace{-0.7mm} &  \mathfrak{0} \hspace{-0.7mm}& \mathfrak{0} \hspace{-0.7mm} & \mathfrak{0}\hspace{-0.7mm}&\mathfrak{0}\hspace{-0.7mm}& \mathfrak{0}  \end{array}\right] }$}}}} +  \hat{E}_{\mbox{{{\tiny $ { \left[ \begin{array}{cccccc}  & & \mathfrak{0} \hspace{-0.7mm}&&& \vspace{ -1.5mm} \\ 6 \hspace{-0.7mm} &  \mathfrak{0} \hspace{-0.7mm}& \mathfrak{0} \hspace{-0.7mm} & \mathfrak{0}\hspace{-0.7mm}&\mathfrak{0}\hspace{-0.7mm}& \mathfrak{0}  \end{array}\right] }$}}}} + \mathcal{O}(\epsilon)\ . \ee
However, we will see in the following that this function does not appear in string theory, similarly as the $\nabla^6 R^4$ threshold function is not described by an Eisenstein series in type IIB supergravity. Nonetheless, some components of this function should appear, as we will argue in the following. 

\subsection{$F^{2k} \nabla^4 R^4$ type invariants}
The $F^2 \nabla^4R^4$ type invariants we have discussed in section \ref{F2D4R4} have a natural generalisation to higher order invariants. Considering the same chiral harmonic superspace defined in terms of the harmonic variables $u^{\hat{r}}{}_i,\, u^r{}_i$ parametrising $SU(8)/S(U(2)\times U(6))$ \cite{Drummond:2003ex}, one can define the G-analytic superfields 
\be \bar F_{\adt\bdt}^{12} = u^1{}_i u^2{}_j  F_{\adt\bdt}^{ij}\  ,\qquad  \bar \chi_\adt^{12r} = u^1{}_i u^2{}_j u^r{}_k \bar \chi_\adt^{ijk} \ . \ee
They do not permit to define directly chiral primary operators of $SU(2,2|8)$, because 
\be \bar D_{\adt t} W^{rs}  = \delta_t^{[r} \bar \chi_\adt^{12s]} \ , \qquad \bar D_{\adt r} \chi_\bdt^{12s} = \delta_r^s \bar F_{\adt\bdt}^{12} \ . \ee
Chiral primary operators are annihilated by the special supersymmetry generators at the origin, \ie 
\be S_{\dot{\gamma}}^r \bar F_{\adt\bdt}^{12} = \varepsilon_{\cdt(\adt} \bar \chi_{\bdt)}^{12r} \ , \qquad S_\adt^r \bar \chi_\bdt^{12s} = \varepsilon_{\adt\bdt} W^{rs} \ , \qquad S_\adt^t W^{rs} = 0 \ . \ee
One can enforce this property by defining a chiral primary as 
\be \mathcal{O}^\ord{k}_{\cF} = (S)^{12} \Scal{  ( \bar F_{\adt\bdt}^{12} \bar F^{\adt\bdt 12} )^{2+k} \cF[W] } \propto \scal{ \varepsilon_{rstuvw} W^{rs} W^{tu} W^{vw}}^2  ( \bar F_{\adt\bdt}^{12} \bar F^{\adt\bdt 12} )^{k-1}  \cF[W]+\dots \label{ChiralPrimary}   \ee
for an arbitrary function $\cF$ of the G-analytic superfield $W^{rs}$. By construction such a chiral primary operator is never short, and defines a non-trivial integrand for the $(8,2,0)$ measure. Because one can consider an arbitrary representative up to a total fermionic derivative, one can as well consider the first term in \eqref{ChiralPrimary} as the integrand. 

So similarly as in section  \ref{F2D4R4}, we get the general class of linearised invariants for an arbitrary positive integer $k$,
\bea&&  \int du \bar D^{16}  D^{12}  \, F(u)^{[0,n_1,0,n_2,0]}_{[0,n_2+2n_3+2k,0,n_1,0,n_2,0]} \, \bar F^{2k-2}\, W^{n_1 + 2n_2 + 3n_3 + 6} |_{[0,n_1,0,n_2,0]}  \\
&\sim&W^{n_1 + 2n_2 + 3n_3}_{\scriptscriptstyle [0,n_2+2n_3,0,n_1,0,n_2,0]} \bar F^{2k}_{\scriptscriptstyle[0,2k,0,0,0,0,0]} \nabla^4 R^4+\dots \CR
&& \hspace{10mm} + W^{n_1+2n_2+n_3-22}_{\scriptscriptstyle [0,n_2+2n_3-8,0,n_1-8,0,n_2-4,0]} \bar F^{2k-2}_{\scriptscriptstyle [0,2k-2,0,0,0,0,0]} \bar \chi^{16}_{\scriptscriptstyle[0,8,0,4,0,0,0]}  \chi^{12}_{\scriptscriptstyle[0,2,0,4,0,4,0]} \ . \nn\eea
We conclude that the corresponding supersymmetry invariants admit the same gradient expansion in 
\be \cL^\ord{k}_\grad820[\cE_\grad820^\ord{k}] =\sum_{n_1\ge 0,n_2\ge0,n_3\ge k}  \cD^{n_1+2n_2+3n_3}_{[0,n_2,0,n_1,0,n_2+2n_3,0]} \cE_\grad820^\ord{k} \, \cL^{k\, [0,n_2+2n_3,0,n_1,0,n_2,0]}_\grad820 \ , \ee
for a function $\cE_\grad820^\ord{k}$ satisfying to \eqref{CubicR}, and $k\ge2$. The coupling at the lowest number of points is then of the kind 
\be \cD^{3k}_{[0,0,0,0,0,2k,0]} \cE_\grad820^\ord{k} \, \cL^{k\, [0,2k,0,0,0,0,0]}_\grad820 = \cD^{3k}_{[0,0,0,0,0,2k,0]} \cE_\grad820^\ord{k} \, \bar F^{2k}_{[0,2k,0,0,0,0,0]} \, \nabla^4 R^4 + \dots \ee
In principle one could expect to have a non-trivial mixing with another class of linearised invariant at the non-linear level, just as the one of the $\bar F^2 \nabla^4 R^4$ type invariant with the $\nabla^6 R^4$ type invariant described in section  \ref{F2D4R4}. However, there is no higher order chiral primary operator that can define a non-trivial  $(8,1,1)$ harmonic superspace integral, and we did not find any linearised invariant with the right structure to define a possible cohomology class as does \eqref{d0Cohom}. Therefore we expect these invariants to have the same structure as the associated linearised invariants, \ie to only contribute to $(4+2k)$-point amplitudes and higher.

Independently of this assumption, the structure of these invariants requires that the action of the derivative $\cD_{ijkl}$ on $\cD^{3k} \cE_\grad820^\ord{k}$ does not generate lower order derivatives of the function. This condition is precisely \eqref{ConsSF2k}, and we conclude therefore that the eigenvalue of the Laplace operator is determined in the same way as
\be \Delta \cE_\grad820^\ord{k} = 3 (k+4)(k-5) \cE_\grad820^\ord{k} \ , \ee
such that the function satisfies to 
\be 12 \cD_{jr[kl} \cD^{irmn} \cD_{pq]mn} \cE_\grad820^\ord{k} =  (k+2)(k-3)  \delta^i_j \cD_{klpq}  \cE_\grad820^\ord{k}  - ( k(k-1) -60)  \delta^i_{[k} \cD_{lpq]j} \cE_\grad820^\ord{k}\ .  \label{CubicK} \ee
It is therefore tempting to conjecture that 
\be  \cE_\grad820^\ord{k} \propto  E_{\mbox{\DEVII{0}00000{4\mbox{+}k}}} \ , \label{EisF2k}  \ee
in the string theory effective action, and we will indeed show in section \ref{StringPerturb} that this function admits a consistent perturbative string theory limit. Moreover, we will see in section \ref{HigherDimensions} that it also admits an appropriate decompactification limit.

\subsection{Wavefront set and Poisson equation source term}
We have seen that there are two classes of $\nabla^6 R^4$ type invariants in four dimensions, that preserve tree-level supersymmetry modulo the classical field equations. However, considering that the effective action already includes an $R^4$ type correction, we must take into account the action of the accordingly modified supersymmetry transformation on the $R^4$ type invariant itself. This is a very difficult task to carry out in practice, but one can nonetheless show general properties on these corrections. We recall that the $R^4$ type invariant admits the following gradient expansion in derivatives of the function $\cE_\grad844$ 
\be \cL_\grad844[\cE_\grad844] = \sum_{n=0}^{12} \cD^n_{[0,0,0,n,0,0,0]}  \cE_\grad844 \cL_\grad844^{[0,0,0,n,0,0,0]} , \ee
with $\cE_\grad844$ satisfying to \eqref{R4Quadra}. The first order modification of the supersymmetry transformations will therefore necessarily admit the same gradient expansion in the function  $\cE_\grad844$, such that
\be \delta = \delta^\ord{0} + \sum_{n=0}^{12}   \cD^n_{[0,0,0,n,0,0,0]}  \cE_\grad844 \delta^{\ord{1}\, [0,0,0,n,0,0,0]} +\dots \ , \ee
where the dots stand for higher order corrections. It follows that the correction at second order will admit the expansion 
\bea && \delta \int \cL_\grad844[\cE_\grad844] \CR
&=&  \int \Scal{ \sum_{n=0}^{12}   \cD^n_{[0,0,0,n,0,0,0]}  \cE_\grad844 \delta^{\ord{1}\, [0,0,0,n,0,0,0]} } \Scal{ \sum_{m=0}^{12} \cD^m_{[0,0,0,m,0,0,0]}  \cE_\grad844 \cL_\grad844^{[0,0,0,m,0,0,0]} }  \CR
&=& \int \sum_{m \, n} \sum_R   \scal{ \cD^n_{[0,0,0,n,0,0,0]}  \cE_\grad844 \cD^m_{[0,0,0,m,0,0,0]}\cE_\grad844  }_R \Psi^R_{m,n} \,  \eea
where the sum over $R$ runs over all irreducible representations of $SU(8)$ in the tensor product  $[0,0,0,n,0,0,0]\otimes [0,0,0,m,0,0,0]$, and $ \Psi^R_{m,n} $ are understood to be $E_{7(7)}$ invariant densities function of the fields and their covariant derivatives in the irreducible representation $R$. One checks that all the appearing irreducible representations $R$ are self-conjugate, \ie  of the type $[n_4,n_3,n_2,n_1,n_2,n_3,n_4]$, by property of the tensor product $[0,0,0,n,0,0,0]\otimes[0,0,0,m,0,0,0] $. The $\bar F^2 \nabla^4 R^4$ type invariant admits a gradient expansion with non-self conjugate irreducible representations, and all its components in self-conjugate representations do in fact coincide with ones appearing in the $(8,1,1)$ $\nabla^6 R^4$ type invariant. It follows that the analysis of the supersymmetry constraints on the $\bar F^2 \nabla^4 R^4$ type invariant is not modified by the presence of the $R^4$ correction, and equations (\ref{CubicR},\ref{Laplace820}) are the exact equations to be solved by the corresponding function $\cE_\grad820$ in the Wilsonian action. 

However, all the irreducible representations that appear in the gradient expansion \eqref{811GradExpand} are included in the tensor product $[0,0,0,n,0,0,0]\otimes[0,0,0,m,0,0,0] $ for $m$ and $n$ running from 1 to twelve, and the differential equations (\ref{CubicC},\ref{Laplace811}) must be modified in the presence of the $R^4$ type correction. Following the analysis carried out in \cite{Green:2005ba,Green:2010kv}, we conclude that 
\be  \Delta  \cE_\grad811 = - 60  \cE_\grad811  - (\cE_\grad844)^2 \ .\label{Laplace811R4} \ee
As explained in \cite{Minimal}, this requires then to modify \eqref{CubicC} to 
\bea \cD_{ijpq} \cD^{pqmn} \cD_{mnkl}  \cE_\grad811 &=& - 9 \cD_{ijkl}   \cE_\grad811 - \frac{1}{2} \cE_\grad844 \cD_{ijkl} \cE_\grad844  \ , \label{CubicCR4}\CR
 \cD^{ijpq} \cD_{pqmn} \cD^{mnkl} \cE_\grad811 &=& -9  \cD^{ijkl} \cE_\grad811 - \frac{1}{2} \cE_\grad844 \cD^{ijkl} \cE_\grad844 \ . \eea
Using in particular the tensor product 
\be {\scriptstyle [0,0,0,2,0,0,0]}\otimes {\scriptstyle[0,0,0,1,0,0,0]}\cong {\scriptstyle [0,0,0,3,0,0,0]}\oplus {\scriptstyle [0,1,0,1,0,1,0] } \oplus {\scriptstyle [1,0,0,1,0,0,1] }\oplus  {\scriptstyle [0,0,1,1,1,0,0]}\oplus  {\scriptstyle [0,0,0,1,0,0,0]} \nn \ee
one shows that 
\be \Scal{ 4 \cD_{ijpq} \cD^{pqmn} \cD_{mnkl} - \cD_{ijkl} \scal{ \Delta + 24} }   (\cE_\grad844)^2 = 0 \ , \ee
whereas
\be \Scal{ 36 \cD_{jr[kl} \cD^{irmn} \cD_{pq]mn} - \delta^i_j \cD_{klpq} ( \Delta + 42)   + \delta^i_{[k} \cD_{lpq]j} ( \Delta-120)}   (\cE_\grad844)^2  \ne 0 \ . \ee
We therefore conclude that no higher derivative correction in $ (\cE_\grad844)^2$ can consistently modify \eqref{CubicCR4} without contradicting \eqref{Laplace811R4}. 

These properties of the source term in the Poisson equation \eqref{Laplace811R4} can also be understood through the structure of the Fourier modes of these functions. In the decompactification limit, the Fourier modes of a function are the coefficients,  functions on $E_{6(6)}/\Sp$, of $e^{2\pi i (q,a)}$, with the axion field $a$ in the ${\bf 27}$ of the $E_{6(6)}$ subgroup. We have shown in \cite{Minimal} that \eqref{R4Quadra} implies then that the associated momenta are rank one vectors, \ie using the cubic Jordan norm $3 \, \det(q) = \tr \, q (q\times q )$,
\be q\times q = 0\ ,\label{qxq} \ee
 consistently with the properties of the $R^4$ threshold function.  If we consider the square of $\cE_\grad844$, it admits by construction Fourier modes of momenta $q_1 + q_2$ where $q_1$ and $q_2$ satisfy to \eqref{qxq}, such that 
 \be \det (q_1+q_2) = \det(q_1) + \tr \, q_1 (q_2\times q_2 )+ \tr \, q_2 (q_1\times q_1 )+ \det(q_2) = 0 \ , \ee
As one can see in \cite{Minimal}, equation \eqref{Laplace811R4}  implies that the Fourier modes of the function $\cE_\grad811$ must indeed carry momenta satisfying to the rank 2 constraint $\det(q)=0$, whereas the Fourier modes of the function $\cE_\grad820$ are generic by construction in the parabolic decomposition. The  nilpotent orbit associated to  $\cE_\grad820$  is indeed defined from the graded decomposition 
\be \mathfrak{e}_{7(7)} \cong \overline{\bf 27}^\ord{-2}\oplus \scal{ \mathfrak{gl}_1\oplus \mathfrak{e}_{6(6)}}^\ord{0} \oplus {\bf 27}^\ord{2}\ , \ee
such that a representative of the nilpotent orbit is a generic element of the grad two component in the ${\bf 27}$. 

Considering instead the string theory limit, the non-abelian Fourier modes are defined over a Heisenberg algebra with $32$ momenta in the positive chirality Weyl spinor representation of $Spin(6,6)$ associated to Ramond--Ramond D-brane charge $Q$, and an additional momentum associated to the Neveu--Schwarz 5-brane charge $N_5$. The nilpotent orbit associated to $\cE_\grad811$ is defined from the associated graded decomposition 
\be \mathfrak{e}_{7(7)} \cong{\bf 1}^\ord{-4}  \oplus {\bf 32}^\ord{-2}\oplus \scal{ \mathfrak{gl}_1\oplus \mathfrak{so}(6,6)}^\ord{0} \oplus {\bf 32}^\ord{2}\oplus {\bf 1}^\ord{4} \ . \label{D6Grad}\ee
A representative of the nilpotent orbit is defined as a generic Weyl spinor in the grad 2 component \cite{SpinorOrbits}, 
\be Q \in Spin(6,6)/ SU(2,4) \ , \quad {\rm or}\quad Q \in Spin(6,6)/ SL(6) \ , \label{OrbitsRank4}  \ee
to which one can add an arbitrary element of the grad 4 component $N_5$. This implies in particular that equation \eqref{CubicCR4} does not imply any constraint on the Fourier modes. Equation \eqref{R4Quadra} implies instead that $Q$ must be a rank 1 spinor, \cite{SpinorOrbits}
\be(Q^2)|_{\bf 66} \ \hat{=}\   (Q\Gamma^{MN} Q) = 0 \ , \qquad Q \in Spin(6,6)/ \scal{SL(6)\ltimes \mathds{R}^{15}} \ , \ee
 as for example the grad 3 singlet in the decomposition 
\bea \mathfrak{so}(6,6) &\cong& \overline{\bf 15}^\ord{-2}\oplus \scal{ \mathfrak{gl}_1\oplus \mathfrak{sl}_6}^\ord{0} \oplus {\bf 15}^\ord{2}\ , \CR 
 {\bf 32}& \cong& {\bf 1}^\ord{-3} \oplus {\bf 15}^\ord{-1} \oplus \overline{\bf 15}^\ord{1} \oplus  {\bf 1}^\ord{3} \ . 
 \eea
A generic rank 1 charge vector can always be rotated to the grad 3 component. Considering the sum of two rank one charges, respectively in the grad -3 and the grad 3 components, one obtains a generic rank 4 spinor of stabilizer $SL(6)\subset Spin(6,6)$. All the rank four charges defined as the sum of two rank 1 charges with a non-trivial symplectic product can be written in this form. Therefore the right-hand-side in \eqref{CubicCR4}  indeed sources generic Fourier modes of $\cE_\grad811$. More precisely, all the Fourier modes with a negative quartic invariant $I_4(Q)\le 0$ (belonging to the second orbit in \eqref{OrbitsRank4}) are sourced by the function $\cE_\grad{8}{1}{1}^{\; 2}$, whereas the Fourier modes with a strictly positive quartic invariant $I_4(Q)$  (belonging to the first orbit in \eqref{OrbitsRank4}) satisfy to a homogeneous equation. 

On the contrary, a representative of the nilpotent orbit associated to $\cE_\grad820$ satisfies that its third power in the adjoint representation vanishes, which according to \eqref{D6Grad} implies that \cite{SpinorOrbits}
\be (Q^3)|_{{\bf 32}}  \ \hat{=}\  ( Q \Gamma^{MN} Q  ) \Gamma_{MN} Q = 0 \ , \quad Q \in Spin(6,6)/\scal{ SL(2)\times Spin(3,4)\ltimes \mathds{R}^{2\times 8 +1}} \ .\label{Q3String}  \ee 
The relation with the Fourier modes is not completely straightforward in the presence of a non-trivial NS5-brane charge, because in that case the nilpotent subgroup is a non-abelian Heisemberg group, such that the corresponding Killing vector 
\be \kappa_\alpha = \frac{\partial\, }{\partial a^\alpha} - \frac{1}{2}C_{\alpha\beta} a^\beta\frac{\partial\, }{\partial b} \ , \qquad k_5 = \frac{\partial\, }{\partial b} \ , \ee
satisfy to
\be [ \kappa_\alpha , \kappa_\beta ] = C_{\alpha\beta} k_5 \ , \ee
where $C_{\alpha\beta}$ is the antisymmetric charge conjugation matrix of $Spin(6,6)$. For a Fourier mode of vanishing NS5-brane charge, $k_5 \cE_{Q,0} = 0 $, and one can define the spinor charge $Q$ such that $\kappa_\alpha   \cE_{Q,0}  = i  Q_\alpha  \cE_{Q,0}$, and $Q$ must satisfy to the same algebraic equations as the representatives of the nilpotent orbits associated to the differential equations. For a non-zero NS5-brane charge the relevant equations are more complicated, but still involve the Killing vector $\kappa_\alpha$ to the third order in the same combination.

Let us now consider the M-theory limit, for which one considers the decomposition 
\be \mathfrak{e}_{7(7)} \cong{\bf 7}^\ord{-4}  \oplus \overline{\bf 35}^\ord{-2}\oplus \scal{ \mathfrak{gl}_1\oplus \mathfrak{sl}_7}^\ord{0} \oplus {\bf 35}^\ord{2}\oplus \overline{\bf 7}^\ord{4} \ , \label{A6Grad}\ee
In this case the nilpotent subgroup also generate a non-abelian Heisenberg type algebra 
\be \kappa_{mnp} = \frac{\partial\, }{\partial a^{mnp} } - \frac{1}{12} \varepsilon_{mnpqrst} a^{qrs}  \frac{\partial\, }{\partial b_t } \ , \qquad k^m =  \frac{\partial\, }{\partial b_m }  \ , \ee
such that 
\be [\kappa_{mnp} ,\kappa_{qrs}] = \frac{1}{6}  \varepsilon_{mnpqrst}  k^t \ . \ee
For a Fourier mode of vanishing M5-brane charge, $k^m  \cE_{q,0} = 0 $, and one can define the M2-brane charge $\kappa_{mnp}  \cE_{q,0}  = i  q_{mnp}  \cE_{q,0}$. For a non-zero M5-brane charge $k^m  \cE_{q,p} = i p^m \cE_{q,p}  $ the relevant equations are more complicated, but still involve the Killing vector in a way similar as does the corresponding nilpotent orbit characteristic equation involves the algebraic charges. For a 1/2 BPS charge satisfying to the quadratic constraint, one obtains \cite{Lu:1997bg}
\be \varepsilon^{mnqrstu} q_{pqr} q_{stu} = 0 \ , \qquad q_{mnp} p^p = 0 \ , \ee
giving $17=13+4$ linearly independent solutions, with typical representative
\be \frac{1}{6} q_{mnp} dy^m \wedge dy^n \wedge dy^p = q_1 dy^1 \wedge dy^2 \wedge dy^3 \ . \ee
The cubic constraint in the adjoint representation implies 
\be \varepsilon^{nrstuvw} q_{rst}q_{uv[p} q_{qm]w} = 0 \ ,  \qquad \varepsilon^{mnqrstu} q_{pqr} q_{stu} p^p = 0 \ , \ee 
that gives $27=21+6$ linearly independent solutions, with typical representative
\be  \frac{1}{6} q_{mnp} dy^m \wedge dy^n \wedge dy^p = dy^1 \wedge \scal{ q_1 dy^2 \wedge dy^3 + q_2 dy^4 \wedge dy^5 + q_3 dy^6 \wedge dy^7}  \ . \label{Rank3} \ee
 The cubic constraint in the fundamental implies instead 
\be  \varepsilon^{mnrstuv} q_{mnr}q_{st(p} q_{q)uv} = 0 \ , \label{Rank4} \ee
that gives $33=26+7$ linearly independent solutions, such that the $SL(7)$ M2-brane charge orbits are either 
\be q \in SL(7,\mathds{R})/ (SL(3,\mathds{C}) \ltimes \mathds{C}^3) \ , \quad {\rm or}\quad q\in SL(7,\mathds{R})/ (SL(3,\mathds{R}) \ltimes \mathds{R}^3)^{\times 2}  \ , \label{SL7orbit} \ee
with typical representative 
\be   \frac{1}{6} q_{mnp} dy^m \wedge dy^n \wedge dy^p = q_1 dy^1 \wedge  dy^2 \wedge dy^3 + q_2 dy^1\wedge  dy^4 \wedge dy^5 + q_3 dy^2 \wedge dy^6 \wedge dy^4 + q_4 dy^3 \wedge dy^5 \wedge dy^6  \ , \ee 
and 
\be  \frac{1}{2\times 12^3}\varepsilon^{pmnrstu}  \varepsilon^{qm^\prime n^\prime r^\prime s^\prime t^\prime u^\prime } q_{mnr}q_{stm^\prime} q_{n^\prime r^\prime u} q_{s^\prime t^\prime u^\prime} =  q_1 q_2 q_3  q_4 \delta^p_7 \delta^q_7 \ , \ee
such that the orbit \eqref{SL7orbit} is determined by the sign of the eigenvalue of this rank one symmetric tensor, $I_4 =  q_1 q_2 q_3 q_4$. The generic sum of two rank one charges takes the form 
\be \frac{1}{6} q_{mnp} dy^m \wedge dy^n \wedge dy^p = q_1 dy^1 \wedge dy^2 \wedge dy^3  + q_2dy^4 \wedge dy^5 \wedge dy^6 \ ,\ee
and is a generic solution to \eqref{Rank4} associated to the second orbit (\ie $I_4<0$), and violates equation \eqref{Rank3}. Therefore we confirm that a quadratic source in $\cE_\grad{8}{4}{4}$ is in contradiction with the cubic equation satisfied by $\cE_\grad{8}{2}{0}$, whereas it is consistent with the one satisfied by $\cE_\grad{8}{1}{1}$.

Let us now argue that all the invariants of the infinite series of $\bar F^{2k} \nabla^4 R^4$ do not get modified at the same order by lower order modifications to the supersymmetry transformations. By power counting, the next order correction to the $R^4$ type invariant and a $\bar F^{2k} \nabla^4 R^4$ type invariant can in principle contribute to a right-hand-side for the classical supersymmetry variation of a $\bar F^{2k+6} \nabla^4 R^4$ type invariant. So in principle one could expect that the function $\cE_\grad820^\ord{k}$ satisfies to a Poisson equation of the kind 
\begin{multline} \Delta \cE_\grad820^\ord{k} =3 (k+4)(k-5) \cE_\grad820^\ord{k}   - a^\ord{k}_3 \cE_\grad844 \, \cE_\grad820^\ord{k-3} \\ - a^\ord{k}_5   \cE_\grad822 \, \cE_\grad820^\ord{k-5} - a^\ord{k}_6 (\cE_\grad844)^2 \, \cE_\grad820^\ord{k-6} - \sum_{p=0}^{k-8} b^\ord{k}_{p}   \cE_\grad820^\ord{k-8-p} \cE_\grad820^\ord{p} + \dots \end{multline}
However, the solutions to the differential equation \eqref{CubicK} admit restricted Fourier modes in the string theory limit, satisfying to \eqref{Q3String} for a vanishing NS5-brane charge. As we have already explained, the product of two functions including non-perturbative corrections admits generic Fourier modes in the string theory limit, because the sum of two pure spinors can be a generic spinor. We see therefore that a source term modifying \eqref{CubicK} would necessarily involve the third order differential operator such as to source these Fourier modes. Such a modification would destroy completely the structure of the equations, which would reduce then to some kind of Poisson equation.

\subsection{String theory perturbation theory}
\label{StringPerturb} 
In order to deduce constraints on the contributions that can possibly appear in perturbative string theory, it is important to solve the differential equations satisfied by the threshold functions in the parabolic gauge with manifest T-duality symmetry \eqref{D6Grad}. In this section we will solve these equations on an ansatz function depending only on the string theory dilaton $e^{2\phi}$ and the scalar fields parametrising $SO(6,6)/(SO(6)\times SO(6))$. We have not computed explicitly the decomposition of the differential equations, but using the manifest covariance, and the known solutions for the $R^4$ and the $\nabla^4 R^4$ threshold functions \cite{Green:2010kv}, we can determine unambiguously all the unknown coefficients. 

We define the covariant derivative $\cD_{a\hat{b}}$ on $SO(6,6)/(SO(6)\times SO(6))$ in tangent frame, such that $a=1$ to $6$ of one $SO(6)$ and $\hat{b}=1$ to $6$ of the other. It is convenient to define the covariant derivative as an $SU(4)\times SU(4)$ tensor
\be \cD_{ij\hat{k}\hat{l}} = \frac{1}{4} \gamma^a{}_{ij} \gamma^{\hat{b}}{}_{\hat{k}\hat{l}} \cD_{a\hat{b}} \ , \ee
with $i=1,\, 4$ of one $SU(4)$ and $\hat{\imath}=1,\, 4$ of the other. 

Calibrating the equations on the known solutions, one obtains that 
\be {\bf D}_{56}^{\;3} \cE = \scal{ s^2 - \tfrac{17}{2} s+6}    {\bf D}_{56} \cE \label{D563sString} \ee
decomposes on $\mathds{R}_+^* \times SO(6,6)/(SO(6)\times SO(6))$ as
\bea\biggl(  \Scal{  \frac{1}{64} \partial_\phi^{\; 3} + \frac{17}{32} \partial_\phi^{\; 2} + \frac{3}{2} \partial_\phi -  \cD_{a\hat{b}} \cD^{a\hat{b}} } \delta_a^b + \Scal{ \frac{3}{4} \partial_\phi + 6 } \cD_{a\hat{c}} \cD^{b\hat{c}} \biggr) \cE &=&\scal{ s^2 - \tfrac{17}{2} s+6}   \ \delta_{a}^b \,  \frac{1}{4} \partial_\phi\cE \ , \qquad \CR
\biggl( \cD_{a\hat{c}} \cD^{d\hat{c}} \cD_{d\hat{b}} + \Scal{ \frac{3}{16} \partial_\phi^{\; 2} + \frac{31}{8} \partial_\phi + 9 } \cD_{a\hat{b}} \biggr) \cE &=&\scal{ s^2 - \tfrac{17}{2} s+6}   \cD_{a\hat{b}} \cE\ ,  \CR
\biggl( 8 \cD_{ip}{}^{\hat{k}\hat{q}} \cD^{pr}{}_{\hat{q}\hat{s}} \cD_{jr}{}^{\hat{l}\hat{s}} + \Scal{ \frac{5}{4} \partial_\phi + 2} 2\cD_{ij}{}^{\hat{k}\hat{l}}\biggr) \cE &=& 2\scal{ s^2 - \tfrac{17}{2} s+6}  \cD_{ij}{}^{\hat{k}\hat{l}} \cE \ , \ 
\eea
whereas 
\be {\bf D}_{133}^{\;3} \cE =s(s-9)   {\bf D}_{133} \cE \label{AdjointEquation}  \ee
gives the components in the ${\bf 32}$ of $Spin(6,6)$
\bea \biggl(\Scal{ \frac{1}{64} \partial_\phi^{\; 3} + \frac{5}{8} \partial_\phi^{\; 2} - \frac{5}{16} \partial_\phi - \cD_{pq{\hat{r}\hat{s}}} \cD^{pq{\hat{r}\hat{s}}} } \delta_i^k \delta_{\hat{\jmath}}^{\hat{l}} + 3 ( \partial_\phi +6)  \cD_{ip\hat{\jmath}\hat{q}} \cD^{kp\hat{l}\hat{q}} \biggr) \cE &=& s(s-9) \ \delta_i^k \delta_{\hat{\jmath}}^{\hat{l}}\,  \frac{1}{4} \partial_\phi \cE\ , \qquad  \CR
\biggl(8  \cD_{ip\hat{\jmath}\hat{q}} \cD^{pr\hat{q}\hat{s}}\cD_{kr\hat{l}\hat{s}} + \Scal{ \frac{3}{16} \partial_\phi^{\; 3} + \frac{31}{8} \partial_\phi}2 \cD_{ik\hat{\jmath}\hat{l}} \biggr) \cE &=&   2 s(s-9)\cD_{ik\hat{\jmath}\hat{l}} \cE \ . \label{D1333sString}
 \eea
The $\nabla^4 R^4$ threshold function solves \eqref{D563sString} for $s=\frac{3}{2}$ and \eqref{D1333sString} for $s=4$. One reads directly from these equations, that a solution of type $e^{a\phi} \cE_{D_6}$  on $\mathds{R}_+^* \times SO(6,6)/(SO(6)\times SO(6))$ must be such that $\cE_{D_6}$ satisfies to the quadratic equations in all fundamental representations (\ie the vector and Weyl spinor of positive and negative chirality), unless $a=-6$ or $a=-8$. The only other solutions are therefore such that $\cE_{D_6}$ is either a constant, or solves \eqref{MinimalD6}. One finds the unique  solution $e^{-10\phi}$.  For the values $a=-8$, the function $\cE_{D_6}$ satisfies to a quadratic equation in the spinor representation, and solve \eqref{VecD6} for $s=4$ (or $1$ which is equivalent). For $a=-6$, $\cE_{D_6}$ satisfies to a quadratic equation in the vector representation, and cubic equations in the two spinor representations, and must therefore satisfy to  \eqref{NextMinimalD6}. We find therefore that supersymmetry and T-duality alone already determine the $\nabla^4 R^4$ type corrections in perturbative string theory, up to three free coefficients, that are given in \cite{Green:2010kv}
\be \frac{1}{2} {E}_{\mbox{\DEVII{\frac{5}{2}}00000{\mathfrak{0}}}}  = \zeta(5) e^{-10\phi}  + \frac{4}{15\pi} e^{-8\phi} E_{\mbox{\WSOXII{4}00000}}  + \frac{2}{3} e^{-6\phi} E_{\mbox{\WSOXII{0}00020}} + \mathcal{O}(e^{-e^{-\phi}})  \ . \label{E4} \ee
This confirms that supersymmetry alone already prevents any perturbative correction to the $\nabla^4 R^4$ threshold function beyond 2-loop in perturbative string theory. 

The functions defining the $\bar F^{2k} \nabla^4 R^4$ solve equation  \eqref{D1333sString} for $s=k+4$. 
Similarly one obtains the general $SO(6,6,\mathds{Z})$ invariant solution 
 \be \cE_\grad820^\ord{k} = c^\ord{k}_1 e^{-2(k+4)\phi} {E}_{\mbox{\WSOXIIT{k\mbox{+}4}000{\mathfrak{0}}0}}   + c_{k+2}^\ord{k} e^{-6\phi} {E}_{\mbox{\WSOXIIT{\mathfrak{0}}000{k\mbox{+}2}0}} +c^\ord{k}_{2k}   e^{2(k-5)\phi} {E}_{\mbox{\WSOXIIT{k}000{\mathfrak{0}}0}} +\mathcal{O}(e^{-e^{-\phi}})  \ .\ee
It is quite remarkable that the only solutions we get all correspond to a strictly positive number of loops in perturbative string theory. After implementing the Weyl rescaling, one obtains indeed that $c^\ord{k}_\ell$ is a coefficient for a $\ell$-loop correction in string theory for the $\bar F^{2k} \nabla^4 R^4$ threshold function. For $k=1$ and $k=2$, equation   \eqref{D1333sString} is exact for the Wilsonian effective action (not taking into account linear corrections associated to logarithms in the complete effective action). $U$-duality therefore implies that $\cE_\grad820$ must be an Eisenstein function as in \eqref{EisF2k}. Assuming that our argumentation in the preceding section is correct, and that equation \eqref{D1333sString} is satisfied for all $k$, we arrive at the conjecture that  the $\bar F^{2k} \nabla^4 R^4$ threshold function is defined by the Eisenstein series $E{\mbox{\DEVII{0}00000{4\mbox{+}k}}}$ for all $k$. It is rather remarkable that this coupling would only get three corrections in perturbation theory, at 1-loop, $k+2$-loop and $2k$-loop.

This Eisenstein function diverges precisely for $k=1$, corresponding to the $\bar F^2 \nabla^4 R^4$ threshold related to the $\nabla^6 R^4$  threshold function by supersymmetry.  One must therefore consider the regularised Eisenstein series 
\begin{multline}  \frac{16}{63} \hat{{E}}_{\mbox{\DEVII000000{5}}} = \frac{16}{63}e^{-10\phi}  \hat{E}_{\mbox{\WSOXIIT{5}000{\mathfrak{0}}0}}-\frac{15\zeta(5)}{4} \phi\,  e^{-10\phi}  + \frac{1}{\pi} e^{-8\phi} \Scal{ 2 \phi  {E}_{\mbox{\WSOXIIT{4}000{\mathfrak{0}}0}} - \partial_s {E}_{\mbox{\WSOXIIT{s}000{\mathfrak{0}}0}} \big|_{s=4} }   \\ + \frac{\pi}{9} e^{-6\phi} \hat{E}_{\mbox{\WSOXIIT{\mathfrak{0}}000{3}0}}  +   \mathcal{O}(e^{-e^{-\phi}}) \ .  \end{multline} 
Here we have fixed all the coefficients by consistency with \eqref{DivergenceF2D4R4} and \eqref{E4}. The logarithm of the dilaton indicates a divergence of the $\nabla^4 R^4$ form factor into  $\nabla^6 R^4$ in supergravity. Note nonetheless that the 3-loop contribution in the last line violate T-duality parity in $O(6,6,\mathds{Z})$, and the string theory effective action must include the same function with opposite chirality. Because it is a three-loop contribution, it cannot come from the completion of the $R^4$ type invariant and it must appear as a solution to equation \eqref{D563sString} for $s=6$. 

Considering the general $SO(6,6,\mathds{Z})$ invariant solution of \eqref{D563sString}, one finds indeed 
\be {\cal E}_\grad811  =  c_{\mbox{-}6}e^{-4s\phi} + c_{\mbox{-}\frac{1}{2}}e^{-(2s+1)\phi}{E}_{\mbox{\WSOXII{0}0000{\mathnormal{s}\mbox{-}\hspace{-0.2mm}\stfrac{1}{2}}}}  +c_1 e^{2(2s-17)\phi}  +c_2 e^{-8\phi} {E}_{\mbox{\WSOXII0{\mathnormal{s}\mbox{-}2}0000}}+ c_3 e^{2(s-9)\phi} {E}_{\mbox{\WSOXII00000{\mathnormal{s}\mbox{-}3}}}  \ee
where the coefficients $c_\ell$ are constants that would correspond to $\ell$-loop contributions for the $\nabla^6 R^4$ threshold function in string theory.  Note that the first two terms do not make sense in perturbative string theory. The corresponding Eisenstein function ${E}{\mbox{\DEVII{s}00000{\mathfrak{0}}}}$ includes generically all these terms, and therefore cannot define the string theory threshold function, consistently with the property that \eqref{D563sString} is corrected by a source term \eqref{CubicCR4}. However, the three-loop contribution is not affected by the source term, and one can take seriously the last contribution, which is precisely the one required to restore  $O(6,6,\mathds{Z})$ invariance  for $s=6$.

This is indeed confirmed by the expression obtained in \cite{Pioline:2015yea} for the $\nabla^6 R^4$ threshold function, and using these results we conclude therefore that the exact threshold function for the $\nabla^6 R^4$ coupling is defined as
\be E_{\gra{0}{1}} = \hat{\cE}_{\grad{8}{1}{1}}  + \frac{32}{189\pi}\hat{{E}}_{\mbox{\DEVII000000{5}}}\ ,  \ee
where the function $ \cE_{\grad{8}{1}{1}}$ solve the differential equation 
\bea\label{E811Equation} \Delta \hat{\cE}_{\grad{8}{1}{1}} &=& - 60 \hat{\cE}_{\grad{8}{1}{1}} - \Scal{ {E}_{\mbox{\DEVII{\mathnormal{\tfrac{3}{2}}}00000{\mathfrak{0}}}}}^2 + \frac{35}{\pi} \Scal{ \frac{1}{2}  {E}_{\mbox{\DEVII{\mathnormal{\tfrac{5}{2}}}00000{\mathfrak{0}}}}}\ \\
 \cD_{ijpq} \cD^{pqmn} \cD_{mnkl}  \hat{\cE}_\grad811 &=& - 9 \cD_{ijkl}   \hat{\cE}_\grad811 - \frac{1}{2} {E}_{\mbox{\DEVII{\mathnormal{\tfrac{3}{2}}}00000{\mathfrak{0}}}} \cD_{ijkl} {E}_{\mbox{\DEVII{\mathnormal{\tfrac{3}{2}}}00000{\mathfrak{0}}}}  + \frac{35}{4\pi}   \cD_{ijkl}  \Scal{ \frac{1}{2}  {E}_{\mbox{\DEVII{\mathnormal{\tfrac{5}{2}}}00000{\mathfrak{0}}}}}\ , \CR
 \cD^{ijpq} \cD_{pqmn} \cD^{mnkl} \hat{\cE}_\grad811 &=& -9  \cD^{ijkl} \hat{\cE}_\grad811 - \frac{1}{2}{E}_{\mbox{\DEVII{\mathnormal{\tfrac{3}{2}}}00000{\mathfrak{0}}}} \cD^{ijkl} {E}_{\mbox{\DEVII{\mathnormal{\tfrac{3}{2}}}00000{\mathfrak{0}}}} +\frac{35}{4\pi}   \cD^{ijkl}  \Scal{ \frac{1}{2}  {E}_{\mbox{\DEVII{\mathnormal{\tfrac{5}{2}}}00000{\mathfrak{0}}}}}\ . \nn
\eea
Here the anomalous right-hand-side is determined such as to coincide with the one obtained in \cite{Pioline:2015yea} for the complete function $ E_{\gra{0}{1}}$. These coefficients can also be directly computed from the properties of the Eisenstein functions and the structure of the differential equations \cite{Axel}. 
\section{Supergravity in higher dimensions}
\label{HigherDimensions}
In this section we will consider the extension of the results of the preceding section in five, six, seven and eight dimensions. We will see that the two $\nabla^6 R^4$ type invariants both lift to higher dimensions, even if they cannot be defined as harmonic superspace integrals in the linearised approximation in general.

\subsection{$\cN =4$ supergravity in five dimensions}
Let us recall in a first place some properties of maximal supergravity in five dimensions. The scalar fields parametrise the symmetric space $E_{6(6)}/Sp(4)_{\scriptscriptstyle \rm c}$. We use  $i, j = 1, ..., 8$ as indices in the fundamental representation of  $Sp(4)$, and $\O^{i j}$ defines the symplectic form with the normalisation $\O^{i k} \O_{j k} = \delta^{i}_{j} $. The covariant derivative in tangent frame $\cD_{i j k l}$ is a symplectic traceless rank four antisymmetric tensor in the representation $[0,0,0,1]$ of $Sp(4)$.

\addtocontents{toc}{\protect\setcounter{tocdepth}{1}}
\subsubsection{Linearised $\nabla^6 R^4$ type invariants}
\addtocontents{toc}{\protect\setcounter{tocdepth}{2}}
In five dimensions there is only one kind of 1/8 BPS harmonic superspace integral that one can define \cite{Bossard:2009sy}. For this purpose, one considers $Sp(4)/(U(1)\times Sp(3))$ harmonic variables $(u^{1}{}_{i}, u^{r}{}_{i}, u^{8}{}_{i})$ with $r = 2, ... ,7$ in the fundamental of  $Sp(3)$, and the decomposition 
\bea
\mathfrak{sp}(4) &\cong& {\bf 6}^\ord{-1} \oplus \scal{ \mathfrak{u}(1)\oplus \mathfrak{sp}(3)}^\ord{0} \oplus {\bf 6}^\ord{1} \CR
{\bf 42} &\cong&  {\bf 14}^{\ord{-1}}_{3} \oplus {\bf 14}^{\ord{0}}_{2} \oplus {\bf 14}^{\ord{1}}_{3} \ . 
\eea
One defines the G-analytic superfield $W^{r s t}$ in the $[0,0,1]$ of $Sp(3)$
\be
W^{r s t} = u^{1}{}_{i}  u^{r}{}_{j} u^{s}{}_{j} u^{t}{}_{k} L^{i j k l} \ , 
\ee
which satisfies the constraint \be
u^1{}_{ i} D_{\alpha}^{i} W^{r s t} = 0 \ . 
\ee
Following the same reasoning as in section \ref{811D6R4}, we consider a general monomial of $W^{rst}$ in an irreducible representation of $Sp(3)$. In this case we obtain equivalently that the monomials are freely generated by $W^{rst}$ in the $[0,0,1]$,  the elements 
\be W^{r t p} W^{s q r} \O_{t q} \O_{p r} \ , \ee
in the $[2,0,0]$,
\be W^{r p q} W^{s u}{}_{p} W^{t}{}_{q u} \, , \ee
in the $[0,0,1]$, and  
\be W^{r] t u} W^{[s}{}_{t u} W^{p] v w} W^{[q}{}_{v w} - {\rm ``symp\ trace"} \ , \qquad W^{r s t} W^{p q}{}_{r} W^{u}{}_{s p} W{}_{t q u}\ , \ee
respectively in the $[0,2,0]$ and the singlet representation. The general linearised invariant takes therefore the form 
\bea
&&  \int du D^{28} F(u)^{[2n_2,2n_4,n_1 + n_3]}_{[2 n_3 + 2 n_{4} + 4 n'_{4}, 2n_2, 2n_4, n_1 + n_3]} W^{4 + n_1 + 2 n_2 + 3 n_3 + 4 n_4 + 4 n'_4}_{[2n_2,2n_4,n_1 + n_3]} \CR
&=&   L^{n_1 + 2 n_2 + 3 n_3 + 4 n_4 + 4 n'_4}_{[2 n_3 + 2 n_{4} + 4 n'_{4}, 2n_2, 2n_4, n_1 + n_3]} \scal{ \nabla^6 R^4  + \dots } + \dots 
\eea
The structure of these linearised invariants suggests that the complete non-linear invariant admits the following gradient expansion 
\be
\cL_{\gra{4}{1}}[ \cE_{\gra{4}{1}}] = \sum_{n_1,n_2,n_3,n_4,n_4^\prime} \cD^{n_1 + 2 n_2 + 3 n_3 + 4 n_4 + 4 n'_4}_{[2 n_3 + 2 n_{4} + 4 n'_{4}, 2n_2, 2n_4, n_1 + n_3]} \cE_{\gra{4}{1}} \cL_{\gra{4}{1}}^{[2 n_3 + 2 n_{4} + 4 n'_{4}, 2n_2, 2n_4, n_1 + n_3]}\ . 
\ee
The consistency of this ansatz requires that the function $\cE_{\gra{4}{1}} $ must be an eigenfunction of the Laplace operator, and that its third order derivative restricted to the $[0,2,0,0]$ is proportional to its second derivative in the same representation. This linearised analysis is consistent with the one of the $(8,1,1)$ type invariant in four dimensions, and we are going to see that the relevant equation is 
\be \cD_{ijpq} \cD^{pqrs} \cD_{rskl}  \cE_{\gra{4}{1}} = \frac{1}{4} \cD_{ijkl} \scal{ 34 + \Delta}  \cE_{\gra{4}{1}}  \ .\ee
However, the $(8,2,0)$ type invariants cannot be defined in the linearised approximation through a harmonic superspace integral, and we shall instead consider the uplift of the general invariant to five dimensions. 

\subsubsection{Decompactification limit from four to five dimensions}
We are therefore going to solve the differential equations  \eqref{CubicR} and \eqref{E811Equation}  for a function depending only of the Levi subgroup $\mathds{R}_*^+\times  E_{6(6)}$ of the parabolic subgroup associated to the decompactification limit, such that 
\be \mathfrak{e}_{7(7)}\cong \overline{\bf 27}^\ord{-2} \oplus \scal{ \mathfrak{gl}_1 \oplus \mathfrak{e}_{6(6)} }^\ord{0} \oplus {\bf 27}^\ord{2} \ , \qquad 
{\bf 56} \cong {\bf 1}^\ord{-3} \oplus {\bf 27}^\ord{-1} \oplus \overline{\bf 27}^\ord{1} \oplus {\bf 1}^\ord{3} \ .\label{56E6}  \ee
For this purpose we use the same conventions as in \cite{Minimal}, such that the coset representative in $E_{7(7)}/SU(8)_{\scriptscriptstyle \rm c}$ is defined as
\be \cV = \left( \begin{array}{cccc} \ e^{3\phi} \ & \ 0 \ & \ 0 \ & \ 0 \\
0 &\ e^{\phi} V_{ij}{}^I \ &0&0\\
0&0& \ e^{-\phi} V^\inv{}_I{}^{ij} \ &0\\
0&0&0& e^{-3\phi} \end{array}\right) \left( \begin{array}{cccc} \ 1 \ & \ a^J \ & \ \tfrac{1}{2} t_{JKL} a^K a^L  \ & \ \tfrac{1}{3} t_{KLP} a^K a^L a^P \\
0 &\ \delta_I^J \ &t_{IJK} a^K&\tfrac{1}{2} t_{IKL} a^K a^L\\
0&0& \delta_J^I \ &a^I \\
0&0&0& 1 \end{array}\right)  \ , \label{corset_elem} \ee
where $V_{i j}{}^{I}$ is the coset representative in $E_{6(6)}$ with the $Sp(4)$ pair $ij$ being antisymmetric symplectic traceless and the index $I$ in the fundamental of $E_{6(6)}$. $t_{I J K}$ is the $E_{6(6)}$ invariant symmetric tensor normalised as in \cite{Minimal}. We have already computed the decomposition of the cubic equation \eqref{Cubic56} in \cite{Minimal}, which is
\bea \Bigl( \frac{1}{64} \partial_\phi^{\; 3} + \frac{21}{32} \partial_\phi^{\; 2} +\frac{9}{2} \partial_\phi - \frac{3}{4} \Delta \Bigr) \cE_\grad{8}{1}{1}   &=& -  \frac{1}{4} \partial_\phi\Scal{9 \cE_\grad{8}{1}{1}  + \frac{1}{4}   \cE_\grad{8}{4}{4}^{\; 2}}  \\
 \biggl(  \cD_{ijpq} \cD^{pq}{}_{rs} \cD^{rskl}  + \cD_{ij}{}^{kl} \Bigl( \frac{1}{48} \partial_\phi^{\; 2} + \frac{27}{24} \partial_\phi + \frac{7}{2} \Bigr)  + \cD_{ijpq} \cD^{klpq} \Bigl( \frac{1}{4} \partial_\phi+3 \Bigr)\hspace{-32mm} && \CR +\delta_{ij}^{kl} \Bigl( \frac{1}{12^3} \partial_\phi^{\; 3} + \frac{5}{96} \partial_\phi^{\; 2} +\frac{1}{6} \partial_\phi - \frac{1}{4} \Delta  \Bigr) 
 \biggr) \cE_\grad{8}{1}{1} &=& -  \Bigl( \frac{1}{12} \delta_{ij}^{kl} \partial_\phi + \cD_{ij}{}^{kl} \Bigr)  \Scal{ 9 \cE_\grad{8}{1}{1}  + \frac{1}{4}   \cE_\grad{8}{4}{4}^{\; 2}} \nn \ ,
 \eea
 where 
 \be
\delta_{ij}^{kl} = \delta_{[i}^{[k} \delta_{j]}^{l]} - \frac{1}{8} \Omega_{ij} \Omega^{k l} \ , \qquad \Delta = \frac{1}{3} \cD_{ijkl} \cD^{ijkl} \ ,
\ee
and indices are raised and lowered with the symplectic matrix $\Omega_{ij}$.  Because of the Weyl rescaling required to stay in Einstein frame, the relevant radius power in the decompactification limit for a $\nabla^{2n} R^4$ threshold function is such that  $\cE_{E_{7}} = e^{-( 6 + 2 n)\phi} \cE_{E_6}$, and because we are interested in the constraint on the $\nabla^{6} R^4$ threshold function, we use the ansatz 
 \be
 \cE_{\grad{8}{1}{1}} = e^{- 12 \phi} \cE_{\gra{8}{1}} \ , \qquad  \cE_\grad844 = e^{-6 \phi}  \cE_{\gra{8}{4}} \label{LaplaceE6}  \ ,
 \ee
where $ \cE_{\gra{8}{1}}$ and $ \cE_{\gra{8}{4}}$ are functions on $E_{6(6)}/Sp(4)_{\scriptscriptstyle \rm c}$. Using this ansatz, one derives 
 \be
\Delta \cE_{\gra{8}{1}} = - 18  \cE_{\gra{8}{1}}  - \cE^{\; 2}_{\gra{8}{4}} \ , 
 \ee
and
\be
\bigl( \cD_{ijpq} \cD^{pqrs} \cD_{rskl}  + 2  \cD_{i jkl} \bigr) \cE_{\gra{8}{1}} = - \frac{1}{4}  \cD_{i jkl} \cE^{\; 2}_{\gra{8}{4}} \ . \label{D6R4E6} 
\ee
These equations are satisfied by the 1/8 BPS threshold functions in the Wilsonian effective action, but the U-duality invariant function appearing in the 1PI effective action satisfied to anomalous equations with additional terms linear in $\cE_\gra{8}{4}$ in the right-hand-side \cite{Pioline:2015yea}. 

We shall now consider the uplift of the  $F^{2 k} \nabla^4 R^4$ type invariants, but for this purpose it will be more convenient to consider directly the decompactification limit of the Eisenstein function  ${E}{\mbox{\DEVII000000{k\mbox{+}4}}}$. We shall only consider the term with the correct power of the compactification radius $r$ to lift to a diffeomorphism invariant in five dimensions, as computed in appendix \ref{E66Eis},
\be \int d^4 x \, \sqrt{-g} \  {E}_{\mbox{\DEVII000000{k\mbox{+}4}}}\,  \nabla^{4+2k} R^4 \rightarrow  \frac{\pi^{\frac{1}{2}} \Gamma(k + \frac{7}{2})}{\Gamma(k+4)}   \int d^5x \, \sqrt{-g}\  E_{\mbox{\DEVI00000{k\mbox{+}\frac{7}{2}}}}   \nabla^{4+2k} R^4 \ . \ee
We conclude in this way that the threshold function  $\cE_{\frac{1}{8}}^\ord{k}$ defining the $F^{2k} \nabla^{4} R^4$ type invariants satisfies to 
\bea
\Scal{ \cD_{ijpq} \cD^{pq}{}_{rs} \cD^{rskl}  - \frac{2k(k+1)-10}{3} \cD_{ij}{}^{kl}} \cE_{\frac{1}{8}}^\ord{k} &=&
\frac{k+2}{2} \biggl(\cD_{i j p q} \cD^{k lp q}  - \frac{2(2 k-5) (2 k+7)}{27}  \delta_{ij}^{kl} \biggr) \cE_{\frac{1}{8}}^\ord{k} \CR
\Delta \cE_{\frac{1}{8}}^\ord{k} = \frac{2}{3} (2 k-5) (2 k+7) \cE_{\frac{1}{8}}^\ord{k} \ , && \qquad \cD^{3}_{[2,0,0,1]} \cE_{\frac{1}{8}}^\ord{k} = 0 \ . \label{F2kD4R4E6}
\eea
It follows from representation theory that such equations are indeed implied by \eqref{AdjointEquation}, and this explicit example permits to determine them uniquely. 

We therefore obtain that the threshold function is the regularised Eisenstein series 
\be \hat{\cE}_{\frac{1}{8}}^\ord{1} =   \frac{5}{108} \hat{E}_{\mbox{\DEVI00000{\mbox{$\frac{9}{2}$}}}} \ , \ee
such that the exact $\nabla^6 R^4$ threshold function $E_{(0,1)}$ is 
\be E_{(0,1)} =  \hat{\cE}_{\gra{8}{1}} +  \frac{5}{108}  \hat{E}_{\mbox{\DEVI00000{\mbox{$\frac{9}{2}$}}}}  \ .\ee
The series $E{\mbox{\DEVI00000{s}}}$ admits a pole at $s=\frac{9}{2}$ proportional to the series $E{\mbox{\DEVI{\mathnormal{\mbox{$\frac{3}{2}$}}}0000{\mathfrak{0}}}}$ defining the $R^4$ threshold, exhibiting that the $R^4$ invariant form factor diverges at two loop into the $\nabla^6 R^4$ form factor associated to the same function. This is in agreement with \cite{Pioline:2015yea}, where the explicit coefficient is computed.

\subsection{$\cN = (2,2)$ supergravity in six dimensions}
We shall now discuss these invariants in $\cN=(2,2)$ supergravity in six dimensions. We recall that the scalar fields parametrise in this case  the symmetric space $SO(5,5) / ( SO(5) \times SO(5) )$. 
\subsubsection{Linearised invariant}
In the linearised approximation, the theory is defined from the scalar superfield $L^{ij \hi \hj}$ in the  $[0,1] \times [0,1]$ of $Sp(2) \times Sp(2)$, where $i,j $ and $\hat{\imath},\hat{\jmath}$ run from $1$ to $4$ in the fundamental of the two respective $Sp(2)$. One can define a $\nabla^6 R^4$ type invariant by considering harmonic variables $u^{1}{}_{i}, u^{r}{}_{i} ,u^{4}{}_{i}$ parametrising $Sp(2)/(U(1)\times Sp(1))$ associated to one $Sp(2)$ factor, with $r=2,3$ of $Sp(1)$, such that \be
\mathfrak{sp}(2)  \cong {\bf 2}^\ord{-1}  \oplus \scal{  \mathfrak{u}(1)  \oplus \mathfrak{sp}(1) }^\ord{0}\oplus {\bf 2}^\ord{1}   \ , \qquad 
{\bf 4}  \cong{\bf  1}^{\ord{-1}} \oplus {\bf 2}^{\ord{0}} \oplus {\bf 1}^{\ord{1}} \ . \ee
One can in this way introduce the G-analytic superfield \cite{Bossard:2009sy}
\be
W^{r \hi \hj} = u^{1}{}_{i} u^{r}{}_{j} L^{ij \hi \hj} \ , 
\ee
that transforms in the fundamental of $Sp(1)$ and as a vector of $SO(5) \cong Sp(2)/\mathds{Z}_2$, and satisfies to the $1/8$ BPS G-analyticity  constraint 
\be
u^1{}_i D^{i}_{\alpha} W^{r, \hi \hj} = 0 \  . 
\ee
A general polynomial in $W^{r, \hi \hj}$ decomposes into irreducible representations of $Sp(1)\times Sp(2)$. Similarly as in lower dimensions, one shows that the latter are freely generated by $W^{r, \hi \hj}$ itself in the $[1]\times[0,1]$ of $SU(2)\times Sp(2)$, the two quadratic monomials \be 
W^{r \hi \hj} W^{s}{}_{\hi \hj} \ , \qquad  W^{r \hi \hk} W_{r}{}^{\hj}{}_{\hk} \ , \ee
in the $[2]$ and the $[2,0]$, respectively, the cubic monomial 
\be
 W^{s \hi \hj }W^{r \hk \hl} W_{s \hk \hl} \ , \ee
in the $[1]\times[0,1]$, and the two quartic monomials 
\be
W^{s \hi \hp} W_{s}{}^{\hj}{}_{\hp}   W^{p}{}_{\hk}{}^{\hq} W_{p \hl \hq} - \frac{1}{6} \delta^{\hi \hj}_{\hk \hl} W^{s \hi \hp} W_{s}{}^{\hj}{}_{\hp}   W^{p}{}_{\hi}{}^{\hq} W_{p \hj \hq} 
\ , \quad  W^{s \hi \hp} W_{s}{}^{\hj}{}_{\hp}   W^{p}{}_{\hi}{}^{\hq} W_{p \hj \hq}  \ , \ee
in the $[0,2]$ and the singlet representation, respectively. One concludes that the most general monomial is labeled by 6 integers, such that 
\bea
 && \int du D^{12} \bar{D}^{16} F(u)^{[n_1 + 2 n_2 + n_3] [2 n'_2, n_1 + n_3 + 2 n_4]}_{[ 2 n'_2 + 2 n_3  + 4 n_4 + 4 n'_4 ,n_1 + 2 n_2 + n_3] [2 n'_2, n_1 + n_3 + 2 n_4]} 
W^{4 + n_1 + 2 n_2 + 2 n'_2 + 3 n_3 + 4 n_4 + 4 n'_4}_{[n_1 + 2 n_2 + n_3] [2 n'_2, n_1 + n_3 + 2 n_4]} \CR
&=&  L^{n_1 + 2 n_2 + 2 n'_2 + 3 n_3 + 4 n_4 + 4 n'_4}_{[ 2 n'_2 + 2 n_3  + 4 n_4 + 4 n'_4 ,n_1 + 2 n_2 + n_3] [2 n'_2, n_1 + n_3 + 2 n_4]}\nabla^{6} R^{4} + \dots 
\eea
The linear analysis therefore suggests the form of the nonlinear invariant 
\bea
&& \cL_{\grad410}[ \cE_{\grad410}] \\
&=& \sum_{\substack{ n_1,n_2,n^\prime_2\\ n_3 ,n_4,n_4^\prime}} \cD^{n_1 + 2 n_2 + 2 n'_2 + 3 n_3 + 4 n_4 + 4 n'_4}_{\scriptscriptstyle [ 2 n'_2 + 2 n_3  + 4 n_4 + 4 n'_4 ,n_1 + 2 n_2 + n_3] [2 n'_2, n_1 + n_3 + 2 n_4]} \cE_{\grad410} 
\cL_{\grad410}^{\scriptscriptstyle [ 2 n'_2 + 2 n_3  + 4 n_4 + 4 n'_4 ,n_1 + 2 n_2 + n_3] [2 n'_2, n_1 + n_3 + 2 n_4]}\CR
&& +  \sum_{\substack{ n_1,n_2,n^\prime_2\\ n_3 ,n_4,n_4^\prime}} \cD^{n_1 + 2 n_2 + 2 n'_2 + 3 n_3 + 4 n_4 + 4 n'_4}_{\scriptscriptstyle[2 n'_2, n_1 + n_3 + 2 n_4] [ 2 n'_2 + 2 n_3  + 4 n_4 + 4 n'_4 ,n_1 + 2 n_2 + n_3] } \cE_{\grad410} 
\cL_{\grad410}^{\scriptscriptstyle [2 n'_2, n_1 + n_3 + 2 n_4][2 n'_2 + 2 n_3  + 4 n_4 + 4 n'_4 ,n_1 + 2 n_2 + n_3] }\nn
\eea
where we consider the possibility of a mixing between the invariant $\cL_{\grad410}$ with its conjugate obtained by exchanging the two $Sp(2)$ factors, according to the observation in \cite{D4R4} for the $\nabla^4 R^4$ type invariant in eight dimensions. 

From this structure one deduces that supersymmetry requires the function $\cE_{\grad410}$ to be an eigenfunction of the Laplace operator, and to satisfy equations of the form 
\be \cD^3_{[2,0],[2,0]} \cE_{\grad410} \propto  \cD^{\; 2}_{[2,0],[2,0]} \cE_{\grad410} \ , \qquad  \cD^3_{[0,1],[0,1]} \cE_{\grad410} \propto  \cD_{[0,1],[0,1]} \cE_{\grad410} \ , \ee
as well has a highest weight constraint 
\be \cD^{2k}_{[0,0],[0,2k]}  \cE_{\grad410} = 0 \ , \label{HighestWeight} \ee
for some integer $k$. We will see in the next section that the standard $\nabla^6 R^4$ type invariant threshold function indeed satisfies to these equations for $k=2$. 
\subsubsection{Decompactification limit from five to six dimensions}
We are now going to solve the differential equations  \eqref{D6R4E6} and \eqref{F2kD4R4E6}  for a function depending only of the Levi subgroup $\mathds{R}_*^+\times  SO(5,5)$ of the parabolic subgroup associated to the decompactification limit, such that 
\be \mathfrak{e}_{6(6)}\cong {\bf 16}^\ord{-3} \oplus \scal{ \mathfrak{gl}_1 \oplus \mathfrak{so}(5,5) }^\ord{0} \oplus \overline{\bf 16}^\ord{3} \ , \qquad 
{\bf 27} \cong {\bf 10}^\ord{-2} \oplus {\bf 16}^\ord{1} \oplus {\bf 1}^\ord{4} \ .\label{27D5}  \ee
The covariant derivative on $E_{6(6)}/\Sp$ acting on such a function takes the block diagonal form 
\be
{\bf D}_{27} = \text{diag} \biggl(\tfrac{1}{6} \partial_{\phi}, {\bf D}_{16}+ \tfrac{1}{24} {\bf\mathds{1}}_{16} \partial_{\phi}, {\bf D}_{10} - \tfrac{1}{12} {\bf\mathds{1}}_{10} \partial_{\phi}  \biggr) \ . 
\ee
To check the differential equations \eqref{D6R4E6} and \eqref{F2kD4R4E6} we need to compute the block diagonal decomposition of the higher order differential operators. In order to do this computation we consider a general ansatz and determine all the free coefficients by consistency with the various differential equations displayed in appendix \ref{TensorEquationsEd}. We obtain in this way 
\begin{multline}
{\bf D}^{\; 2}_{27} = \text{diag} \biggl(\tfrac{1}{6^2} \partial^{\; 2}_{\phi} + \tfrac{1}{2} \partial_{\phi}\,  ,\   {\bf D}^{\; 2}_{16} + \tfrac{1}{2}  {\bf D}_{16} \bigl(\tfrac{1}{6} \partial_{\phi}+1\bigr)  + \tfrac{1}{16}  {\bf \mathds{1}}_{16}  \bigl( \tfrac{1}{6^2}  \partial^{\; 2}_{\phi} + 3 \partial_{\phi} \bigr) \, , \\
 {\bf D}^{\; 2}_{10} -  {\bf D}_{10} \bigl(  \tfrac{1}{6} \partial_{\phi} +1 \bigr) + \tfrac{1}{4} {\bf\mathds{1}}_{10} \bigl(  \tfrac{1}{6^2} \partial^{\; 2}_{\phi} + \partial_{\phi} \bigr) \biggr)
\end{multline}
and
\begin{multline}
{\bf D}^{\; 3}_{27} = \text{diag} \biggl(\tfrac{1}{6^3} \partial^{\; 3}_{\phi} + \tfrac{3}{16} \partial^{\; 2}_{\phi} + \tfrac{5}{4} \partial_{\phi} - \tfrac{1}{2} \Delta \,,\\
 {\bf D}^{\; 3}_{16}
+\tfrac{3}{4}  {\bf D}^{\; 2}_{16} \bigl(\tfrac{1}{6} \partial_{\phi} +2 \bigr)+  {\bf D}_{16} \bigl( \tfrac{3}{16}\scal{ \tfrac{1}{6^2} \partial^{\; 2}_{\phi} +2  \partial_{\phi}}+1 \bigr)+ \tfrac{1}{64} {\bf\mathds{1}}_{16} \bigl( \tfrac{1}{6^3}  \partial^{\; 3}_{\phi} + \tfrac{1}{2} \partial^{\; 2}_{\phi} - 8 \Delta \bigr) \, , \\  {\bf D}^{\; 3}_{10} - \tfrac{3}{2}  {\bf D}^{\; 2}_{10} \bigl( \tfrac{1}{6} \partial_{\phi}  +2 \bigr) 
 + {\bf D}_{10} \bigl(\tfrac{3}{4}  \scal{ \tfrac{1}{6^2} \partial^{\; 2}_{\phi} + \partial_{\phi} }+\tfrac{5}{2} \bigr)- \tfrac{1}{8} 
 {\bf\mathds{1}}_{10} \bigl( \tfrac{1}{6^3} \partial^{\; 3}_{\phi} +\tfrac{1}{4} \partial^{\; 2}_{\phi} +\partial_{\phi}-2\Delta \bigr)  \biggr) \ ,
\end{multline}
where $\Delta \equiv \text{Tr}\,  {\bf D}_{10}^{\; 2}$. In order to determine the constraints on the threshold function is six dimensions, we consider an ansatz with the appropriate power of the radius modulus $e^{-3\phi}$ such as to compensate for the Weyl rescaling to Einstein frame, \ie 
\be
\cE_{\gra84} = e^{-6 \phi}\cE_{\grad422}\ ,  \qquad \cE_{\gra81} = e^{-12 \phi} \cE_{\grad410} \ . 
\ee
The singlet component of \eqref{D6R4E6} gives directly the Poisson equation 
\be \label{E6D_laplace}
\Delta \cE_{\grad410}  = - \cE^{\; 2}_{\grad422} \ , 
\ee
which is indeed consistent with \eqref{LaplaceE6}. Working out spinor and the vector equations, using the Poisson equation \eqref{E6D_laplace}, one obtains similarly 
\be \label{E6D}
\biggl( {\bf D}^{\; 3}_{16} - \frac{3}{4} {\bf D}_{16} \biggr) \cE_{\grad410}  = - \frac{1}{4} {\bf D}_{16} \cE^{\; 2}_{\grad422} \ , \qquad 
\biggl( {\bf D}^{\; 3}_{10} -\frac{3}{2} {\bf D}_{10} \biggr) \cE_{\grad410}   = - \frac{1}{4}  {\bf D}_{10} \cE^{\; 2}_{\grad422} \ . 
\ee
The only Eisenstein function that solves this homogenous equation (for $\cE_{\grad422} =0$) is the Eisenstein series $\hat{E}{\text{\DSOX0{\mbox{$\frac{7}{2}$}}000}}$, but its  expansion in the string theory limit is inconsistent with perturbation theory. As is computed in appendix \ref{DnAdj}, one has moreover
\be  D^d{}_{\hat{a}}\scal{  \cD_{(a}{}^{\hat{b}} \cD_{b|\hat{b}} \cD_{c}{}^{\hat{c}} \cD_{d)\hat{c}} }|_{[0,4]} \hat{E}_{\text{\DSOX0{\mbox{$\frac{7}{2}$}}000}} = 0 \ , \ee
which defines an integrability condition for the function to decompose into the sum of two functions satisfying to \eqref{HighestWeight} and its conjugate obtained by exchange of the two $Sp(2)$ for $k=2$.

Let us now consider the differential equation \eqref{F2kD4R4E6} for the function $\cE^{(k)}_{\frac{1}{8}\, E_6}$ defining the  $F^{2k} \nabla^{4} R^4$ threshold function in five dimensions. Diffeomorphism invariance in six dimensions requires an ansatz of the form 
\be
\cE^{(k)}_{\frac{1}{8}\, E_{6}} = e^{- 2(k+5)\phi} \cE^{(k)}_{\frac{1}{8}\, D_5} \ , 
\ee
and using this ansatz, one obtains from the singlet component of \eqref{F2kD4R4E6} the Laplace equation 
\be
\Delta \cE^{(k)}_{\frac{1}{8}} = \frac{5}{2}  (k+3)(k-1) \cE^{(k)}_{\frac{1}{8}} \ , 
\ee
where we removed the $D_5$ label for simplicity. The spinor and the vector equations give then
\bea
\biggl( {\bf D}^{\; 3}_{16} -\frac{13  (k+3) (k-1)+24}{16} {\bf D}_{16}  \biggr) \cE^{(k)}_{\frac{1}{8}}   &=& - \frac{3(k+1)}{4}  \Scal{ {\bf D}^{\; 2}_{16} -\frac{5 (k+3) (k-1)}{16}   \mathds{1}_{16} } \cE^{(k)}_{\frac{1}{8}} \ , \CR
 {\bf D}^{\; 2}_{10}  \, \cE^{(k)}_{\frac{1}{8}} &=&  \frac{ (k+3) (k-1)  }{4} \mathds{1}_{10} \, \cE^{(k)}_{\frac{1}{8}}   \ ,
\eea
where we used that the even and odd powers of ${\bf D}_{10}$ lie in different irreducible representations of $Sp(2)\times Sp(2)$, and must therefore vanish separately. The unique Eisenstein function satisfying to this equation is 
\be
\cE^{(k)}_{\frac{1}{8}} \propto E_{\mbox{\DSOX{0}000{\mbox{$k$+$3$}}}}\, , 
\ee
consistently with the decompactification limit of the five-dimensional function 
\be   E_{\mbox{\DEVI00000{k\mbox{+}\frac{7}{2}}}}    =2 \zeta(2k+7)  e^{-4(2k+7) \phi }   + \frac{\pi^\frac{1}{2} \Gamma(k+3)}{\Gamma(k + \frac{7}{2})} e^{- (2k+10)\phi} E_{\mbox{\DSOX{0}000{\mbox{$k$+$3$}}}} + \dots \ee
We conclude therefore that the exact $\nabla^6 R^4$ function is defined as
\be E_{(0,1)}= \hat{\cE}_\grad{4}{1}{0}+ \hat{\cE}_\grad{4}{0}{1} + \frac{8}{189} \hat{E}_{\mbox{\DSOX{0}000{\mbox{$4$}}}}\ , \ee
where $\hat{\cE}_\grad{4}{1}{0}$ satisfies to \eqref{E6D}, and to an anomalous Poisson equation with an additional constant source term. This function is consistent with \cite{Pioline:2015yea}, where the second Eisenstein function appears with this normalisation, and the 2-loop five dimensional threshold function must indeed solve \eqref{E6D}, because the equation is parity invariant with respect to $O(5,5)$.

\subsection{$\cN = 2$ supergravity in seven dimensions}
None of the $\nabla^6 R^4$ type invariants can be defined in the linearised approximation as harmonic superspace integrals in seven dimensions. We will therefore consider the uplift of the four-dimensional invariants in the decompactification limit. In seven dimensions the scalar fields parametrize the symmetric space $SL(5)/SO(5)$, and the covariant derivative $\cD_{ij}$ transforms as a symmetric traceless tensor of $SO(5)$, with  $i, j = 1,\dots,5$ of $SO(5)$. We consider therefore the parabolic subgroup of $E_{7(7)}$ of semi-simple Levi subgroup $SL(5) \times SL(3)$ associated to the decomposition 
\bea \label{133Graded}
\mathfrak{e}_{7(7)} &\cong&  \bar{{\bf 5} }^\ord{-6} \oplus  ( {\bf 3} \otimes {\bf 5})^\ord{-4}  \oplus   ( \bar{ \bf 3} \otimes {\bf \overline{10}})^\ord{-2}  \oplus \scal{  \mathfrak{gl}_1 \oplus \mathfrak{sl}_3 \oplus \mathfrak{sl}_5 }^\ord{0} \oplus ( { \bf 3} \otimes { \bf 10})^\ord{2}
\oplus (\bar{ \bf 3} \otimes \bar{ \bf 5})^\ord{4} \oplus { \bf 5}^\ord{6} \CR
{\bf 56} &\cong&\overline{\bf 3}^\ord{-5} \oplus {\bf 10}^\ord{-3} \oplus ({\bf 3}\otimes  \overline{\bf  5})^\ord{-1} \oplus (\overline{\bf 3}\otimes {\bf 5})^\ord{1} \oplus \overline{\bf 10}^\ord{3} \oplus {\bf 3}^\ord{5}\ . 
\eea
We will use the same conventions as in \cite{D4R4}, where the decompactification limit of equations  \eqref{CubicR} and \eqref{E811Equation}  is already discussed in details. We consider the ansatz 
 \be
 \cE_{\grad{8}{1}{1}} = e^{- 36 \phi} \cE_{\frac{1}{8}} \ , \qquad    \cE_{\grad{8}{4}{4}} = e^{- 18 \phi} \cE_{\gra{4}{2}} \ , 
 \ee
with $ \cE_{\frac{1}{8}}$ and $\cE_{\gra{4}{2}} $ defined on $SL(5)/SO(5)$, and the appropriate power of the volume modulus $e^{-3\phi}$ required by diffeomorphism invariance in seven dimensions. The ${\bf 3}^\ord{3}$ component of the equation reduces to the Poisson equation 
\be\label{Laplace7D} 
\Delta \cE_{\frac{1}{8}} \equiv  2 \cD^{ij} \cD_{ij} \cE_{\frac{1}{8}}  = \frac{42}{5} \cE_{\frac{1}{8}} - \cE^{\; 2}_{\gra{4}{2}} \ . 
\ee
Using this equation in the $ (\overline{\bf 3}\otimes {\bf 5})^\ord{1} $ component of \eqref{E811Equation}, one obtains 
\be\label{Vec7D} 
\biggl( \cD_{i}{}^{k} \cD_{k}{}^{l}  \cD_{l}{}^{j} + \frac{1}{5} \cD_{i}{}^{k} \cD_{k}{}^{j} - \frac{1053}{400} \cD_{i}{}^{j} +\frac{177}{500} \delta_{i}^{j} \biggr) \cE_{\frac{1}{8}} =
\frac{1}{4} \biggl(\frac{1}{10} \delta_{i}^{j} - \cD_{i}{}^{j} \biggr) \cE^{\; 2}_{\gra{4}{2}}  \ . 
\ee
Using moreover these equations to simplify the $ \overline{\bf 10}^\ord{3}$ components, one obtains 
\be \label{Tens7D}   \Scal{ 6\cD^{[i}{}_{[k} \cD^{j]}{}_{p} \cD^{p}{}_{l]}  + \frac{3}{10} \cD^{[i}{}_{[k} \cD^{j]}{}_{l]}} \cE_{\frac{1}{8}} = \frac{3}{5}  \delta^{[i}_{[k} \biggl(  \cD^{j]}{}_p \cD^p{}_{l]}  -\frac{71}{20}\cD^{j]}{}_{l]} + \frac{9}{25}  \delta_{l]}^{j]} \biggr) \cE_{\frac{1}{8}}  \ . \ee
The solution to the homogenous equation (with $ \cE_{\gra{4}{2}} =0$) can be written as the Eisenstein function $E_{\scriptscriptstyle [3, 0,0,\frac{5}{2}]}$, using the formulae of Appendix \ref{SL5Review}. 

We consider now the $F^{2k} \nabla^4 R^4$ threshold function, with the ansatz 
\be
\cE_{\grad820}^\ord{k} = e^{-6(k +5) \phi}\cE_{\frac{1}{8}}^\ord{k} \ . 
\ee
Appropriate linear combinations of the grad 6 and 4 components of \eqref{CubicR} in \eqref{133Graded} give the two equations 
\bea \biggl( \cD_{i}{}^{k} \cD_{k}{}^{l}  \cD_{l}{}^{j}  + \frac{1}{2}  \cD_{i}{}^{k} \cD_{k}{}^{j}  \biggr) \cE_{\frac{1}{8}}^\ord{k}  &=&\biggl( \frac{16 k(7k+20) + 75}{400} \cD_{i}{}^{j} +\frac{3 k(k+5)(2k+5)}{125} \delta_{i}^{j}- \frac{1}{40}  \delta_{i}^{j} \Delta  \biggr) \cE_{\frac{1}{8}}^\ord{k}   \CR
k \,   \cD_{i}{}^{k} \cD_{k}{}^{j}  \cE_{\frac{1}{8}}^\ord{k}  &=& \biggl(k \frac{4k+5}{20} \cD_{i}{}^{j} +\frac{3k(k+5)(2k+5)}{25} \delta_{i}^{j} - \frac{1}{2}  \delta_{i}^{j} \Delta \biggr)     \cE_{\frac{1}{8}}^\ord{k}   \ .  \eea
For $k\ge 1$ we conclude that the function must satisfy to the quadratic equation
\be  \label{fundamental_1}
\cD_{i}{}^{p} \cD_{p}{}^{j} \cE_{\frac{1}{8}}^\ord{k} = \frac{4 k + 5}{20} \cD_{i}{}^{j} \cE_{\frac{1}{8}}^\ord{k} + \frac{3 k (2k + 5)}{25} \delta_{i}^{j} \cE_{\frac{1}{8}}^\ord{k} \ , 
\ee
such that 
\be
\Delta \cE_{\frac{1}{8}}^\ord{k} = \frac{6 k (2k + 5) }{5} \cE_{\frac{1}{8}}^\ord{k} \ . 
\ee
One can then check that all the other equations implied by  \eqref{CubicR} are indeed satisfied provided that \eqref{fundamental_1} is.  This equation is satisfied by the Eisenstein function $E_{\scriptscriptstyle [0,0,k + \frac{5}{2},0]}$, which appears in the decompactification limit of the corresponding Eisenstein function on $E_{7(7)}/ \SU$, \ie 
\be
E_{\mbox{\DEVII000000{k\mbox{+}4}}} =e^{- 10(k+4) \phi} E_{\scriptscriptstyle [k+4,0]}  +  \frac{\pi^{\frac{3}{2}}\Gamma(k+\frac{5}{2})}{\Gamma(k+4)}  e^{- 6(k+5) \phi} E_{\scriptscriptstyle [0,0,k + \frac{5}{2},0]} + \dots 
\ee
where the Eisenstein series is normalised with an extra  $2\zeta(2s)$ factor with respect to the Langlands normalisation. We conclude therefore that the exact $\nabla^6 R^4$ threshold function is 
\be  E_{(0,1)}  = \cE_{\frac{1}{8}}  + \frac{5 \pi}{378} E_{\scriptscriptstyle [0,0,\frac{7}{2},0]}  \ ,\ee
where $\cE_{\frac{1}{8}}$ is a solution to (\ref{Laplace7D},\ref{Vec7D},\ref{Tens7D}) in agreement with \cite{Green:2010wi}.\footnote{Note that the normalisation of the Eisenstein function $E_{\scriptscriptstyle [0,0,s,0]} $ does not include the additional factor of $\zeta(2s-1)$ as in  \cite{Green:2010wi}.}

\subsection{$\mathcal{N} = 2$ supergravity in eight dimensions}
We shall now consider the oxidation of the  seven-dimensional  $\nabla^6 R^4$ and $F^{2n} \nabla^4 R^4$ type invariants to eight dimensions. Because there is a 1-loop divergence in eight dimensions, the exact $R^4$ threshold function differs from the Wilsonian effective action function. In the dimensional reduction, the divergence appears to be absorbed into the infinite sum of Kaluza--Klein states over the circle such that the function is finite in seven dimensions, but involves a logarithm of the radius modulus in the decompactification limit \cite{Green:2010sp}. In order to consider the non-analytic terms in eight dimensions, we will take these logarithms into account in the decompactification limit.

 We shall use the same conventions as in \cite{D4R4}, \ie the complex scalar field $\tau$ parametrises the coset representative $v_\alpha{}^j \in SL(2) / SO(2)$, with $\alpha, \beta = 1,2$ of $SO(2)$ and $i,j = 1,2$ of $SL(2)$, whereas the five real scalar fields $t$ parametrise the coset representative $V^{a}{}_J \in SL(3)/SO(3)$, with $a, b = 1,2,3$ of $SO(3)$ and $I, J = 1,2,3$ of $SL(3)$. The corresponding covariant derivative in tangent frame are then traceless symmetric tensors $\cD_{\alpha \beta}$ and  $\cD_{a b}$, respectively. In the decompactification limit, one writes the  $SL(5)/SO(5)$ coset element in the parabolic gauge 
 \be
\cV = \left( \begin{array}{cc} \ e^{-3\phi} v^\inv_j{}^\alpha  \ &  \ 0 \ \\ 
\  e^{2\phi} V^a{}_K a_j^K & \ e^{2\phi} V^{a}{}_J \ \end{array} \right) \ , 
\ee
associated to the graded decomposition 
\be \mathfrak{sl}_5 \cong ( {\bf 2}\otimes \overline{\bf 3})^\ord{-5} \oplus \scal{ \mathfrak{gl}_1 \oplus \mathfrak{sl}_2 \oplus \mathfrak{sl}_3}^\ord{0} \oplus  ( {\bf 2}\otimes {\bf 3})^\ord{5} \ . \ee
In this way one computes that the covariant derivative over $SL(5)/SO(5)$ in tangent frame acts on a function of $\phi, \tau,t$ as \cite{D4R4}
\be {\bf D}_{5}  =\text{diag} \left( -  \tfrac{1}{20} \partial_\phi  \delta_\alpha^\beta - {\cD}_\alpha{}^\beta,  \tfrac{1}{30} \partial_\phi \delta_a^b +\cD_a{}^b \right) \  .  \ee
One computes that the higher order derivative operators obtained as products of ${\bf D}_{5} $ in the ${\bf 5}$ of $SL(5)$ decompose similarly as
\begin{multline} {\bf D}^{\; 2}_{5} = \text{diag} \biggl(  {\cD}_\alpha{}^\beta \scal{\tfrac{1}{10} \partial_\phi+ \tfrac{3}{4} }+ \scal{  \tfrac{1}{400} \partial_\phi^{\; 2} + \tfrac{1}{16} \partial_\phi + \tfrac{1}{2} \cD_{\gamma\delta} \cD^{\gamma\delta}  }   \delta_\alpha^\beta , \hspace{3 cm} \\ 
  \cD_a{}^c \cD_c{}^b + \cD_a{}^b \scal{ \tfrac{1}{15} \partial_\phi+\tfrac{1}{2}  } +\scal{ \tfrac{1}{900}  \partial_\phi^{\; 2} + \tfrac{1}{24} \partial_\phi  }  \delta_a^b \biggr) \ , \end{multline}
and
\begin{multline} {\bf D}^{\;3}_{5} = \text{diag} \biggl( - \tfrac{1}{4}  {\cD}_\alpha{}^\beta \scal{ \tfrac{3}{10^2} \partial_{\phi}^{\; 2}+ \tfrac{7}{10} \partial_\phi  +\tfrac{9}{4} +2  \cD_{\gamma \delta} \cD^{\gamma \delta}} \\
\hspace{10mm} - \tfrac{1}{8}  \scal{ \tfrac{1}{10^3} \partial_\phi^{\; 3} +\tfrac{1}{30} \partial_\phi^{\; 2}+\tfrac{1}{8} \partial_\phi + 6  \cD_{\gamma \delta} \cD^{\gamma \delta} \scal{ \tfrac{1}{10}  \partial_\phi +1} -2 \cD_{a b} \cD^{a b}}   \delta_\alpha^\beta\ ,  \\ 
  \cD_a{}^c \cD_c{}^d  \cD_d{}^b +\cD_a{}^c \cD_c{}^b \scal{\tfrac{1}{10} \partial_{\phi} + 1} + \tfrac{1}{3} \cD_a{}^b \scal{ \tfrac{1}{10^2} \partial_\phi^{\; 2} +  \tfrac{11}{40}\partial_\phi + \tfrac{3}{4} }\\ + \tfrac{1}{8} \scal{ \tfrac{1}{15^3}  \partial_\phi^{\; 3} + \tfrac{1}{180} \partial_\phi^{\; 2} -\tfrac{1}{12} \partial_\phi -2 \cD_{\alpha \beta} \cD^{\alpha \beta} }  \delta_a^b \biggr) \ . \end{multline}
We can now solve equation \eqref{Vec7D} in the docompactification limit, with $ \cE_{\frac{1}{8}}(\phi,\tau,t)$ and \cite{Green:2010wi}
\be  \cE_{\gra{4}{2}}  = e^{-6\phi} \Scal{ 2 \hat{E}_{[1]}(\tau) + \hat{E}_{[\stfrac{3}{2},0]}(t) - 20 \pi ( \phi - \phi_0)} \ , \ee
where $\phi_0$ is a constant that depends on the renormalisation scheme. Using the property that 
\be \cD_{ac} \cD^{bc}   \hat{E}_{[\stfrac{3}{2},0]}(t)  = - \frac{1}{4} \cD_a{}^b   \hat{E}_{[\stfrac{3}{2},0]}(t)  + \frac{2\pi}{3} \ , \qquad 2 \cD_{\alpha\beta} \cD^{\alpha\beta}  \hat{E}_{[1]}(\tau) = \pi \ , \ee
one shows that the general solution to \eqref{Vec7D}, as a function of $\phi, \tau$ and $t$ takes the general form 
\begin{multline}  \cE_{\frac{1}{8}}(\phi,\tau,t) = e^{-12\phi} \biggl(\hat{\cF}_{[4]}(\tau)  +\hat{\cF}_{[4,-2 ]}(t) + \frac{1}{3} \hat{E}_{[1]}(\tau) \hat{E}_{[\stfrac{3}{2},0]}(t) + \frac{\pi}{18} \Scal { \hat{E}_{[1]}(\tau)  + \frac{19\pi}{12}} \biggr . \\ \biggl .  - \frac{10\pi}{3} ( \phi - \phi_0 ) \Scal{2 \hat{E}_{[1]}(\tau) + \hat{E}_{[\stfrac{3}{2},0]}(t)  + \frac{\pi}{3}} + \frac{100\pi^2}{3} ( \phi- \phi_0)^2\biggr)  \ ,  \end{multline} 
where $\cF_{[4]}(\tau)$ and $\cF_{[4,-2 ]}(t)$ are solutions to 
\bea \Delta\hat{\cF}_{[4]}(\tau) &=& 12\hat{\cF}_{[4]}(\tau)  - ( \hat{E}_{[1]}(\tau))^2 \ , \qquad  \Delta\hat{\cF}_{[4,-2 ]}(t)  = 12\hat{\cF}_{[4,-2 ]}(t) -(  \hat{E}_{[\stfrac{3}{2},0]}(t) )^2 \ , \CR
\cD_{ac} \cD^{cd} \cD_{db} \hat{\cF}_{[4,-2 ]}(t)  &=& \frac{49}{16} \cD_{ab}\hat{\cF}_{[4,-2 ]}(t) - \frac{3}{2} \delta_{ab}\hat{\cF}_{[4,-2 ]}(t) - \frac{1}{2}   \hat{E}_{[\stfrac{3}{2},0]}(t)  \cD_{ab}   \hat{E}_{[\stfrac{3}{2},0]}(t)  \CR
&& \hspace{30mm} + \frac{1}{2} \delta_{ab} \Scal{ \frac{1}{4} (  \hat{E}_{[\stfrac{3}{2},0]}(t) )^2 + \frac{\pi}{9}  \hat{E}_{[\stfrac{3}{2},0]}(t)  + \frac{\pi^2}{27} } \ . \eea
Here the notation is used to emphasize that these solutions are defined modulo the homogeneous solutions $E_{[4]}(\tau)$ and $E_{[4, -2]}(t)$, respectively, as one can see using the formulae of Appendix \ref{SL3Equa}. Note nonetheless that these homogeneous solutions are inconsistent with the string  theory perturbation expansion, and the exact threshold function is uniquely determined by these equations and consistency with string theory \cite{Basu:2007ck}. 

The structure of the threshold function exhibits that there is a 1-loop divergence of the $R^4$ type invariant form factor proportional to the $\nabla^6 R^4$ type invariant. This implies the presence of an addition renormalisation scheme ambiguity in the definition of the analytic part of the effective action. It appears that the renormalisation scheme used in \cite{Green:2010wi,Basu:2007ck}, cannot be obtained by simply neglecting the terms in $\phi-\phi_0$, but one finds nonetheless that the threshold function only differs by terms proportional to the linear and the quadratic term in  $\phi-\phi_0$, \ie 
\begin{multline}  \hat{ \cE}_{\frac{1}{8}}(\tau,t) = \hat{\cF}_{[4]}(\tau)  +\hat{\cF}_{[4,-2 ]}(t) + \frac{1}{3} \hat{E}_{[1]}(\tau) \hat{E}_{[\stfrac{3}{2},0]}(t) + \frac{\pi}{18} \Scal { \hat{E}_{[1]}(\tau)  + \frac{19\pi}{12}}  \\ +\frac{\pi}{36}  \Scal{2 \hat{E}_{[1]}(\tau) + \hat{E}_{[\stfrac{3}{2},0]}(t)  - \frac{5\pi}{2}} \ . \end{multline}

Let us now consider the oxidation of the $F^{2k} \nabla^4 R^4$ type invariants, \ie solve the differential equation \eqref{fundamental_1} for a function of the form $e^{- (10 + 2k) \phi}\cE_{\frac{1}{8}}^\ord{k}(\tau, t)$, as required by diffeomorphism invariance in eight dimensions. One obtains straightforwardly 
\bea
  2 \cD_{\alpha \beta} \cD^{\alpha \beta} \cE^\ord{k}_{\frac{1}{8}}(\tau,t)  &=& (1 + k) (2 + k)\cE^\ord{k}_{\frac{1}{8}}(\tau,t)   \ ,  \CR
\cD_{a}{}^{c} \cD_{c}{}^{b}\cE^\ord{k}_{\frac{1}{8}}(\tau,t)  &=& \frac{(5 + 4 k)}{12} \cD_{a}{}^{b}  \cE^\ord{k}_{\frac{1}{8}}(\tau,t)+ \frac{(2 + k)(1 + 2 k)}{9} \delta_{a}^{b} \cE^\ord{k}_{\frac{1}{8}}(\tau,t) \ . 
\eea
Using the results of Appendix \ref{SL3Equa} one obtain that the solution can be written in terms of Eisenstein functions as
\be
\cE^\ord{k}_{\frac{1}{8}}(\tau, t) \propto  E_{[k + 2]}(\tau) E_{[0, k+2]}(t) \ , 
\ee
consistently with the decompactification limit of the $SL(5)/SO(5)$ Eisenstein function \cite{D4R4},
 \begin{multline}  E_{[0,0,k + \stfrac{5}{2},0]} =2 \zeta( 2k+5)\zeta(2k+4)  e^{-6(2k+5) \phi} + \frac{2\pi^2 \zeta(2k+2)}{(2k+3)(k+1)}  e^{8k \phi} E_{[k+\stfrac{3}{2},0]}(t) \\  +\frac{\sqrt{\pi} \Gamma(k+2)}{2\Gamma(k + \stfrac{5}{2})}  e^{-2(k+5) \phi} E_{[k+2]}(\tau) E_{[0,k+2]}(t)  + \mathcal{O}(e^{- e^{-5 \phi}}) \ . \end{multline}
The sum of the two functions reproduces correctly the threshold function obtained in \cite{Green:2010wi,Basu:2007ck}.

\section*{Acknowledgment}
We would like to thank Axel Kleinschmidt and Boris Pioline for useful discussions, and are particularly grateful to Axel Kleinschmidt for providing some explicit relations between Eisenstein series.

 \appendix

\section{$E_{d(d)}$ Eisenstein series, and tensorial differential equations}
\label{TensorEquationsEd} 
In this appendix we collect the differential equations satisfied by the Eisenstein functions that are relevant in the analysis of BPS threshold functions in string theory, and their related coadjoint nilpotent orbits. We write them in terms of the covariant derivative in tangent frame valued in the Lie algebra in some particular representations, which are specified by their dimension.
\addtocontents{toc}{\protect\setcounter{tocdepth}{1}}
\subsection{$E_{8(8)}$}
\addtocontents{toc}{\protect\setcounter{tocdepth}{2}}
The Eisenstein function in the adjoint representation is associated to the nilpotent orbit of Dynkin label \DEVIII00000002, with $D_8$ Dynkin label \DSOXVI02000000, and satisfies in general to the differential equation 
\be \scal{ \cD \Gamma_{i[jk}{}^r \cD} \scal{ \cD \Gamma_{lpq]r} \cD } E_{\mbox{\DEVIII0000000{\mathnormal{s}}}} = - \delta_{i[j} \scal{ \cD \Gamma_{klpq]} \cD} ( 2s(2s-29)+48  ) E_{\mbox{\DEVIII0000000{\mathnormal{s}}}} \ . \ee
For the following three special values of $s$, the function is associated to lower dimensional nilpotent orbits, and satisfies moreover to
\bea \scal{ \cD \Gamma_{ijpq}  \cD} \scal{ \cD \Gamma^{klpq} \cD}  E_{\mbox{\DEVIII0000000{\mathnormal{7}}}} &=& 648 \scal{\cD \Gamma_{ij}{}^{kl} \cD}  E_{\mbox{\DEVIII0000000{\mathnormal{7}}}}+105840 \,  \delta_{ij}^{kl}  E_{\mbox{\DEVIII0000000{\mathnormal{7}}}} \ , \CR
 \Gamma^{kl}    \cD \scal{ \cD\Gamma_{ijkl} \cD} E_{\mbox{\DEVIII0000000{\mbox{$\frac{9}{2}$}}}}&=& - 168\,  \Gamma_{ij} \cD E_{\mbox{\DEVIII0000000{\mbox{$\frac{9}{2}$}}}}\ , \CR
\scal{ \cD\Gamma_{ijkl} \cD} E_{\mbox{\DEVIII0000000{\mbox{$\frac{5}{2}$}}}}&=& 0 \ . 
 \eea
 \addtocontents{toc}{\protect\setcounter{tocdepth}{1}}
\subsection{$E_{7(7)}$}
\addtocontents{toc}{\protect\setcounter{tocdepth}{2}}
The Eisenstein function in the adjoint representation is associated to the nilpotent orbit of Dynkin label \DEVII200000{\mathfrak{0}}, with $A_7$ Dynkin label  [$\scriptstyle \mathfrak{2}\mathfrak{0}\mathfrak{0}\mathfrak{0}\mathfrak{0}\mathfrak{0}\mathfrak{2}$], and satisfies in general to the differential equation 
\be  {\bf D}_{56}^{\;3}   E_{\mbox{\DEVII{\mathnormal{s}}00000{\mathfrak{0}}}} = \Scal{ \frac{s(2s-17)}{2}+6}   {\bf D}_{56} E_{\mbox{\DEVII{\mathnormal{s}}00000{\mathfrak{0}}}}  \ , \qquad \Delta  E_{\mbox{\DEVII{\mathnormal{s}}00000{\mathfrak{0}}}} = 2s(2s-17)  E_{\mbox{\DEVII{\mathnormal{s}}00000{\mathfrak{0}}}} \ .  \ee
For the following three special values of $s$, the function is associated to lower dimensional nilpotent orbits, and satisfies moreover  to
\bea \cD^{mrs(i} \cD_{nrs(k} \cD_{l)mpq} \cD^{j)npq}  E_{\mbox{\DEVII{\mathnormal{4}}00000{\mathfrak{0}}}}  =-\frac{13}{4} \delta^{(i}_{(k} \delta^{j)}_{l)}    E_{\mbox{\DEVII{\mathnormal{4}}00000{\mathfrak{0}}}}   \ , \CR
{\bf D}_{133}^{\; 3} E_{\mbox{\DEVII{\mbox{$\frac{5}{2}$}}00000{\mathfrak{0}}}} = - 20 {\bf D}_{133} E_{\mbox{\DEVII{\mbox{$\frac{5}{2}$}}00000{\mathfrak{0}}}}  \ , \CR
  {\bf D}_{56}^{\; 2}  E_{\mbox{\DEVII{\mbox{$\frac{3}{2}$}}00000{\mathfrak{0}}}} = - \frac{9}{2} \mathds{1}_{56} E_{\mbox{\DEVII{\mbox{$\frac{3}{2}$}}00000{\mathfrak{0}}}}  \ . 
 \eea
The Eisenstein function in the fundamental representation is associated to the nilpotent orbit of Dynkin label \DEVII0000002, with $A_7$ Dynkin label  [$\scriptstyle \mathfrak{0}\mathfrak{2}\mathfrak{0}\mathfrak{0}\mathfrak{0}\mathfrak{0}\mathfrak{0}$] and its  conjugate, and satisfies in general to the differential equation 
\be  {\bf D}_{133}^{\; 3} E_{\mbox{\DEVII000000{s}}}  = s(s-9) {\bf D}_{133} E_{\mbox{\DEVII000000{s}}}  \ , \qquad \Delta E_{\mbox{\DEVII000000{s}}}  = 3s(s-9) E_{\mbox{\DEVII000000{s}}}  \ .  \ee
The function moreover satisfies to highest weight representations constraints for integral $s$, and is associated to lower dimensional nilpotent orbits for $s=2$ and $4$. The relation of these Eisenstein functions with nilpotent orbits can be summarised in the following truncated closure diagram \cite{E7Djo}
\begin{figure}[htbp]
\begin{center}
 \begin{tikzpicture}
 \draw (\xmin,\ymin) node{\textbullet};
 \draw (\xmin,\ymin - 1) node{\textbullet};
  \draw (\xmin,\ymin - 2)  node{\textbullet};
  \draw (\xmin,\ymin + 2)  node{\textbullet};
  \draw (\xmin,\ymin + 4)  node{\textbullet};
  \draw (\xmin + 1,\ymin + 2.5)  node{\textbullet};
   \draw (\xmin - 1,\ymin + 3.5)  node{\textbullet};
    \draw (\xmin - 1 ,\ymin + 1) node{\textbullet};
   
  \draw (\xmin + 0.9,\ymin - 2.2) node{};
  \draw (\xmin - 0.8,\ymin - 1.2) node{$E_{\mbox{\DEVII0000002}}$}; \draw (\xmin + 0.9,\ymin - 1.2) node{$E_{\mbox{\DEVII{\mbox{$\frac{3}{2}$}}00000{\mathfrak{0}}}} $};

  \draw (\xmin - 0.8,\ymin - 0.2) node{$E_{\mbox{\DEVII0000004}}$}; \draw (\xmin + 0.9,\ymin - 0.2) node{$E_{\mbox{\DEVII{\mbox{$\frac{5}{2}$}}00000{\mathfrak{0}}}} $};
  \draw (\xmin - 1.8,\ymin + 1) node{$E_{\mbox{\DEVII000000{\mbox{s}}}}$};
  
  \draw (\xmin + 0.9,\ymin + 1.7) node{$E_{\mbox{\DEVII{\mathnormal{4}}00000{\mathfrak{0}}}}$};
  
  \draw (\xmin + 1.9,\ymin + 2.4) node{$E_{\mbox{\DEVII{\mbox{s}}00000{\mathfrak{0}}}}$};
  
  \draw[-,draw=black,very thick](\xmin,\ymin) -- (\xmin,\ymin + 2 );
   \draw[-,draw=black,very thick](\xmin - 1,\ymin + 1) -- (\xmin - 1,\ymin + 3.5);
    \draw[-,draw=black,very thick](\xmin - 1,\ymin + 3.5) -- (\xmin,\ymin + 2);
    \draw[-,draw=black,very thick](\xmin,\ymin + 2) -- (\xmin + 1,\ymin + 2.5);
    \draw[-,draw=black,very thick](\xmin + 1,\ymin + 2.5) -- (\xmin,\ymin + 4);
    \draw[-,draw=black,very thick](\xmin - 1,\ymin + 3.5) -- (\xmin,\ymin + 4);
     \draw[dashed,draw=black,very thick](\xmin,\ymin + 4) -- (\xmin,\ymin + 4.5);
  \draw[-,draw=black,very thick](\xmin,\ymin) -- (\xmin - 1,\ymin + 1);
\draw[-,draw=black,very thick] (\xmin,\ymin - 1) -- (\xmin,\ymin);
\draw[-,draw=black,very thick] (\xmin,\ymin - 2) -- (\xmin,\ymin - 1);
\draw[<-,draw=black,thick] (\xmin - 3-1,\ymin + 5) -- (\xmin - 3 - 1,\ymin - 2);
\draw (\xmin - 3 + 0.2 - 1,\ymin - 2) node{$0$};
\draw (\xmin - 3- 1,\ymin - 2) node{-};
\draw (\xmin - 3- 1,\ymin - 1) node{-};
\draw (\xmin - 3- 1,\ymin) node{-};
\draw (\xmin - 3- 1,\ymin + 2) node{-};
\draw (\xmin - 3- 1,\ymin + 1) node{-};
\draw (\xmin - 3- 1,\ymin + 2.5) node{-};
\draw (\xmin - 3- 1 ,\ymin + 3.5) node{-};
\draw (\xmin - 3 - 1,\ymin + 4) node{-};
\draw (\xmin - 3 + 0.3 - 1,\ymin - 1) node{$34$};
\draw (\xmin - 3 + 0.3 - 1,\ymin) node{$52$};
\draw (\xmin - 3 + 0.3 - 1,\ymin + 1) node{$54$};
\draw (\xmin - 3 + 0.3 - 1,\ymin + 2) node{$64$};
\draw (\xmin - 3 + 0.3 - 1,\ymin + 2.5) node{$66$};
\draw (\xmin - 3 + 0.3 - 1,\ymin + 3.5) node{$70$};
\draw (\xmin - 3 + 0.3 - 1,\ymin + 4) node{$76$};
\draw (\xmin - 3 + 0.5 - 1,\ymin + 5) node{dim};
\end{tikzpicture}
\end{center}
\caption{\small Nilpotent orbits associated to Eisenstein series in the $E_{7(7)}$ closure diagram}
\label{ClosureDiag}
\end{figure}
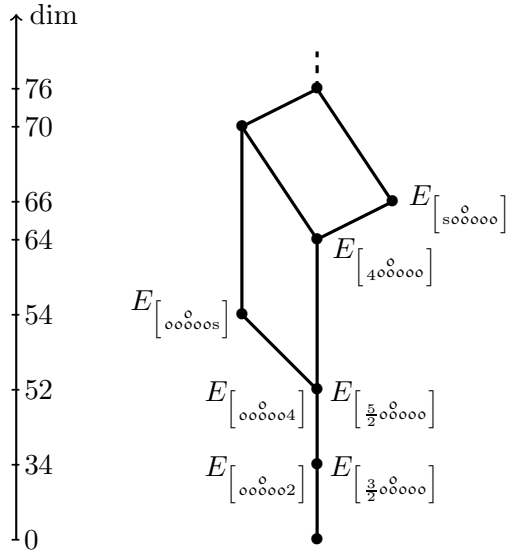
 \addtocontents{toc}{\protect\setcounter{tocdepth}{1}}
\subsection{$E_{6(6)}$}
\addtocontents{toc}{\protect\setcounter{tocdepth}{2}}
The Eisenstein function in the adjoint representation is associated to the nilpotent orbit of Dynkin label \DEVI02000{\mathfrak{0}}, with $C_4$ Dynkin label $[2,0,0,0]$ \cite{E6Djo}, and satisfies in general to the differential equations 
\be   {\bf D}_{27}^{\; 3}  E_{\text{\DEVI0{\mathnormal{s}}000{\mathfrak{0}}}} =  \frac{1}{2} (s-5) (2 s-1)     {\bf D}_{27} E_{\text{\DEVI0{\mathnormal{s}}000{\mathfrak{0}}}} \ , \qquad \Delta   E_{\text{\DEVI0{\mathnormal{s}}000{\mathfrak{0}}}} = 2s(2s-11) E_{\text{\DEVI0{\mathnormal{s}}000{\mathfrak{0}}}} \ .  \ee
The function is associated to the next nilpotent orbit in the closure diagram for $s=\frac{5}{2}$, to the next to minimal nilpotent orbit for $s=\frac{3}{2}$ and to the minimal nilpotent orbit for $s=1$. However, there is a 1-parameter family of Eisenstein functions associated to the  next to minimal nilpotent orbit. It is the Eisenstein function in the fundamental representation, that satisfies to 
\be \label{E1}\Scal{   {\bf D}_{27}^{\; 3}  - \Scal{ \frac{2s(s-6)}{3} + \frac{5}{2}}  {\bf D}_{27} } E_{\text{\DEVI{\mathnormal{s}}0000{\mathfrak{0}}}}  = (3-s) \Scal{  {\bf D}_{27}^{\; 2} - \frac{8}{27} s (s-6) \mathds{1}_{27} } E_{\text{\DEVI{\mathnormal{s}}0000{\mathfrak{0}}}}   \ , \ee
and its third order derivative restricted to the $[2,0,0,1]$ of $Sp(4)$ vanishes. It is functionally related to the Eisenstein function in the anti-fundamental representation at $6-s$, and reduces to the unique Eisenstein function associated to the minimal nilpotent orbit at $s=\frac{3}{2}$. 
 \addtocontents{toc}{\protect\setcounter{tocdepth}{1}}
\subsection{$SO(5,5)$}
\addtocontents{toc}{\protect\setcounter{tocdepth}{2}}
The Eisenstein function in the adjoint representation is associated to the nilpotent orbit of Dynkin label \DSOX{0}{2}000, with $C_2\times C_2$ Dynkin label $[2,0]\times[0,0]$ and $[0,0]\times[2,0]$, and satisfies in general to the differential equations 
\bea {\bf D}_{16}^{\; 3} E_{\mbox{\DSOX{0}{\mathnormal{s}}000}}&=&  \frac{2s(2s-7)+3}{4} {\bf D}_{16} E_{\mbox{\DSOX{0}{\mathnormal{s}}000}} \ ,\qquad   {\bf D}_{10}^{\; 3}  E_{\mbox{\DSOX{0}{\mathnormal{s}}000}}= \frac{s(2s-7)+3}{2} {\bf D}_{10}E_{\mbox{\DSOX{0}{\mathnormal{s}}000}}\ , \CR
\Delta  E_{\mbox{\DSOX{0}{\mathnormal{s}}000}}&=&2s(2s-7) E_{\mbox{\DSOX{0}{\mathnormal{s}}000}}\ . \eea
The function is associated to lower dimensional nilpotent orbits for $s=\frac{3}{2},\, 1,\, \frac{1}{2}$. The Eisenstein function in the Weyl spinor representation is associated to the largest next to minimal nilpotent orbit, and satisfies in general to the differential equations 
\bea \biggl( {\bf D}_{16}^{\; 3}-\frac{13s(s-4)+24}{16} {\bf D}_{16} \biggr)  E_{\mbox{\DSOX{0}000{\mathnormal{s}}}} &=& - \frac{3(s-2)}{4} \Scal{  {\bf D}_{16}^{\; 2}  -  \frac{5s(s-4)}{16} \mathds{1}_{16} } E_{\mbox{\DSOX{0}000{\mathnormal{s}}}} \ ,  \CR
 {\bf D}_{10}^{\; 2}  E_{\mbox{\DSOX{0}000{\mathnormal{s}}}} &=& \frac{s(s-4)}{4} \mathds{1}_{10}  E_{\mbox{\DSOX{0}000{\mathnormal{s}}}}  \ . \eea
It is functionally related to the Eisenstein function in the conjugate representation at $4-s$. The Eisenstein function in the vector representation  is associated to the smallest next to minimal nilpotent orbit, and satisfies in general to the differential equations 
\be {\bf D}_{16}^{\; 2} E_{\mbox{\DSOX{\hspace{-0.5mm}{\mathnormal{s}}}0000}} = \frac{s(s-4)}{4} \mathds{1}_{16}  E_{\mbox{\DSOX{\hspace{-0.5mm}{\mathnormal{s}}}0000}} \   , \qquad  {\bf D}_{10}^{\; 3} E_{\mbox{\DSOX{\hspace{-0.5mm}{\mathnormal{s}}}0000}} = (s-1)(s-3) {\bf D}_{10}   E_{\mbox{\DSOX{\hspace{-0.5mm}{\mathnormal{s}}}0000}} \ . \ee
 \addtocontents{toc}{\protect\setcounter{tocdepth}{1}}
\subsection{$SL(5)$}
\addtocontents{toc}{\protect\setcounter{tocdepth}{2}}
\label{SL5Review} 
The Eisenstein function in the adjoint representation is associated to the nilpotent orbit of weighted Dynkin diagram $[2,0,0,2]$, and depends on two parameters. Weyl group symmetry implies functional relations between the functions 
\be E_{[s,t,0,0]}  \propto E_{[1-s,s+t-\stfrac{1}{2},0,0]} \propto  E_{[t,0,0,\stfrac{5}{2}-s-t]} \propto E_{[s+t-\stfrac{1}{2},0,0,2-t]} \ , \ee 
and the former satisfies to the differential equations 
\bea  &&  2\cD^{[i}{}_{[k} \cD^{j]}{}_{p} \cD^{p}{}_{l]}  E_{\scriptscriptstyle [ s,t,0,0]}  + \frac{2s+4t-5}{10} \cD^{[i}{}_{[k} \cD^{j]}{}_{l]}E_{\scriptscriptstyle [ s,t,0,0]}  \CR
&=& \delta^{[i}_{[k} \biggl(  \frac{2s+4t-5}{5} \cD^{j]}{}_p \cD^p{}_{l]}  + \Scal{  \Scal{ \frac{2s+4t-5}{5}}^2  - \frac{3}{4} } \cD^{j]}{}_{l]} \biggr . \CR
&& \hspace{40mm} \biggl .  - 3 \frac{2s+4t-5}{40} \Scal{\Scal{ \frac{2s+4t-5}{5}}^2  -1}  \delta_{l]}^{j]} \biggr)  E_{\scriptscriptstyle [ s,t,0,0]}  \ , \CR
&&\biggl( \cD^i{}_k \cD^k{}_l \cD^l{}_j  + \frac{2s+4t-5}{5} \cD^i{}_k \cD^k{}_j -\Scal{ \frac{3( 2s+4t-5)^2}{400} + \frac{2s^2-2s-3}{8}} \cD^i{}_j\biggr)  \,   E_{\scriptscriptstyle [ s,t,0,0]}  \CR
&=&  \frac{2s+4t-5}{160} \Scal{ \frac{9( 2s+4t-5)^2}{25} - 4 s^2 + 4 s-9} \delta^i_j\,   E_{\scriptscriptstyle [ s,t,0,0]} \ , \label{st00}
\eea
as well as to the Laplace equation
 \be
\Delta E_{\scriptscriptstyle [s,t,0,0]}  = \Scal{ s(s-1) + \frac{3}{20} (2s+4t-5)^2-\frac{15}{4}}  E_{\scriptscriptstyle [s,t,0,0]}  \ . 
 \ee
The antisymmetric tensor Eisenstein function is associated to the next to minimal nilpotent orbit, and satisfies to the differential equation
\be \cD_i{}^k \cD_k{}^j \,  E_{\scriptscriptstyle [ 0,0,s,0]}  =   \frac{4s-5}{20} \cD_i{}^j  \,  E_{\scriptscriptstyle [ 0,0,s,0]}  + \frac{3s(2s-5)}{25}  \delta_i^j \,  E_{\scriptscriptstyle [ 0,0,s,0]} \ , \label{0010} \ee
whereas the vector Eisenstein function is associated to the minimal nilpotent orbit, and satisfies to both 
\bea \cD_i{}^k \cD_k{}^j \, E_{\scriptscriptstyle [ s,0,0,0]}  &=&- \frac{3(4s-5)}{20} \cD_i{}^j  \,  E_{\scriptscriptstyle [ s,0,0,0]}  + \frac{2s(2s-5)}{25}  \delta_i^j \,  E_{\scriptscriptstyle [ s,0,0,0]} \ ,  \CR
\cD_{[i}{}^{[k} \cD_{j]}{}^{l]} \, E_{\scriptscriptstyle [ s,0,0,0]} &=&  \frac{4s-5}{10}  \delta_{[i}^{[k} \cD_{j]}{}^{l]} \, E_{\scriptscriptstyle [ s,0,0,0]} - \frac{s(2s-5)}{50} \delta_{ij}^{kl} \, E_{\scriptscriptstyle [ s,0,0,0]} \label{1000}\ .  \eea
 \addtocontents{toc}{\protect\setcounter{tocdepth}{1}}
\subsection{$SL(3)$}
\addtocontents{toc}{\protect\setcounter{tocdepth}{2}}
\label{SL3Equa}
There are only two nilpotent orbits of $SL(3)$, the general Eisenstein function satisfies to 
\bea 
{\bf D}_3^{\; 3} E_{[s,t]} &=& \frac{2 s^2+(s+t) (2 t-3)+ \frac{3}{8}}{6}     {\bf D}_3 E_{[s,t]} + \frac{(s - t) (4 s+2 t-3) (2 s+4 t-3)}{108}  \mathds{1}_3 E_{[s,t]}\CR
\Delta  E_{[s,t]}  &=&   \frac{2\scal{2 s^2+(s+t) (2 t-3)}}{3} E_{[s,t]} \ ,
\eea
and the Eisenstein function associated to the minimal nilpotent orbit satisfies 
\be {\bf D}_3^{\; 2}  E_{[s,0]} = - \frac{4 s -3}{12} {\bf D}_3 E_{[s,0]}  + \frac{s ( 2 s - 3)}{9}  \mathds{1}_3  E_{[s,0]} \ . \ee
\section{Some additional computations on Eisenstein series} 
\subsection{$E_{6(6)}$ Eisenstein series in the fundamental representation}
\label{E66Eis} 
In the decompactification limit, the series definition \eqref{E56s} of the Eisenstein series ${E}{\mbox{\DEVII000000s}}$ as a sum over the rank 1 charges in the ${\bf 56}$ or $E_{7(7)}$, decomposes into the four components \eqref{56E6}  $p^{0}, p^{I}, q_{I}, q_{0}$ of grad $-3, -1, 1, 3$ respectively, with the rank one constraint 
\be \label{constrain}
\frac{1}{2} t^{I J K} q_{J} q_{K} =  q_0 p^I \ , \qquad t^{IKP} t_{JLP} \, q_K p^L - p^I q_J=  \delta^I_J q_0 p^0 \ , \qquad \frac{1}{2} t_{IJK} p^J p^K  =   p^0 q_I \ . \ee 
The $E_{7(7)}$ invariant norm then reads 
\begin{multline} 
Z(\Gamma)^2 = e^{6 \phi} \biggl( q_{0} + a^{I} q_{I} + \frac{1}{2} t_{I J K} a^{I} a^{J} p^{K} + \frac{1}{6} t_{I J L} a^{I} a^{J} a^{L} p^{0} \biggr)^{2}\\ + e^{2 \phi} \Bigl| Z\Scal{q_{I} + t_{I J K} a^{J} p^{K} + \frac{1}{2} t_{I J K} a^{J} a^{K} p^{0}} \Bigr|^{2}   + e^{- 2 \phi} \bigl| Z(p^{I} + a^{I} p^{0}) \bigr|^{2} + e^{- 6 \phi} (p^{0})^{2} \ , 
\end{multline}
where $e^{-2\phi}$ is the radius moduli, whereas $|Z(q)|^2$ now represents the $E_{6(6)}$ invariant norm. At large $e^{-2\phi}$ we will only consider the sum over the maximal weight charges $q_0, \, q_I$
\bea
&& \sum_{\substack{\Gamma \in \mathds{Z}^{56}_* \\ \Gamma \times \Gamma = 0 }}  |Z(\Gamma)|^{- 2s} \CR
&=& 2 \zeta(2 s) e^{- 6 s \phi} +  \sum_{\substack{q \in \mathds{Z}^{27}_* \\ q \times q = 0 }} \sum_{q^{0} \in \mathds{Z}}  \frac{\pi^{s}}{\Gamma(s)} \int_{0}^{\infty} \frac{dt}{t^{1 + s}} e^{- \frac{\pi}{t} \bigl( e^{6 \phi}(q_{0} + a^{I} q_{I})^{2} + e^{2 \phi} |Z(q)|^{2} \bigr)} + \dots \CR
 &=& 2 \zeta(2 s) e^{- 6 s \phi} +  \sum_{\substack{q \in \mathds{Z}^{27}_* \\ q \times q = 0 }}  \sum_{\tilde{q}^{0} \in \mathds{Z}} \frac{\pi^{s}}{\Gamma(s)} \int_{0}^{\infty} \frac{dt}{t^{\frac{1}{2} + s}} e^{- 3 \phi} e^{- \frac{\pi}{t}  e^{2 \phi} |Z(q)|^{2} - \pi t e^{-6 \phi} \tilde{q}_{0} + 2 \pi i \tilde{q}_{0} q_{I} a^{I}} +\dots \CR
&=& 2 \zeta(2 s) e^{- 6 s \phi} + \frac{\pi^{\frac{1}{2}} \Gamma(s - \frac{1}{2})}{\Gamma(s)} e^{- 2 (s + 1) \phi}  \   \sum_{\substack{q \in \mathds{Z}^{27}_* \\ q \times q = 0 }} |Z(q)|^{- 2 (s - \frac{1}{2})} +\dots
\eea
The other terms are more complicate to obtain explicitly, but they follow the same pattern such that the perturbative terms reduce to sum over the charges of grad $-1$ and $-3$ after Poisson resumation. The complete perturbative expansion in $e^{-2\phi}$ is then determined by compatibility with the Langlands functional identity to be 
\begin{multline}
E_{\mbox{\DEVII000000s}}  = 2 \zeta(2 s) e^{- 6 s \phi} +  \frac{\pi^{\frac{1}{2}} \Gamma(s - \frac{1}{2})}{\Gamma(s)} e^{- 2(s + 1) \phi} E_{\mbox{\DEVI00000{{\mbox{$s$-$\frac{1}{2}$}}}}}  \\+\frac{\pi^5 \Gamma(s-\frac{9}{2}) \Gamma(s-\frac{5}{2}) \zeta(2s-9)}{\Gamma(s-2)\Gamma(s)\zeta(2s-4)} e^{2(s-10)\phi} E_{\mbox{\DEVI{{\mbox{$s$-$\frac{5}{2}$}}}0000{\mathfrak{0}}}}   
\\+ 2 \frac{\pi^{\frac{27}{2}} \Gamma(s-\frac{17}{2}) \Gamma(s-\frac{13}{2}) \Gamma(s-\frac{9}{2}) \zeta(2s-17) \zeta(2s-13) \zeta(2s-9) }{\Gamma(s-4) \Gamma(s-2) \Gamma(s) \zeta(2s-8) \zeta(2s-4) } e^{6(s-9) \phi} \end{multline}
The generating character of the function $E_{\mbox{\DEVI00000{s}}}$ is defined in terms of the central charge of $q_I$ 
\be Z_{ij}(q) = V_{ij}{}^I q_I \ , \qquad |Z(q)|^2 = Z_{ij}(q) Z^{ij}(q) \ , \ee
and the quadratic constraint $t^{IJK} q_J q_K= 0 $ is equivalent to 
\be  Z_{ik}(q) Z^{jk}(q) = \tfrac{1}{8} \delta_i^j |Z(q)|^2 \ . \ee
The covariant derivative in tangent frame acts on the central charge as
\be \cD^{ijkl} Z_{pq}(q) = 3 \Scal{ \delta_{pq}^{[ij} Z^{kl]}(q) - \Omega^{[ij} \delta_{[p}^{k} Z_{q]}{}^{l]}(q) - \frac{1}{4} \Omega_{pq} \Omega^{[ij} Z^{kl]}(q) - \frac{1}{12}    \Omega^{[ij} \Omega^{kl]} Z_{pq}(q) } \ , \ee
such that 
\be
\cD_{i j k l} |Z(q)|^2 = \frac{1}{4} Z_{[i j} Z_{k l]} + \frac{1}{96} \O_{[i j} \O_{k l]}  |Z(q)|^2 \ . 
\ee
One computes that 
\be \label{DonZs}
\cD_{i j k l} |Z(q)|^{-2s} = s \biggl(\frac{1}{4} Z(q)_{[i j} Z(q)_{k l]} + \frac{1}{96} \O_{[i j} \O_{k l]}  |Z(q)|^2\biggr) |Z(q)|^{-2 (s + 1)} \ , 
\ee
and
\begin{multline}  \label{D2onZs}
\cD^{i j p q}\cD^{k l}{}_{p q} |Z(q)|^{-2s} = \biggl( - \frac{2}{3} s (3 - 2 s) \bigl( Z(q)^{i j} Z(q)^{k l} - Z(q)^{i [k} Z(q)^{l] j} \bigr) \hspace{2 cm} \\
 + \frac{1}{36} s (63 - 10 s) \O^{i [k} \O^{l] j} |Z(q)|^2 + \frac{1}{36} s (9 - 2 s) \O^{i j} \O^{k l} |Z(q)|^2  \biggr) |Z(q)|^{-2 (s + 1)} \ . 
\end{multline} 
The right-hand-side decomposes into an $Sp(4)$ singlet and a tensor in the irreducible represention $[0,2,0,0]$\be
 Z(q)^{i j} Z(q)^{k l} \bigr|_{[0,2,0,0]} = \frac{1}{2} \Scal{  Z(q)^{i j} Z(q)^{k l} - Z(q)^{i [k} Z(q)^{l] j} - \frac{1}{72} \O^{i j} \O^{k l} |Z(q)|^2 +  \frac{1}{72} \O^{i [k} \O^{l] j} |Z(q)|^2}
\ee
One deduces the Laplace equation
\bea 
\Delta |Z(q)|^{-2s} = \frac{1}{3} \cD^{i j p q}\cD_{i j p q} |Z(q)|^{-2s} = \frac{8}{3} s (s-6)  |Z(q)|^{-2s} \ , 
\eea
and one gets at third order 
\bea \label{D3onZs}
&& \cD^{i j r s} \cD_{r s}{}^{p q}\cD^{k l}{}_{p q} |Z(q)|^{-2s} \\
&=& \biggl( -\frac{1}{3} s \left(8 s^2-42 s+33\right) Z(q)^{i j} Z(q)^{k l} - 
 \frac{1}{108} s \left(14 s^2-81 s+54\right) \O^{i [k} \O^{l] j} |Z(q)|^2 \CR && \quad -\frac{1}{216} \left(8 s^3-54 s^2+27 s\right) \O^{i j} \O^{k l} |Z(q)|^2 - \frac{2}{3} s \left(2 s^2-15 s+6\right) Z(q)^{i [k} Z(q)^{l] j} \biggr)  |Z(q)|^{-2 (s + 1)} \nn \ . 
\eea
The right hand side can be expressed in terms of lower oder derivatives of $|Z(q)|^{- 2 s}$ using the relations \eqref{DonZs} and \eqref{D2onZs}, such that 
\bea
&& \cD^{i j r s} \cD_{r sp q}\cD^{klp q} |Z(q)|^{-2s} \CR
&=&\left(\frac{2}{3}  s(s-6) + \frac{5}{2}\right) \cD^{i j k l} |Z(q)|^{-2s}\CR
& &\qquad + (3 - s) \biggl(\cD^{i j p q} \cD^{k l}{}_{p q}  + \frac{1}{27} \bigl( \O^{i [k} \O^{l] j} + \frac{1}{8} \O^{i j} \O^{k l} \bigr) \cD^{p q r s} \cD_{p q r s}\biggr) |Z(q)|^{-2s} \ . 
\eea
Moreover, one straightforwardly works out that the third order derivative projected to the $[2,0,0,1]$ irreducible representation of $Sp(4)$ vanishes 
\be
\cD^{3}_{[2,0,0,1]}  |Z(q)|^{-2s}  = 0 \ . 
\ee

\subsection{$SO(6,6)$ Eisenstein series}
We define first the series associated to anti-chiral spinors. The associated `central charge' is $Z_i{}^{\hat{\imath}}$ and its complex conjugate  $Z^i{}_{\hat{\imath}}$, where $i=1,\, 4$ of one $SU(4)$ factor and ${\hat{\imath}}=1,\, 4$ of the other. The rank one constraint on the spinor is the pure spinor constraint, that reads
\be Z^i{}_{\hat{k}} Z_j{}^{\hat{k}} = \stfrac{1}{4} \delta^i_j Z^k{}_{\hat{l}} Z_k{}^{\hat{l}} \ , \qquad   Z^k{}_{\hat{\imath}} Z_k{}^{\hat{\jmath}} = \tfrac{1}{4} \delta^{\hat{\jmath}}_{\hat{\imath}}  Z^k{}_{\hat{l}} Z_k{}^{\hat{l}} \ , \qquad \frac{1}{2} \varepsilon_{ijpq} Z^p{}_{\hat{k}} Z^q{}_{\hat{l}} = \frac{1}{2} \varepsilon_{\hat{k}\hat{l}\hat{p}\hat{q}} Z_i{}^{\hat{p}} Z_j{}^{\hat{q}} \ . \ee
One computes then that the covariant derivative over $SO(6,6)/(SO(6)\times SO(6))$ acts on the central charge as
\be \cD_{ij\hat{k}\hat{l}} Z^p{}_{\hat{q}} = \frac{1}{2}  \varepsilon_{\hat{k}\hat{l}\hat{q}\hat{r}}  \delta^p_{[i} Z_{j]}{}^{\hat{r}} \ , \qquad  \cD_{ij\hat{k}\hat{l}} Z_p{}^{\hat{q}} = \frac{1}{2}  \varepsilon_{ijpr}  \delta^{\hat{q}}_{[\hat{k}} Z^{r}{}_{\hat{l}]} \ . \ee
Considering a homogeneous function of $|Z|^2 = Z^i_{\hat{\jmath}} Z_i{}^{\hat{\jmath}}$, one has 
\be \cD_{ij\hat{k}\hat{l}} |Z|^{-2s} = - s\,   \varepsilon_{\hat{k}\hat{l}\hat{p}\hat{q}} Z_i{}^{\hat{p}} Z_j{}^{\hat{q}} |Z|^{-2s-2} \ , \ee
and more generally in the vector representation (note that $\cD^{ij}{}_{\hat{k}\hat{l}} = \tfrac{1}{2} \varepsilon^{ijpq} \cD_{pq\hat{k}\hat{l}}$ and etc...)
\be \cD_{ij\hat{p}\hat{q}} \cD^{kl\hat{p}\hat{q}} |Z|^{-2s} = \frac{s(s-5)}{4} \delta^{ij}_{kl} \,  |Z|^{-2s}  \ ,\ee
in the chiral spinor representation 
\bea \cD^{ip\hat{k}\hat{q}} \cD_{jp\hat{l}\hat{q}}  |Z|^{-2s}  &=& - \frac{s(s-1)}{2} Z^i{}_{\hat{l}} Z_j{}^{\hat{k}} |Z|^{-2s-2} + \frac{s(s-4)}{8} \delta^i_j \delta^{\hat{k}}_{\hat{l}} |Z|^{-2s} \ , \CR
\cD_{ip\hat{k}\hat{q}} \cD^{pr\hat{q}\hat{s}}\cD_{jr\hat{l}\hat{s}}  |Z|^{-2s}  &=& \frac{2s^2-10s + 5}{8} \cD_{ij\hat{k}\hat{l}} |Z|^{-2s}  \ ,  \eea
and in the anti-chiral representation 
\bea \cD_{ip}{}^{\hat{k}\hat{q}} \cD^{jp}{}_{\hat{l}\hat{q}} |Z|^{-2s} &=& \frac{s(s-1)}{2} Z_i{}^{\hat{k}} Z^j{}_{\hat{l}} |Z|^{-2s-2} + \frac{s(s-7)}{16} \delta_i^j \delta_{\hat{l}}^{\hat{k}} |Z|^{-2s}\ , \CR
 \cD_{ip}{}^{\hat{k}\hat{q}} \cD^{pr}{}_{\hat{q}\hat{s}} \cD_{jr}{}^{\hat{l}\hat{s}} |Z|^{-2s} &=& - \frac{3s(s-1)(s-2)}{8} Z_i{}^{\hat{k}} Z_j{}^{\hat{l}} |Z|^{-2s-2} + \frac{s^2-11s+4 }{16} \cD_{ij}{}^{\hat{k}\hat{l}} |Z|^{-2s} \ . \hspace{10mm} \eea
As in the preceding section, one can define the series 
\be  E_{\mbox{\WSOXII0000{\mathnormal{s}}0}} = \sum_{\vspace{-2mm}\begin{array}{c}\scriptstyle \vspace{-4mm}  \Lambda \in \mathds{Z}^{32} \vspace{2mm}\\\scriptstyle \Lambda \Gamma_{MN} \Lambda =0\end{array}} \scal{ Z(\Lambda)^i{}_{\hat{\jmath}} Z(\Lambda)_i{}^{\hat{\jmath}} }^{-s} \ . \label{E32s} \ee
The series only converges for $s>5$, and satisfies to 
\be E_{\mbox{\WSOXII0000{\mathnormal{s}}0}}  = \pi^\frac{15}{2} \frac{ \Gamma(s-\tfrac{9}{2}) \Gamma(s-\tfrac{7}{2}) \Gamma(s-\tfrac{5}{2}) \zeta(2s-9) \zeta(2s-7) \zeta(2s-5)}{\Gamma(s-2) \Gamma(s-1) \Gamma(s) \zeta(2s-4)\zeta(2s-2)\zeta(10-2s)} \,  E_{\mbox{\WSOXII0000{\mbox{$5$-$s$}}0}} \ . \ee 
The first critical function is $ E_{\mbox{\WSOXII000010}}$, which  solves a quadratic equation in all three fundamental representation,
\bea \cD_{ij\hat{p}\hat{q}} \cD^{kl\hat{p}\hat{q}} |Z|^{-2} &=&-  \delta^{ij}_{kl} \,  |Z|^{-2}  \ ,\CR
\cD^{ip\hat{k}\hat{q}} \cD_{jp\hat{l}\hat{q}}  |Z|^{-2}  &=& - \frac{3}{8} \delta^i_j \delta^{\hat{k}}_{\hat{l}} |Z|^{-2} \ , \CR
\cD_{ip}{}^{\hat{k}\hat{q}} \cD^{jp}{}_{\hat{l}\hat{q}} |Z|^{-2} &=& - \frac{3}{8} \delta_i^j \delta_{\hat{l}}^{\hat{k}} |Z|^{-2}\  \label{MinimalD6}  .  
\eea
and is in fact proportional to $ E_{\mbox{\WSOXII000001}}$ and $ E_{\mbox{\WSOXII200000}}$. This function is associated to the minimal unitary representation of $SO(6,6)$. The next one is $ E_{\mbox{\WSOXII000020}}$, which solves a quadratic vector equation and two cubic spinor equations 
\bea \cD_{ij\hat{p}\hat{q}} \cD^{kl\hat{p}\hat{q}} |Z|^{-4} &=&- \frac{3}{2}   \delta^{ij}_{kl}   \,  |Z|^{-4}  \ ,\CR
\cD_{ip\hat{k}\hat{q}} \cD^{pr\hat{q}\hat{s}}\cD_{jr\hat{l}\hat{s}}  |Z|^{-4}  &=& - \frac{7}{2}  \cD_{ij\hat{k}\hat{l}} |Z|^{-4}  \ , \CR
 \cD_{ip}{}^{\hat{k}\hat{q}} \cD^{pr}{}_{\hat{q}\hat{s}} \cD_{jr}{}^{\hat{l}\hat{s}} |Z|^{-4} &=& - \frac{7}{2} \cD_{ij}{}^{\hat{k}\hat{l}} |Z|^{-4} \  .  \label{NextMinimalD6} 
\eea
It is equal to  $ E_{\mbox{\WSOXII00000{\mathnormal{2}}}}$. The divergent Eisenstein series are 
\bea  E_{\mbox{\WSOXII00000{{\mathnormal{3}}\mbox{+}\epsilon}}} &=& \frac{45}{2\pi\, \epsilon }   E_{\mbox{\WSOXII00000{\mathnormal{2}}}} +   \hat{E}_{\mbox{\WSOXII00000{\mathnormal{3}}}} + \mathcal{O}(\epsilon) \ , \CR
E_{\mbox{\WSOXII00000{{\mathnormal{4}}\mbox{+}\epsilon}}} &=& \frac{14\, 175 \, \zeta(3)}{8\pi^3\, \epsilon }   E_{\mbox{\WSOXII00000{\mathnormal{1}}}} +   \hat{E}_{\mbox{\WSOXII00000{\mathnormal{4}}}} + \mathcal{O}(\epsilon) \ , \CR
E_{\mbox{\WSOXII00000{{\mathnormal{5}}\mbox{+}\epsilon}}} &=& \frac{1\, 488\, 375 \, \zeta(3)\zeta(5)}{256\pi^{5}\, \epsilon }   +   \hat{E}_{\mbox{\WSOXII00000{\mathnormal{5}}}} + \mathcal{O}(\epsilon) \ ,
\eea

We will now consider a charge $Q$ in the vector representation, satisfying $\langle Q,Q\rangle =0$. In this case it is convenient to use vector indices, with the definition 
\be \cD_{ij\hat{k}\hat{l}} = \frac{1}{4} \gamma^a{}_{ij} \gamma^{\hat{b}}{}_{\hat{k}\hat{l}} \cD_{a\hat{b}} \ . \ee
The real `central charges' $Z_a$ and $Z_{\hat{a}}$  then satisfy to the constraint 
\be Z_a Z^a = Z_{\hat{a}} Z^{\hat{a}} \ , \ee
and 
\be \cD_{a\hat{b}} Z_c = \frac{1}{2} \delta_{ac} Z_{\hat{b}} \ , \qquad  \cD_{a\hat{b}} Z_{\hat{c}} = \frac{1}{2} \delta_{\hat{b}\hat{c}} Z_{a} \  . \ee 
One computes then that a homogeneous function of $Z_a Z^a$ satisfies to 
\bea \cD_{a\hat{b}} (Z_c Z^c)^{-s} &=& -  s Z_a Z_{\hat{b}} (Z_c Z^c)^{-s} \ , \CR
\cD_{[a}{}^{[\hat{c}} \cD_{b]}{}^{\hat{d}]}  (Z_e Z^e)^{-s} &=& 0 \ , \CR
\cD_a{}^{\hat{c}} \cD_{b \hat{c}} (Z_d Z^d)^{-s} &=& s(s-2) Z_a Z_b (Z_c Z^c)^{-s-1} - \frac{s}{2} \delta_{ab} (Z_c Z^c)^{-s}\ , \CR
 \cD_{a\hat{c}} \cD^{d\hat{c}} \cD_{b \hat{d}} (Z_e Z^e)^{-s} &=& ( s^2 - 5s+5) \cD_{a\hat{b}} \, (Z_c Z^c)^{-s} \ . \label{VecD6} 
\eea
The second equation implies that this function always satisfies to a quadratic equation in the two spinor representations, whereas it only satisfies to a quadratic constraint in the vector representation for the critical value $s=2$.

As in the preceding section, one can define the series 
\be  E_{\mbox{\WSOXII{s}00000}} = \sum_{\vspace{-2mm}\begin{array}{c}\scriptstyle \vspace{-4mm} Q \in \mathds{Z}^{12} \vspace{2mm}\\\scriptstyle \langle Q,Q\rangle =0\end{array}} \scal{Z(Q)_a Z(Q)^a}^{-s} \ . \label{E12s} \ee
The series only converges for $s>5$, and satisfies to 
\be E_{\mbox{\WSOXII{\mathnormal{s}}00000}}  = \pi^{5} \frac{ \Gamma(s-\tfrac{9}{2})  \Gamma(s-\tfrac{5}{2}) \zeta(2s-9) \zeta(2s-5)}{\Gamma(s-2) \Gamma(s) \zeta(2s-4)\zeta(10-2s)} \,  E_{\mbox{\WSOXII{{\mathnormal{5}}\mbox{-}\mathnormal{s}}00000}} \ . \ee 
The divergent series are 
\bea E_{\mbox{\WSOXII{{\mathnormal{3}}\mbox{+}\epsilon}00000}} = \frac{3}{2\, \epsilon} E_{\mbox{\WSOXII{{\mathnormal{2}}}00000}} +\hat{E}_{\mbox{\WSOXII{{\mathnormal{3}}}00000}} + \mathcal{O}(\epsilon) \ , \CR
E_{\mbox{\WSOXII{{\mathnormal{5}}\mbox{+}\epsilon}00000}} = \frac{945\, \zeta(5)}{128\, \epsilon} +\hat{E}_{\mbox{\WSOXII{{\mathnormal{5}}}00000}} + \mathcal{O}(\epsilon) \ . \eea
However the function is finite at $s=4$ and 
\be E_{\mbox{\WSOXII{{\mathnormal{4}}}00000}} = \frac{15\, \zeta(3)}{2} E_{\mbox{\WSOXII{{\mathnormal{1}}}00000}}\ . \ee

\subsection{$SO(n,n)$ Eisenstein series in the adjoint}
\label{DnAdj} 
For $SO(n,n)$ the adjoint representation decomposes with respect to $SO(n)\times SO(n)$ with $a$ running from $1$ to $n$ of the first $SO(n)$ and $\hat{a}$ running from $1$ to $n$ of the second. We decompose therefore the adjoint into the coset component $X_{a\hat{b}}$ and the two antisymmetric tensors $\Lambda_{ab}$ and $\Lambda_{\hat{a}\hat{b}}$. The minimal representation is such that a charge $Q\in \mathfrak{so}(n,n)$ is nilpotent in all three fundamental representations, which reads explicitly 
\be\begin{split} \Lambda_a{}^c \Lambda_{bc} &= X_a{}^{\hat{c}} X_{b\hat{c}} \ , \\ 
\Lambda_{[ab} \Lambda_{cd]} &= 0  \ , 
\end{split}\hspace{5mm}\begin{split}
\Lambda_a{}^c X_{c\hat{b}} &= - X_a{}^{\hat{c}} \Lambda_{\hat{c}\hat{b}} \ , \\
\Lambda_{[ab} X_{c]\hat{d}} &= 0 \ , 
\end{split}\hspace{5mm}\begin{split}
\Lambda_{\hat{a}}{}^{\hat{c}} \Lambda_{\hat{b}\hat{c}} &= X^c{}_{\hat{a}} X_{c\hat{b}}\ ,  \\
X_{a[\hat{b}} \Lambda_{\hat{c}\hat{d}]} &= 0\ , 
\end{split}\hspace{5mm}\begin{split}
\Lambda_{ab} \Lambda^{\hat{a}\hat{b}} &= - 2 X_{[a}{}^{[\hat{c}} X_{b]}{}^{\hat{d}]}\ ,   \\
\Lambda_{[\hat{a}\hat{b}} \Lambda_{\hat{c}\hat{d}]} &= 0 \ . 
\end{split}  \label{AdjointSOnn} \ee 
They satisfy to 
\be \cD_{a\hat{b}} X_{c\hat{d}} = \frac{1}{2} \delta_{ac} \Lambda_{\hat{b}\hat{d}} + \frac{1}{2} \delta_{\hat{b}\hat{d}} \Lambda_{ac} \ , \qquad \cD_{a\hat{b}} \Lambda_{cd} = \delta_{a[c} X_{d]\hat{b}} \ , \qquad \cD_{a\hat{b}} \Lambda_{\hat{c}\hat{d}} = \delta_{\hat{b}[\hat{c}} X_{a|\hat{d}]} \ . 
\ee
Using this one computes 
\be \cD_{a\hat{b}} \cD^{c\hat{b}} |X|^{-2s} = s(2s-n+3) X_{a\hat{b}} X^{c\hat{b}}|X|^{-2s-2} - s\,  \delta_a^c \, |X|^{-2s} \ , \ee
such that 
\be \Delta  |X|^{-2s} \equiv 2  \cD_{a\hat{b}} \cD^{a\hat{b}} |X|^{-2s} = 2s(2s-2n +3) |X|^{-2s}  \ . \ee
Note that the case $s=\frac{n-3}{2}$ is special, and reduces then to a spinor representation Eisenstein function. In general one has still 
\be\cD_{a\hat{c}} \cD^{d \hat{c}} \cD_{d\hat{b}}|X|^{-2s} = \Scal{ \frac{s(2s-2n+3)}{2} + \frac{(n-2)(n-3)}{4}} \cD_{a\hat{b}} |X|^{-2s} \ . \ee
One computes moreover that 
\be \cD_{[a}{}^{[\hat{c}} \cD_{b]}{}^{\hat{d}]} |X|^{-2s} = s(2s-1) X_{[a}{}^{[\hat{c}} X_{b]}{}^{\hat{d}]} |X|^{-2s-2} \ , \ee
such that the representation $s=\frac{1}{2}$ is special, and then reduce to a vector representation Eisenstein function. Using representation theory, one straightforwardly check that there is no possible rank 3 antisymmetric tensor that one can write, such that 
\be  \cD_{[a}{}^{[\hat{d}} \cD_{b}{}^{\hat{d}} \cD_{c]}{}^{\hat{f}]}  |X|^{-2s} = 0 \ . \ee
This implies that in particular an equation of the type 
\be {\bf D}_{2^{n-1}}^{\; 3}  |X|^{-2s}  = a_s  {\bf D}_{2^{n-1}}  |X|^{-2s} \ , \ee
in the spinor representation, for a coefficient that can straightforwardly be fixed.

Moreover 
\be \cD_{c\hat{b}} \scal{ n X_{a\hat{d}} X^{c\hat{d}}  |X|^{-2s-2} - \delta_a^c  |X|^{-2s} } = -\frac{(n-2)(2s-n)}{2}  \Lambda_a{}^{c} X_{c\hat{b}} |X|^{-2s-2} \  , \ee
suggesting that the function 
\be E_{\alpha_2,\frac{n}{2}} = \sum_{\substack{Q\in \mathfrak{so}(n,n)\\ Q^2 = 0} } |X(Q)|^{-2s} \ee
at $s=\frac{n}{2}$ decomposes into the sum 
\be E_{\alpha_2,s} =  \cE_{\alpha_2,s} + \bar \cE_{\alpha_2,s} \ , \ee
satisfying moreover to 
\be  \cD_{a}{}^{\hat{c}} \cD_{b\hat{c}}\,  \cE_{\alpha_2,\frac{n}{2}} =  \frac{3-n}{2} \delta_{ab} \cE_{\alpha_2,\frac{n}{2}}  \ . \ee 
Similarly as for the $E_{7(7)}$ Eisenstein series in the fundamental representation, we expect this property to generalise to  $s = \frac{n}{2}+k$ for any integer $k$, such that $\cD^{2+2k} \cE_{\alpha_2,\frac{n}{2}+k}$ restricted to the symmetric rank $2+2k$ representation of $SO(n)$ vanishes. For $k=1$ on computes indeed that 
\bea &&  D^d{}_{\hat{a}} \Scal{ X_{(a}{}^{\hat{b}} X_{b|\hat{b}} X_{c}{}^{\hat{c}} X_{d)\hat{c}} |X|^{-2s-4} - \frac{4}{n+4} \delta_{(ab} X_c{}^{\hat{b}} X_{d)\hat{b}} |X|^{-2s-2} + \frac{2}{(n+2)(n+4)} \delta_{(ab} \delta_{cd)} |X|^{-2s}} \CR
&=& \frac{n(n+2-2s)}{2(n+4)} \scal{ X_{(a}{}^{\hat{b}} X_{b|\hat{b}} X_{c)}{}^{\hat{c}} \Lambda_{\hat{a}\hat{c}}  |X|^{-2s-4} -\frac{3}{n+2} \delta_{(ab} X_{c)}{}^{\hat{b}} \Lambda_{\hat{a}\hat{b}}|X|^{-2s-2}  } \ .  \eea

 \end{document}